\documentclass[preprint,5p,times,twocolumn,authoryear]{elsarticle}

\usepackage{hyperref}
\usepackage{mathptmx}
\usepackage{txfonts}
\usepackage{amsmath}
\usepackage[T1]{fontenc}
\usepackage{amssymb}	
\usepackage{newtxtext,newtxmath}
\usepackage{inputenc}

\journal{Astroparticle Physics}

\begin{document}

\begin{frontmatter}

\author{\textbf{The Cherenkov Telescope Array Consortium{~}}\fnref{fn1}}

\author[1,2]{E.O.~Angüner}
\ead{oguzhan.anguner@tubitak.gov.tr}

\author[3]{G.~Spengler}
\ead{spengler@physik.hu-berlin.de}

\author[2]{H.~Costantini}
\ead{costantini@cppm.in2p3.fr}
\author[4]{P.~Cristofari}
\author[2]{T.~Armstrong}
\author[5]{L.~Giunti}

\fntext[fn1]{The full list of authors and institutions can be found after the references.}

\affiliation[1]{country={TÜBİTAK Research Institute for Fundamental Sciences, 41470 Gebze, Kocaeli, Turkey{~}}}

\affiliation[2]{country={Aix-Marseille Université, CNRS/IN2P3, CPPM, 163 Avenue de Luminy, 13288 Marseille cedex 09, France{~}}}

\affiliation[3]{country={Department of Physics, Humboldt University Berlin, Newtonstr. 15, 12489 Berlin, Germany{~}}}

\affiliation[4]{country={LUTH, GEPI and LERMA, Observatoire de Paris, CNRS, PSL University, 5 place Jules Janssen, 92190, Meudon, France{~}}}

\affiliation[5]{country={Universit{\'e} Paris Cité, CNRS, Astroparticule et Cosmologie, F-75013 Paris, France{~}}}

\title{Sensitivity of the Cherenkov Telescope Array to spectral signatures of hadronic PeVatrons with application to Galactic Supernova Remnants}

\begin{abstract}
The local Cosmic Ray (CR) energy spectrum exhibits a spectral softening at energies around 3~PeV. Sources which are capable of accelerating hadrons to such energies are called hadronic PeVatrons. However, hadronic PeVatrons have not yet been firmly identified within the Galaxy. Several source classes, including Galactic Supernova Remnants (SNRs), have been proposed as PeVatron candidates. The potential to search for hadronic PeVatrons with the Cherenkov Telescope Array (CTA) is assessed. The focus is on the usage of very high energy $\gamma$-ray spectral signatures for the identification of PeVatrons. Assuming that SNRs can accelerate CRs up to knee energies, the number of Galactic SNRs which can be identified as PeVatrons with CTA is estimated within a model for the evolution of SNRs. Additionally, the potential of a follow-up observation strategy under moonlight conditions for PeVatron searches is investigated. Statistical methods for the identification of PeVatrons are introduced, and realistic Monte--Carlo simulations of the response of the CTA observatory to the emission spectra from hadronic PeVatrons are performed. Based on simulations of a simplified model for the evolution for SNRs, the detection of a $\gamma$-ray signal from in average 9 Galactic PeVatron SNRs is expected to result from the scan of the Galactic plane with CTA after 10 hours of exposure. CTA is also shown to have excellent potential to confirm these sources as PeVatrons in deep observations with $\mathcal{O}(100)$ hours of exposure per source.
\end{abstract}




\end{frontmatter}


\section{Introduction}
\label{intro}

The term ``PeVatron'' is now widely used to designate astrophysical accelerators which energize particles (electrons, protons, and nuclei) up to the PeV ($10^{15}$ eV) energy range. The interest in these objects is directly linked to the unsolved problem of the origin of cosmic rays (CRs) detected on Earth. More than a century of experiments have provided detailed measurements of the CR energy spectrum. For protons, accounting for $\sim90$\% of Galactic CRs, the spectrum follows a power--law in energy with an index of $\sim-2.7$ up to the "knee" at $\sim$3 PeV energies \citep{BLUMER2009}, where the index steepens to $\sim-3.0$. The ARGO$-$YBJ experiment has reported that the knee of the cosmic hydrogen and helium spectrum is measured below 1 PeV \citep{argo1PeV}. Magnetic effects can confine CRs with energies below the knee within the Galaxy \citep{magnetic_confinement}. The observation of Galactic CRs up to at least PeV energies motivates the search for their source, i.e. ``Galactic\,PeVatrons''. The search for PeVatrons has been conducted across a wide range of multi-messenger observations, from radio to X/$\gamma$ rays, as well as investigating neutrino emission from potential PeVatron candidates \citep{miroslav}. Several source classes, e.g. supernova (SN) Remnants (SNRs)~\citep{bell1978}, massive stars and stellar clusters~\citep{aharonian2019}, core--collapse SNe~\citep{tatischeff2009,bell2013,zirakashvili2016}, pulsar winds \citep{pulsarsElena1,pulsarsElena2,pulsarsClaire3}, star formation regions (SFRs)~\citep{sfr_bykov}, microquasars~\citep{microquasars} and superbubbles~\citep{higdon2003,binns2005}, have been proposed as potential PeVatrons. \\
SNRs have long been the preferred candidates since several strong arguments support the SNR hypothesis~\citep{blasi2013,blasi2019,gabici2019}. For example, the conversion of a reasonable fraction of the total explosion energy of SNRs into CRs can explain the measured CR energy density. Additionally, the detection of $\gamma$-ray emission from numerous SNRs confirms that SNRs accelerate particles efficiently and diffusive shock acceleration can somewhat account for the measured slope of the CR spectrum \citep{cristofari2021}, although the exact spectral index of particles accelerated at SNR shocks, and injected in the ISM is still a matter of active debate \citep{malkov2001,amato2006,recchia2018,dsa1,recchia2018, complexSNR1,caprioli2020, pierreSNRCR,complexSNR2}.\\ 
Galactic PeVatrons were indeed detected, whether it is the Galactic center~\citep{hessGCPeV}, the Crab Nebula~\citep{LHAASOCrab} or a population of Galactic PeVatrons recently revealed by instruments like the Large High Altitude Air Shower Observatory (LHAASO)~\citep{cao2021}, High Altitude Water Cherenkov Observatory (HAWC) ~\citep{hawc_E56TeV} or the High Energy Stereoscopic System (H.E.S.S.)~\citep{HESS1702}.
None of these detected\,PeVatrons, however, is obviously related to any SNR. The Crab nebula is now a widely accepted example of a leptonic\,PeVatron, where electrons are being accelerated to energies above 10$^{15}$~eV~\citep{CrabSpectLHAASO}. However, it is likely not responsible for the acceleration of hadronic CRs up to the knee~\citep{amato2003,amato2021}. Reasons for the non-detection of PeVatron SNR could be that the duration of the PeVatron phase is relatively brief, limited to few tens of years after the SN explosion, or that only a fraction of SNRs ($<$ 1$\%$) are PeVatrons~\citep{cristofari2020}. This would imply only a small number of active SNR PeVatrons in the Galaxy, and current instruments may, especially at energies of tens of TeV and above, not provide enough sensitivity and energy resolution to trace the spectra of these rare objects into the PeV domain.\\
However, it is also possible that SNRs do not accelerate particles up to the PeV range~\citep{lagage1983}, the maximum energy might not go above a few hundreds of TeV \citep{bell2013,schure2013,cardillo2015,cristofari2021, ccSNR1} and that other sources are the Galactic hadronic PeVatrons. This hypothesis is supported by two recent results. First, the detection of a spatial distribution of $\gamma$-rays around massive stellar clusters, which is compatible with a constant injection of CRs in time, suggests that massive stellar clusters could be major contributors to CRs~\citep{aharonian2019}. Second, the detection of Galactic PeVatrons that seem to not be associated with SNRs~\citep{cao2021}. The nature of the majority of these PeVatrons, and the details of the mechanisms at work are not yet understood. For most of the other PeVatron candidates, the discussion is still open. \\
The purpose of this work is to discuss the ability of the planned Cherenkov Telescope Array (CTA) to detect and clearly identify hadronic PeVatrons. As a consequence, the term PeVatron refers in the following always to hadronic PeVatrons, when not stated otherwise. Leptonic PeVatrons are not discussed, since the focus is on the question of where hadronic CRs are accelerated to PeV energies. Motivated by the CR proton spectrum measured on Earth, a PeVatron is, in the following, assumed to be an accelerator of protons whose energy spectrum follows a power-law up to at least 1 PeV. In particular, the spectrum of the proton population of a PeVatron must not have an energy cutoff below 1\,PeV. An important aspect of CTA is its improved angular resolution compared to the current generation of air shower arrays. This improvement will, for example, increase the ability to detect possible spatial correlations between $\gamma$-ray and molecular line emission regions, i.e. $^{12}$CO, $^{13}$CO, and neutrino emission, thereby help to decide whether or not a $\gamma$-ray signal is of hadronic origin from proton-proton (pp) interactions. While the improved angular resolution of CTA will contribute to the identification of mechanisms at work in PeVatrons, such identification is not the focus of this paper as only hadronic PeVatrons are considered.\\
The paper is structured as follows. Methods for PeVatron searches with $\gamma$-ray detectors are summarized in Sec.~\ref{section2}. They are based on the PeVatron definition given above. A statistical test to decide whether a given $\gamma$-ray source is a PeVatron or not is also introduced. General information on the CTA experiment and the data simulation and analysis in this work is provided in Sec.~\ref{section3}. The general ability of CTA to detect a spectral $\gamma$-ray energy cutoff feature and identify PeVatron sources is quantified in Sec.~\ref{section4}. The more specific scenario of SNRs PeVatrons is addressed in Sec.~\ref{snr_pevatrons}. Here, the expected number of SNR\,PeVatrons which can be detected by CTA is estimated. Section \ref{moonlight_sec} is a technical discussion of the potential of a CTA subarray to perform PeVatron candidate follow-up observations under non-standard conditions with respect to ambient light, namely moonlight observations. Finally, conclusions are summarized in Sec.~\ref{conclusions}. Detailed complementary information and discussions on the derivation of lower limits on the spectral cutoff (\ref{lower_limits}), the treatment of multiple hypothesis testing (\ref{multiple_testing_appendix}) and expected systematic uncertainties on the reconstruction of energy cutoffs (\ref{systematics}) are provided as appendices. A final appendix (\ref{newIRFs}) compares the results expected for different CTA telescope configurations.

\section{PeVatron searches with $\gamma$-ray detectors}
\label{section2}
The deflection of CRs by Galactic magnetic fields prevents the localization of PeVatrons by means of the measurement of the incoming direction of CRs on Earth. However, if target nuclei are present at or close to the accelerating site, secondary $\gamma$-rays, together with neutrinos, are generated in the interaction of accelerated CRs with these target nuclei. The study of Galactic $\gamma$-ray sources can therefore identify the location of PeVatrons.\\
Spectral models for the $\gamma$-ray emission of PeVatrons are discussed in Sec. \ref{s2_models}. A definition of a test statistic to decide whether a $\gamma$-ray source is or is not a PeVatron is proposed in Sec. \ref{s2_detection_definition}.

\subsection{Spectral models}
\label{s2_models}
The energy spectrum of a very-high-energy (VHE, E$>$ 0.1 TeV) $\gamma$-ray source is frequently modeled as a power-law with exponential cutoff (ECPL)

\begin{equation}
\label{eq2}
\Phi_\mathrm{ECPL}(E)=\phi_{0} \cdot \left(\frac{E}{E_{0}}\right)^{-\Gamma}
\cdot \exp{\left(-\lambda E \right)}\,\mathrm{.}
\end{equation}
Here, $E$ denotes the $\gamma$-ray energy, $\lambda$ is the inverse of the
$\gamma$-ray energy cutoff $E_{c,\;\gamma}$, $\Gamma$ is the spectral index and $\phi_0$ is the source flux normalization at the reference energy $E_0$. A reference energy of $E_0=1$ TeV is used in the following. A pure power-law (PL), $\Phi_\mathrm{PL}(E)$, is a special case of Eq. \ref{eq2} where $\lambda=0$.\\
A likelihood function $L(\lambda)=L(\lambda,\phi_0,\Gamma|\,\mathrm{D})$ for the parameters $\lambda,\,\phi_0,\,\Gamma$ given observed data $D$ connects spectral models and experimental data. A PL model can be discriminated from the more general ECPL model by means of the likelihood ratio test statistic

\begin{equation}
\label{ts_lambda}
\mathrm{TS}_\lambda=-2\ln\frac{\hat{L}(\lambda=0)}{\hat{L}(\lambda)}
\end{equation}
where $\hat{L}(\lambda)$ and $\hat{L}(\lambda=0)$ are, respectively, the maximum likelihood over the full parameter space $(\lambda,\,\phi_0,\,\Gamma)$, which includes all real values for $\lambda$, and the restricted space  $(\lambda=0,\,\phi_0,\,\Gamma)$. Using Wilks' theorem \citep{wilks} and a sign convention, which transforms the square-root of a $\chi^2$-distributed random variable with one degree of freedom into a standard normal distributed random variable, the asymptotic significance of a cutoff detection is calculated as

\begin{equation}
\label{S_lambda}
S_\lambda=\mathrm{sign}(\hat\lambda)\sqrt{\mathrm{TS_\lambda}}
\end{equation}
where $\hat\lambda$ is the maximum likelihood inverse cutoff parameter.\\

If target nuclei are present at or close to a PeVatron site, secondary $\gamma$-ray emission is created in interactions between target nuclei and accelerated hadrons. However, for $\gamma$-ray emission created in interactions between target nuclei and accelerated hadrons, one is primarily interested in the spectrum of the underlying proton population. Following \cite{expcutoff_ref}, the $\gamma$-ray flux is assumed to be generated by hadronic (proton) CRs with spectrum

\begin{equation}
N(E_\mathrm{P})\sim E_\mathrm{P}^{-\Gamma_\mathrm{P}}\cdot \exp(-\lambda_\mathrm{P} E_\mathrm{P})
\end{equation}
where $E_\mathrm{P}$ is the proton energy, $\Gamma_\mathrm{P}$ is the proton spectral index and $\lambda_\mathrm{p}=1/E_\mathrm{c,\; p}$ is the inverse proton energy cutoff. The generated $\gamma$-ray flux is in the following, denoted as $\Phi_\gamma(E)=\Phi_\gamma(E,\,\lambda_\mathrm{p},\,\Gamma_\mathrm{p},\phi_0)$ and calculated with the {\tt Naima} package~\citep{naima}. The flux normalization $\phi_0$ for  
$\Phi_\gamma(E,\,\lambda_\mathrm{p},\,\Gamma_\mathrm{p},\,\phi_0)$ refers to the $\gamma$-ray flux of the source at the reference energy $E_0=1$ TeV. This convention simplifies the interpretation of the instrumental sensitivity to $\gamma$-ray fluxes within hadronic emission models.\\
Similar to Eq. \ref{ts_lambda}, a test statistic

\begin{equation}
\label{TS_proton}
\mathrm{TS}_\mathrm{p}=-2\ln\frac{\hat L(\lambda_\mathrm{p}=0)}{\hat L (\lambda_\mathrm{p})}
\end{equation}
is used to calculate the asymptotic statistical significance of a cutoff in an underlying proton population,
\begin{equation}
\label{S_p}
S_\mathrm{p}=\mathrm{sign}(\hat\lambda_\mathrm{p})\sqrt{\mathrm{TS}_\mathrm{p}}\,.
\end{equation}

The derivation of upper limits on the inverse energy cutoff parameters $\lambda$
and $\lambda_\mathrm{p}$ is discussed in detail in \ref{lower_limits}.

\subsection{Detection of PeVatron sources}
\label{s2_detection_definition}
Given the PeVatron definition in Sec. \ref{intro}, it can be excluded that a source is a PeVatron if a proton energy cutoff $1/\lambda_\mathrm{p}\leq 1$ PeV is detected. Alternatively, the detection of a $\gamma$-ray energy 
cutoff $1/\lambda\leq100$ TeV can serve as a rough criterion to exclude that a given source is a PeVatron. The translation from the proton cutoff threshold  $E_{c,\;\mathrm{p}}=1/\lambda_\mathrm{p}=1$ PeV to the $\gamma$-ray cutoff $1/\lambda=100$\,TeV relies on an analysis of the contribution of $\pi^0$- and $\eta$-meson decays to the secondary $\gamma$-ray emission, which results from the interaction of hadrons with target nuclei, as discussed in \cite{kelner_aharonian}.\\
When no spectral cutoff below $1/\lambda_\mathrm{p}=1$ PeV is detected, lower limits on the energy cutoff $1/\lambda$ or $1/\lambda_\mathrm{p}$ are frequently derived in the context of PeVatron analyses. The lower limit on the spectral energy cutoff can serve two purposes. First, it quantifies the sensitivity of the analysis to an energy cutoff. Second, a lower limit above the threshold of 1 PeV might be regarded as an indication for the detection of a Pevatron. While the detection of spectral cutoffs below PeV energies and, consequently, the rejection of a PeVatron hypothesis is performed with a test at high levels of statistical significance, the confidence level (CL) for lower limits on the energy cutoff is usually much lower. For example, a $95\%$ CL lower limit of $\sim400$ TeV is derived within a hadronic emission model for the diffuse $\gamma$-ray emission from the vicinity of the Galactic Center \citep{hessGCPeV}. Even if a lower limit on the hadronic energy cutoff larger than 1 PeV could be derived \citep{PacmanGC2,HWCJ1825_134,HESS1702}, a CL which is much larger than $95\%$ would be required to claim a firm PeVatron detection. In other recent\,PeVatron analyses, the detection of a significant $\gamma$-ray flux above, for example, $100$ TeV is considered as indicator for a PeVatron source \citep{hawc_E56TeV, TibetASg_PeV, cao2021}. However, it is unclear whether the energy spectrum of this emission still follows a power-law model above energies of $100$ TeV.\\
In this work, the confirmation and rejection of the hypothesis that a $\gamma$-ray source is a PeVatron is based on a unified test. Instead of a lower limit on the hadronic energy cutoff at a predefined CL, the CL of the deviation of the energy cutoff from the threshold of 1 PeV is quantified. The method is based on the PeVatron Test Statistic (PTS)

\begin{equation}
\label{eq_PTS}
\mathrm{PTS}=-2\ln\frac{\hat
L(\lambda_\mathrm{p}=1\,\mathrm{PeV^{-1}})}{\hat L(\lambda_\mathrm{p})}\,\mathrm{,}
\end{equation}
where $\hat{L}(\lambda_\mathrm{p})=L(\hat{\lambda}_\mathrm{p})$ is the maximum likelihood over all $\lambda_\mathrm{p}$. The PTS is constructed as a likelihood ratio test and quantifies the PeVatron definition given in Sec. \ref{intro}. The null hypothesis, $\lambda_\mathrm{p}=1\,\mathrm{PeV^{-1}}$, corresponds to the threshold model which, by definition, separates PeVatron and non-PeVatron sources. Wilks' theorem assures that the  $\mathrm{PTS}$ follows a $\chi^2$\,$\text{-}$distributed random variable with one degree of freedom if the threshold model is true. Additionally, if the threshold model is true, the likelihoods for positive and negative $\Delta:=1\,\mathrm{PeV^{-1}}-\hat{\lambda_\mathrm{p}}$ are both equal to $0.5$ because the maximum likelihood estimator $\hat\lambda_\mathrm{p}$ is asymptotically unbiased.
Then, it follows that the statistic

\begin{equation}
\label{S_PTS}
S_\mathrm{PTS}=\mathrm{sign}(\Delta)\sqrt{\mathrm{PTS}}
\end{equation}
is asymptotically distributed like a standard normal random variable when the threshold model is true. Conversely, $S_\mathrm{PTS}$ can be interpreted as the asymptotic significance of the deviation from the threshold model. For $S_\mathrm{PTS}<-5$, a PeVatron source can be excluded with a CL corresponding to at least $5\sigma$. For $|S_\mathrm{PTS}|<5$, the data are insufficient to decide between the PeVatron and the non-PeVatron hypothesis. 
Finally, if $S_\mathrm{PTS}>5$, a PeVatron detection can be claimed with a CL corresponding to at least $5\sigma$ under the assumption that the detected $\gamma$-ray emission is generated in interactions of hadrons with target nuclei. This assumption must, however, be confirmed using independent measurements. Possibilities are the detection of high-energy neutrino emission or the spatial correlation of the VHE $\gamma$-ray emission with molecular line emission as a tracer of the target nuclei. Additionally, the detection of the pion-decay signature between 100\,MeV and 1\,GeV $\gamma$-ray energies, together with spectral modeling between MeV and TeV, may be considered as supporting argument. \\
In general, the PTS can be applied with two limitations. First, the overall fit quality of the spectral flux points to the spectral model must be assured, e.g. with a goodness–of–fit test. Once a good fit is achieved, one can then apply Wilks' theorem. Second, special care is necessary in the degenerate case where the significance of the energy cutoff is found to be significantly negative, e.g. (using equation~\ref{S_p}) $S_P<-5$. This degenerate case corresponds to a spectral upturn, an exponential flux enhancement instead of an exponential flux cutoff in the energy spectrum, and might indicate a problem with the data or the data model. It might also indicate a second underlying hard spectral component as it is seen in the recently published spectrum of Crab Nebula at $\sim$PeV energies \citep{CrabSpectLHAASO}. \\
The confirmation of PeVatrons with the PTS is compared to traditional measures for the characterization of PeVatrons candidates, such as the high energy $\gamma$-ray flux and the lower limit on the energy cutoff, in Sec. \ref{pts_discussion}. The PTS can also be used to test a leptonic PeVatron hypothesis when the hadronic cutoff parameter, $\lambda_\mathrm{p}$, is replaced with the corresponding parameter in a leptonic emission model. 

\section{The Cherenkov Telescope Array}
\label{section3}

Observations with the current generation of Imaging Atmospheric Cherenkov Telescopes (IACTs) such as H.E.S.S. \citep{hessCrab}, the Major Atmospheric Gamma-Ray Imaging Cherenkov (MAGIC) telescopes \citep{magicCrab} and the Very Energetic Radiation Imaging Telescope Array System (VERITAS) \citep{veritas} led to the discovery and characterization of close to two hundred Galactic and extra-galactic astrophysical sources\footnote{On 02.08.2022, 197 VHE sources were reported in TeVCat \citep{tevcat}.} of VHE $\gamma$-radiation. CTA is the next-generation IACT system \citep{science_with_cta}. It will consist of two arrays located at the southern Paranal Observatory (Chile) and northern Roque de los Muchachos Observatory (Spain), therefore it will be able to observe the entire sky. Its energy range will extend from 20 GeV to more than 200 TeV, with a sensitivity improving by an order of magnitude depending on the energy range with respect to the current IACT systems. The improvement of the sensitivity and the energy range over current IACTs is expected to lead to the discovery of many more astrophysical sources, and a better understanding of already discovered sources. The angular resolution of southern CTA array, which is expressed as the 68$\%$ containment radius of reconstructed gamma rays, is 0.06$^{\circ}$ at 1 TeV and will approach 0.02$^{\circ}$ at $\sim$100 TeV energies. Along with a large field of view reaching 5$^{\circ}$ from the center of the camera at the highest energies, and its improved energy resolution above 1 TeV of $\sim$7$\%$ \citep{cta_mc}, these characteristics make CTA ideally suited to perform large surveys and detailed PeVatron studies.

The recent discovery of ultra-high-energy (UHE, E$>$0.1~PeV) $\gamma$-ray emission from Galactic sources by particle detector arrays such as HAWC ~\citep{hawc_E56TeV}, Tibet-As-$\gamma$ ~\citep{TibetASg_PeV} and LHAASO ~\citep{cao2021}, has established the direct detection of extended air showers for the exploration of $\gamma$-ray sources above 100 TeV. As discussed in \cite{sciascio, knodlseder}, current and planned air shower arrays have a higher sensitivity than CTA at energies above a few tens of TeV. However, given the PeVatron definition discussed in Sec. \ref{intro}, the detection of $\gamma$-ray emission in the few 10~TeV to multiple 100~TeV energy range is only an indication for the presence of a PeVatron. This indication is a necessary but insufficient condition for the robust identification of a source with a PeVatron.\\

With regard to the search of Galactic PeVatrons, which is a key science project for \cite{science_with_cta}, it has been proposed that CTA acquires data with three main objectives. First, the improved angular resolution enables the search for multi-wavelength counterparts and studies of the energy-dependent source morphology. Second, CTA will be the key instrument to cover the large energy range from a few tens of GeV up to more than 200~TeV. This provides a spectral link from the VHE to the UHE range, which, by means of spectral modeling, can again help to disentangle the hadronic and leptonic nature of potential\,PeVatron candidates. Most of the existing operational ground-based $\gamma$-ray facilities are situated in the Northern hemisphere. Furthermore, there is currently no particle detector array in the Southern Hemisphere capable of efficiently measuring gamma-ray emissions with energy greater than 100 TeV. CTA will have a remarkable high energy sensitivity to scan large parts of the Galactic plane, helping in the search for PeVatrons. Synergies between CTA and upcoming particle detector array experiments, such as the Southern Wide-Field Gamma-ray Observatory (SWGO) \citep{swgo2022} and the Andes Large-area PArticle detector for Cosmic-ray physics and Astronomy (ALPACA) \citep{alpaca}, will be necessary to understand nature of PeVatron sources.

\subsection{Simulation and analysis of CTA data}
\label{cta_sim_ana}
The simulation and analysis of CTA data in this work are based on the instrument response functions (IRFs) for the full CTA south array\footnote{The IRFs for this configuration are officially named "prod3b-v2" and correspond to the 'Omega' configuration} \citep{prod3b}. For technical reasons, all used CTA IRFs assign a vanishing effective area to events with $\gamma$-ray energies larger than $160$ TeV. Therefore, e.g. estimations of integral fluxes are biased towards lower values, since contributions from energies larger than $160$ TeV are neglected.\\
While this study was being completed, the CTA Consortium published updated IRFs\footnote{The IRFs for this configuration are officially named "prod5-v0.1" and correspond to the 'Alpha' configuration.} \citep{prod5}, which consider a possible modification of the southern CTA layout configuration and, in particular, a reduction of the number of Small--Sized--Telescopes (SSTs) from 70 to 37. The effects of this change on the main results discussed below are summarized in \ref{newIRFs}.\\
The implementation of the {\tt Naima} package in the {\tt gammapy} framework \citep{gammapy, cdeil} is used to simulate leptonic and hadronic $\gamma$-ray emission processes. The $\gamma$-ray emission source model is convolved with CTA IRFs to calculate the expected $\gamma$-ray signal event distribution in spatial coordinates and energy. The morphology of extended $\gamma$-ray sources is modelled using 2D symmetric Gaussians throughout the paper, and source extensions are given as the width ($\sigma$) of the Gaussian. The possible effects of source variability is out of the scope of this paper and not taken into account. The expected background is modeled through two components. Residual CR background events, $B_\mathrm{CR}$,  are obtained from the CR background model for the CTA southern array provided in CTA IRFs. Following \cite{Remy:2021QS}, the Galactic diffuse $\gamma$-ray emission $B_\mathrm{Diffuse}$ is modeled through a template based on the {\tt DRAGON} cosmic-ray propagation code \citep{Evoli2017jcap,Evoli2018jcap} and the non-thermal $\gamma$-ray emission computed with the {\tt HERMES} code \citep{Dundovic:2021ryb}.
Binned in spatial coordinates and energy, the sum of the expectation of the signal and the background components is the expectation for the number of events detected with CTA. Simulated CTA event data are drawn from Poisson-distributed random variables around their bin-wise expectation. The assumed zenith and offset angle between the source and the pointing direction for CTA observations are $20^\circ$ and 0.7$^{\circ}$, respectively.\\
A binned 3D-likelihood analysis \citep{mohrmann2019} in the framework of {\tt gammapy} is performed in this work. Event count data are binned in two spatial and one energy dimensions. A maximum likelihood fit of the parameters of a multi-component model for the binned data is performed. The total background model $B_\Sigma$ is the sum of the background components for residual CR background events and the diffuse emission, $B_\Sigma=\alpha B_\mathrm{CR}+\beta B_\mathrm{Diffuse}$. The normalization parameters $\alpha$ and $\beta$ are optimized in the likelihood fit. The total event count model is the sum of the background model $B_\Sigma$ and a source count model. If multiple sources are considered, the source count model is the sum of all source model components. The $\gamma$-ray emission model $\Phi_S=\Phi_S(\phi_0,\vec\theta)$ has parameters $\phi_0$ and $\vec{\theta}$. The first parameter, $\phi_0$, is the flux normalization. No constraint on this parameter is applied in the likelihood fit. In particular, the best fit flux normalization can be negative. Other parameters, which describe the spatial and spectral setup, are summarized as $\vec{\theta}$.\\
The likelihood ratio test statistic

\begin{equation}
\mathrm{TS}_\mathrm{Det}=-2\ln\frac{\hat{L}(\phi_0=0,\,\vec{\theta})}{\hat{L}(\phi_0,\,\vec{\theta})}\,\mathrm{,}
\label{ts_detection}    
\end{equation}
is frequently used to test whether a source is detected or not. The statistic compares the maximum likelihood for the null hypothesis $\hat{L}(\phi_0=0,\,\vec{\theta})$, where no source is present, with the maximum likelihood for the alternative hypothesis $\hat{L}(\phi_0,\,\vec{\theta})$. If the null hypothesis is true, TS$_\mathrm{Det}$ is expected to be distributed like a $\chi^2$-distributed random variable with one degree of freedom, following \cite{wilks}. The notation $\mathrm{TS}_\mathrm{Det}^\mathrm{T}$ is used for the detection test statistic above a specific energy threshold T.

\section{Sensitivity of CTA to PeVatrons and spectral cutoff features}
\label{section4}

The ability of CTA to detect spectral cutoff features and identify PeVatron sources is quantified in this section. For both cases, "detection probability maps", which are in principle probability maps for the detection of spectral cutoff features, are derived for point-like sources. Results are initially derived for 10 h of simulated CTA data, corresponding to the point-like source equivalent exposure expected for large parts of the inner Galaxy from the CTA Galactic Plane Survey (GPS) \citep{Remy:2021QS}. Majority of the CTA GPS observations will performed from the southern site of CTA, therefore corresponding CTA south IRFs are used in the simulations. The generalization to extended sources is briefly discussed in Sec. \ref{gammaray_scdm}. Subsequently, detection probability maps for CTA for deeper observations are derived in Sec. \ref{deep_sens}.\\
The concept of detection probability maps can also be used to quantify, for example, the spectral cutoff detection probability of other $\gamma$-ray experiments. They can therefore be used to compare the respective sensitivities and, in addition, to optimize the performance of different detector configurations in the design and construction phase of an experiment.

\subsection{Spectral $\gamma$-ray cutoff detection}
\label{gammaray_scdm}
\begin{figure*}[ht!]
\centering
\includegraphics[width=15cm]{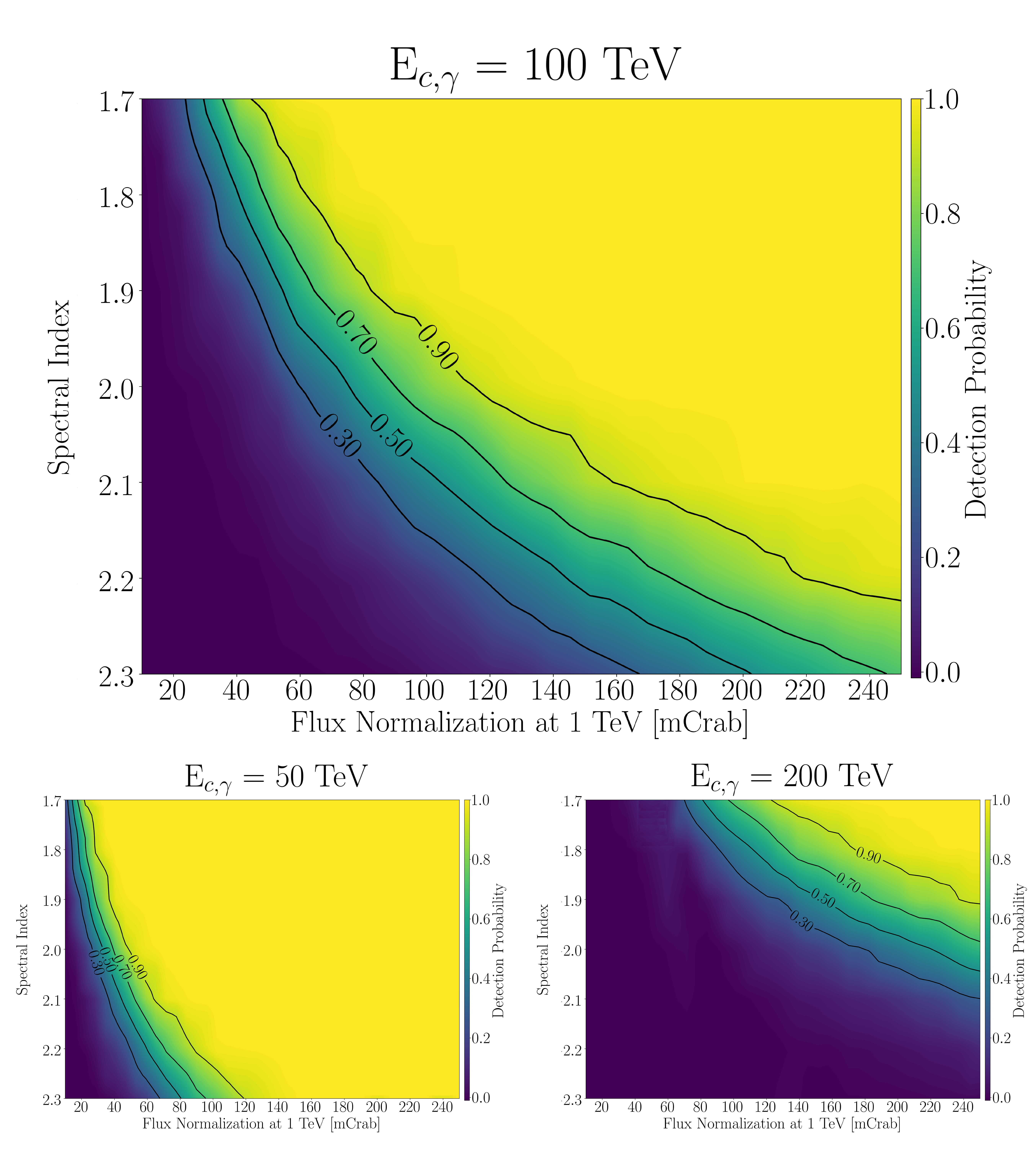}
\caption{Spectral cutoff detection probability maps for 10 h of CTA observations of point-like sources. The abscissa shows the true flux normalization, $\phi_0$, at 1 TeV and the ordinate shows the true spectral index, $\Gamma$, of an ECPL model. The maps are generated for true cutoff energies of E$_{c,\gamma}$ = 50 TeV (lower left panel), E$_{c,\gamma}$ = 100 TeV (upper panel) and E$_{c,\gamma}$ = 200 TeV (lower right panel). A $\mathrm{TS}_\lambda$ threshold of 25 is assumed. The color code shows the $\gamma$-ray spectral cutoff detection probability, while the black lines are the cutoff detection probability contours at 0.3, 0.5, 0.7 and 0.9 levels. For comparison, 100 mCrab and 200 mCrab values correspond to differential flux level at 1 TeV of 3.84~$\times$ 10$^{-12}$ cm$^{-2}$ s$^{-1}$ TeV$^{-1}$ and 7.68~$\times$ 10$^{-12}$ cm$^{-2}$ s$^{-1}$ TeV$^{-1}$, respectively.}
\label{gamma_spectral_cutoff_maps}
\end{figure*}

\begin{figure}[ht!]
\includegraphics[width=\hsize]{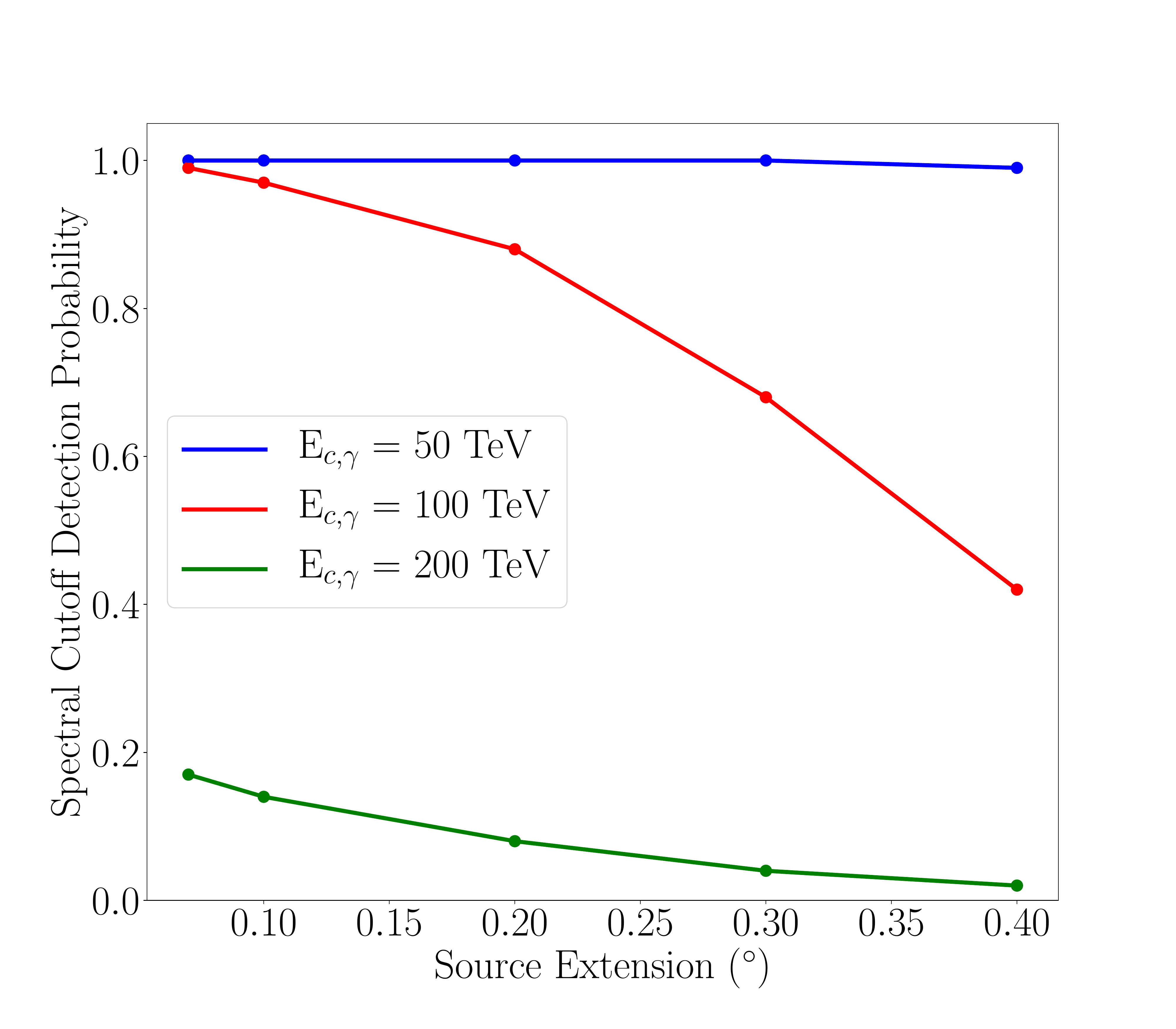}
\caption{Spectral cutoff detection probability as a function of the true source extension. The spectral cutoff detection probability is obtained from MC simulations of $\gamma$-ray ECPL spectral models with flux normalization $\phi_{0}$ (1 TeV) = 150 mCrab, spectral index $\Gamma$ = 2.1 and cutoff energies E$_{c,\gamma}$ = 50 TeV (blue), E$_{c,\gamma}$ = 100 TeV (red) and E$_{c,\gamma}$ = 200 TeV (green) for 10 h exposure.}
\label{extension_scdm}
\end{figure}

Given a $\gamma$-ray source, the ability of CTA to detect spectral cutoffs depends on the source properties, such as the flux normalization and the spectral index. A probability map for the detection of spectral cutoff features illustrates this relationship between spectral parameters for point-like sources and the probability for the detection of spectral cutoffs. The axes are defined as the true flux normalization at 1 TeV ($\phi_{0}$, abscissa) and the true spectral index ($\Gamma$, ordinate) of $\gamma$-rays. The color code of the maps shows the detection probability, i.e. the expected fraction of sources for which TS$_\lambda$ is above a threshold. A $\mathrm{TS}_\lambda$ threshold of 25, corresponding to $5\sigma$, is applied in the following.\\
The $\gamma$-ray spectral cutoff detection maps for three different true cutoff energy values, E$_{c,\gamma}$ = 50~TeV, 100~TeV and 200~TeV, covering the range between $\phi_{0}$ = [10, 250] mCrab\footnote{Throughout the paper, Crab unit is assumed as the differential Crab flux at 1 TeV of 3.84~$\times$ 10$^{-11}$ cm$^{-2}$ s$^{-1}$ TeV$^{-1}$, taken from Table 6 of \cite{hess_crab}.} and $\Gamma$\,=\,[1.7,\,2.3], are presented in Fig. \ref{gamma_spectral_cutoff_maps}. To average out the effect of Galactic diffuse emission, source simulations are performed at different Galactic coordinates. Galactic latitude coordinates are randomly generated between [-0.5$^{\circ}$, 0.5$^{\circ}$], while the Galactic longitude coordinates are randomly generated between [5.0$^{\circ}$, 60.0$^{\circ}$] and [$-$5.0$^{\circ}$,\,$-$60.0$^{\circ}$]. This range of Galactic coordinates is compatible with the inner Galactic region, for which the CTA observation time will be significantly larger than for other regions \citep{science_with_cta}.\\
Figure \ref{gamma_spectral_cutoff_maps} shows that the cutoff detection probability is strongly dependent on the spectral index and the brightness of the source. The $70\%$ detection probability contour lines indicated in Fig. \ref{gamma_spectral_cutoff_maps} provide a reasonable threshold for sources where a spectral cutoff will likely be detected with 10 h of CTA data. For example, the detection of a spectral cutoff at 50 TeV is feasible for the majority of point-like sources with flux normalization $\phi_{0}$\,$\geq$\,100\,mCrab, given 10 h of observational data from CTA. However, given the same observation time, the detection of a spectral cutoff at 100 TeV or larger is only possible for bright sources with a hard spectral index.\\
CTA sensitivity degrades for extended sources. To quantify this effect, sources are simulated with an intrinsic Gaussian angular extension between 0.1$^{\circ}$ and 0.4$^{\circ}$, and the spectral cutoff detection probability is calculated for each case. The sensitivity degradation with increasing source extension can be clearly seen in Fig. \ref{extension_scdm}. For example, while a 100 TeV $\gamma$-ray spectral cutoff from a strong point-like source with $\phi_0=150$ mCrab and spectral index of $\Gamma=2.1$ is likely to be detected given 10 h of observational data from CTA, a spectral cutoff in a source with an extension of $0.4^\circ$ and otherwise identical parameters will only be detected with a probability of $\sim$ 40$\%$.\\
The probability maps shown in Fig.~\ref{gamma_spectral_cutoff_maps} serve as a valuable tool for investigating CTA's capabilities at high energies and demonstrating its abilities to detect spectral cutoffs at high energies. However, they don't offer conclusive evidence regarding the detection of PeVatron sources. Detection of high energy cutoffs, particularly beyond 100~TeV, can be potentially linked to PeVatron detection if the gamma-ray emission originates from hadronic interactions. On the other hand, non-detection of such cutoffs, which could be due to a lack of sensitivity, does not necessarily rule out the PeVatron nature of sources. Therefore, the link between "non-detection of a cutoff" and "detection of a PeVatron" is not always straightforward. The probability maps in Fig.~\ref{gamma_spectral_cutoff_maps} facilitate discussions on the distinctions between "spectral cutoff" and "PeVatron" detection concepts. To enable statements regarding PeVatron detection, this paper introduces the concept of PTS in Sect. \ref{s2_detection_definition}, which enables conclusions about the detection or exclusion of PeVatron sources, not solely based on the detection or non-detection of $\gamma$-ray spectral cutoffs, but instead based on a specific reference value of the proton cutoff, assumed to be 1~PeV throughout this paper.

\subsection{PeVatron detection and rejection probability}
\label{pev_det_power_maps}
\begin{figure*}[ht!]
\centering
\includegraphics[width=11cm]{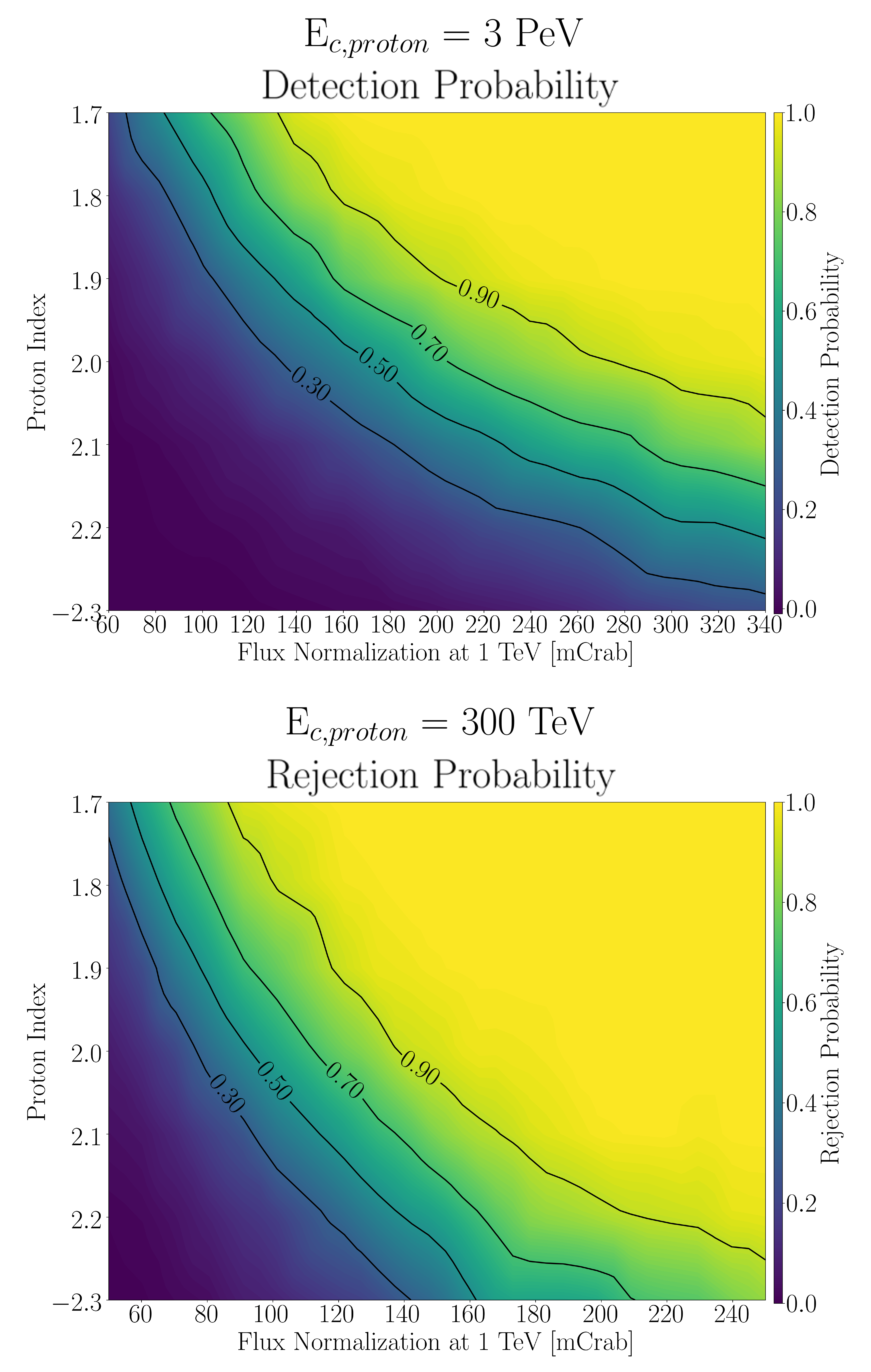}
\caption{PeVatron detection and rejection probability maps for 10 h of CTA observations of point-like sources. The abscissa shows the true observed $\gamma$-ray flux normalization at 1 TeV originating from pp interactions observed from Earth, while the ordinate shows true proton spectral index. The color code shows the PeVatron detection (rejection) probabilities, while the black lines indicate the PeVatron detection (rejection) probability contours at 0.3, 0.5, 0.7, and 0.9 levels. Upper panel: PeVatron detection (PTS$>$25) probability map for a true proton energy cutoff at 3 PeV. Lower plot: PeVatron rejection (PTS$<$-25) probability map for a true proton energy cutoff at 300 TeV. For comparison, 100 mCrab, 200 mCrab and 300 mCrab values correspond to differential flux level at 1 TeV of 3.84~$\times$ 10$^{-12}$ cm$^{-2}$ s$^{-1}$ TeV$^{-1}$, 7.68~$\times$ 10$^{-12}$ cm$^{-2}$ s$^{-1}$ TeV$^{-1}$ and 1.15~$\times$ 10$^{-11}$ cm$^{-2}$ s$^{-1}$ TeV$^{-1}$, respectively.}
\label{pts_power_maps}
\end{figure*}

Probability maps for the detection and rejection of PeVatron sources are presented in this section. Using the PTS instead of TS$_\lambda$, the maps are derived similarly to the procedure explained in Sec. \ref{gammaray_scdm}. The simulation of $\gamma$-ray spectra resulting from hadronic sources follows the discussion in Sec. \ref{cta_sim_ana}. Map axes are the true $\gamma$-ray flux normalization at 1 TeV, resulting from proton-proton (pp) interactions and observed from Earth (abscissa), and the true proton spectral index (ordinate). Since the results are shown as a function of the $\gamma$-ray flux normalization observed from Earth, the position of contour levels is independent of the source distance and target gas density values. These parameters only affect the observed $\gamma$-ray flux level, not the spectral shape of the observed $\gamma$-ray spectrum.\\
Figure \ref{pts_power_maps} shows the probability maps for the detection of\,PeVatrons with CTA with 10 h of data. The upper plot, corresponding to an intrinsic proton cutoff at 3\,PeV, shows the probability for a significant, i.e. 5$\sigma$, PeVatron detection. The lower plot, corresponding to an intrinsic proton cutoff at 300\,TeV, shows the probability to reject the PeVatron hypothesis.\\
Simulation studies of extended sources using true proton models show that the source extension has a very similar effect on probability maps for the detection of\,PeVatrons as for the detection of $\gamma$-ray spectral cutoffs, shown in Fig. \ref{extension_scdm}. Both PeVatron detection and rejection probabilities degrade with increasing source extension.

\subsection{Deep observations} 
\label{deep_sens}
\begin{figure*}[ht!]
\centering
\includegraphics[width=15cm]{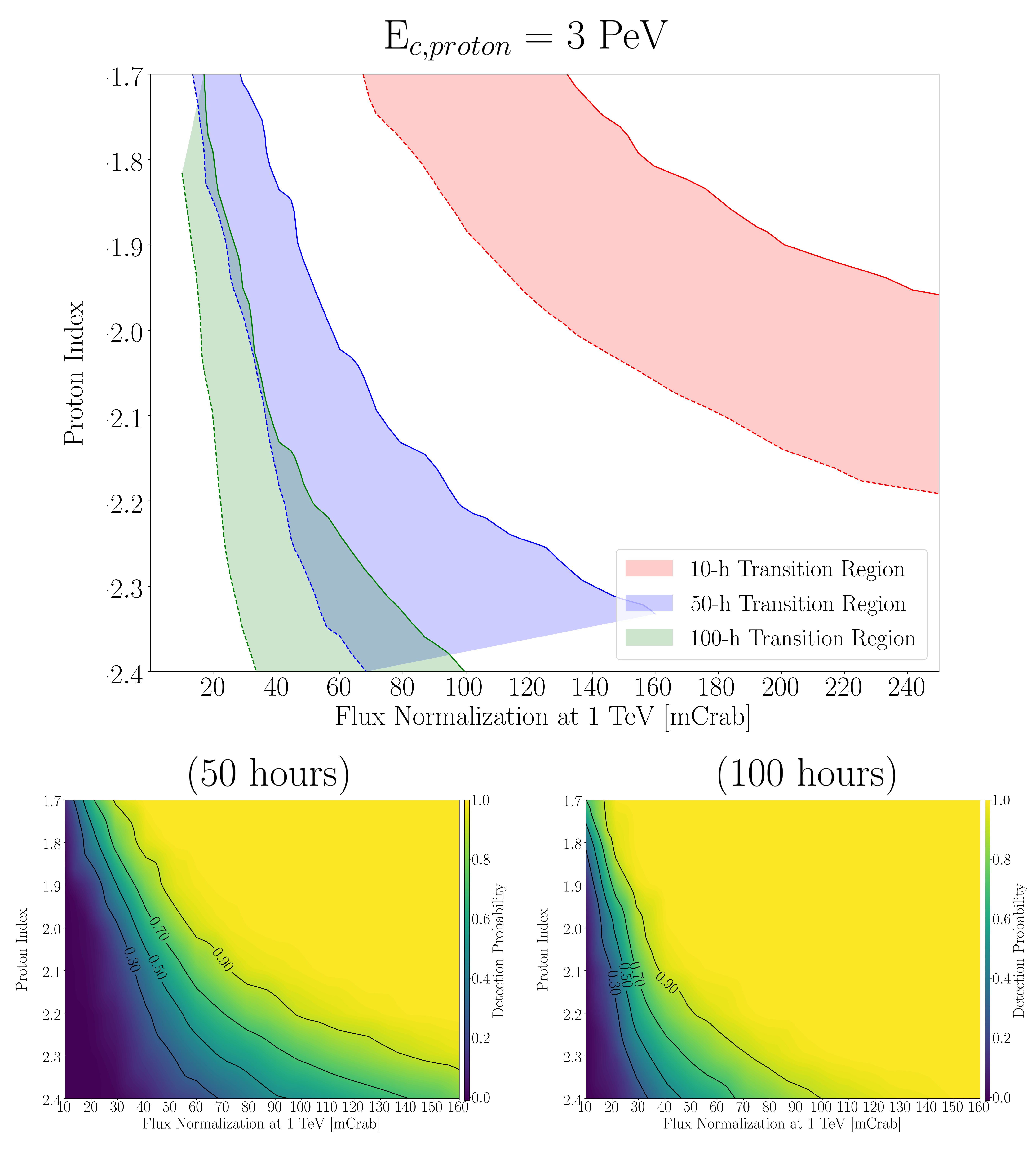}
\caption{Deep observation sensitivity of CTA to point-like PeVatrons. The top panel shows the comparison of the "transition regions", i.e. the region between the detection probability contours of $30\%$ (dashed lines) and $90\%$ (solid lines), for a PeVatron with true proton cutoff at 3 PeV for 10 h, 50 h, and 100 h of CTA data. The corresponding PeVatron detection maps for 50 h and 100 h of observations with CTA are given at the bottom left and right panels, respectively.}
\label{followup_pts_maps}
\end{figure*}

\begin{figure*}[ht!]
\centering
\includegraphics[width=15cm]{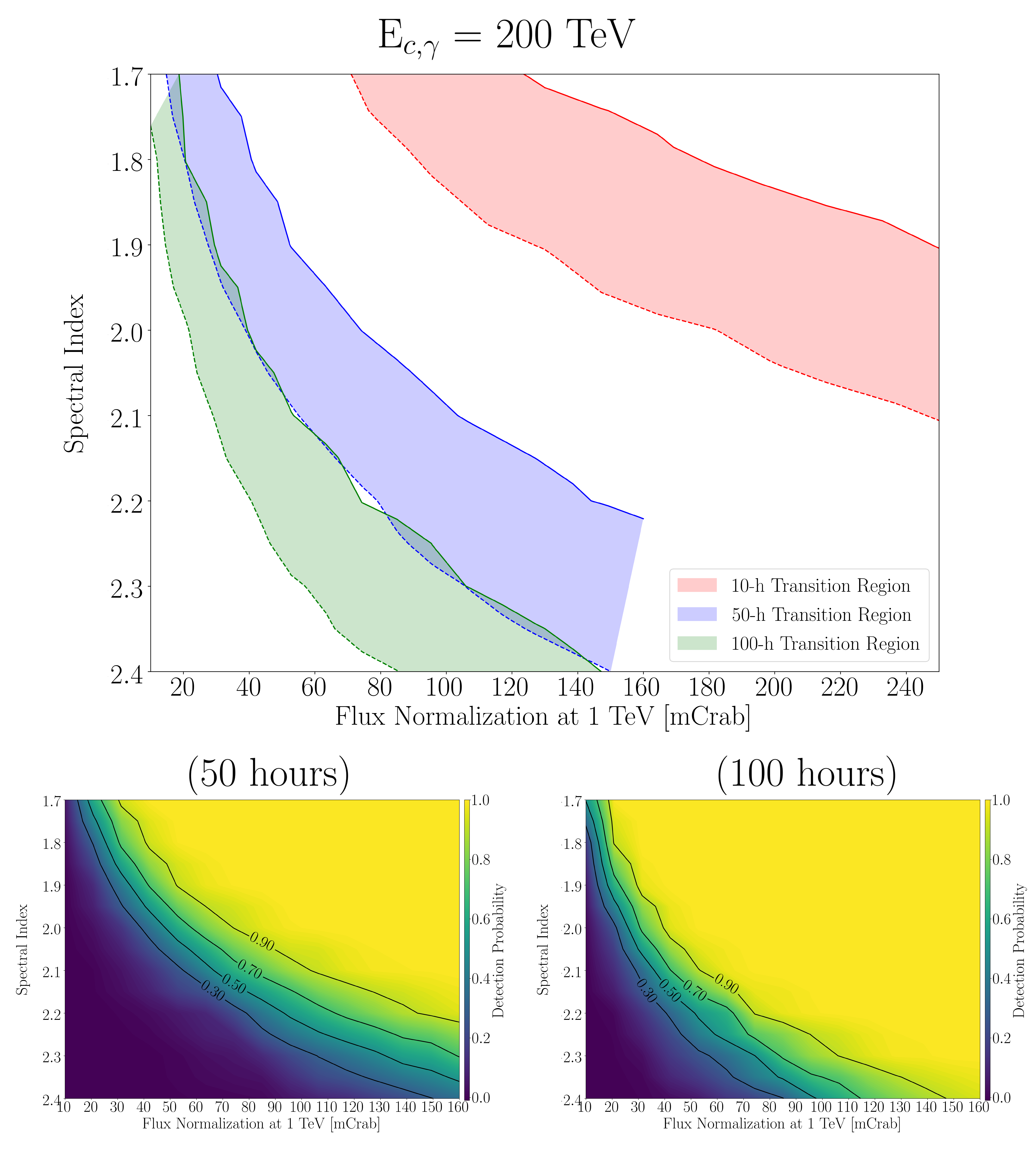}
\caption{Deep observation sensitivity of CTA to spectral cutoff features of point-like sources. As per Fig. \ref{followup_pts_maps}, the top panel shows "transition region" for the detection of a spectral cutoff with true value of $200$ TeV. The performance corresponding to 10 h, 50 h, and 100 h of CTA data is shown in red, blue and green. The corresponding spectral cutoff detection maps for 50 h and 100 h observations are given at the bottom left and right panels, respectively.}
\label{followup_scdm_maps}
\end{figure*}

The performance of CTA with respect to the detection of\,PeVatrons and spectral cutoffs increases with observation time. Figure \ref{followup_pts_maps} shows in the lower panels PeVatron detection probability maps for 50 h and 100 h of observation time with the southern CTA array to quantify the expected sensitivity. The upper panel of Fig. \ref{followup_pts_maps} shows comparison between the "transition regions", i.e. the region encompassing the $30\%$ and $90\%$\,PeVatron detection probability contours, for different CTA observation times. As discussed in Sec. \ref{pev_det_power_maps}, Fig. \ref{followup_pts_maps} shows that the expected PeVatron detection probability for $10$ h CTA data is only non-negligible for hard and bright sources. However, as shown in the upper panel of Fig. \ref{followup_pts_maps}, for deep observations with $100$ h of CTA data, almost the complete relevant parameter space with proton spectral indices $\Gamma_\mathrm{p}<2.4$ and flux normalizations larger than $40$\,mCrab can be tested. A similar result holds for the sensitivity to $\gamma$-ray spectral cutoffs and is shown in Fig.~\ref{followup_scdm_maps}. As previously stated, the CTA's energy coverage will exceed 200 TeV. With only $\mathcal{O}(10)$ hours of limited exposure, the impact of the IRFs stopping at 160 TeV does not significantly affect the detection of spectral cutoffs beyond 100 TeV for the survey data analysis. However, it is expected that this effect becomes more noticeable in deeper observations with $\mathcal{O}(100)$ hours of exposure, especially for hard ($\Gamma_{\gamma}$ $\leq$ 2.0) sources.

\section{Search for SNR PeVatrons with CTA}
\label{snr_pevatrons}

As discussed in Sec. 1, one of the leading hypotheses for the origin of PeV CRs is the acceleration by SNRs. An order of magnitude for the number of SNR PeVatrons which can be detected with CTA is estimated in this section. The estimation is based on the simulation of Pevatron populations in the framework of a simplified model for the emission of $\gamma$-rays by SNRs. While based on a simplified model, the simulation of PeVatron populations allows to test the reliability of the PeVatron search methods which are discussed in Sec. \ref{s2_detection_definition}, and the comparison to traditionally used Pevatron search methods, like the lower limit on the energy cutoff and the significance of the $\gamma$-ray emission at energies above 100~TeV. The investigation of complex morphological models for SNRs, as for example discussed in \citep{complexSNR5}, is beyond the scope of this work. The potential of CTA to detect and spatially resolve young SNRs is discussed in detail in \cite{ACERO2013276, snr_det_CTA, AlisonSNRCTA}. Due to its dependence on a simplified model for the distribution, evolution and $\gamma$-ray emission of SNRs in the Galaxy, the result for the number of Galactic SNRs which can be detected by CTA is considered only as a benchmark result which is valid only within the assumptions of the model.

\subsection{Modelling the $\gamma$-ray emission of Galactic SNR PeVatrons}
\label{sec_modeling}
The population of Galactic SNRs is simulated with a Monte Carlo approach, in which the distribution of SNe in time and space is randomly drawn in multiple samples. The method is briefly summarized below, further details can be found in~\cite{cristofari2013,cristofari2017,cristofari2018}. As detailed in \cite{cristofari2018}, the spatial distribution of SNe follows the explanation in ~\cite{lorimer2006,faucher2006} and a Galactic SNe rate of 3/century is assumed. Following ~\cite{smartt2009,ptuskin2010}, one of four types, either thermonuclear (TN), core--collapse\,high-ejecta-mass (CC-HEM), core--collapse\,low-ejecta-mass (CC-LEM) or core--collapse\,high-explosion-energy (CC-HEE) is assigned randomly to each SN with relative rates 32\% (TN), 44\% (CC-HEM), 22\% (CC-LEM) and 2\% (CC-HEE). The described population of SNRs emphasized here is a toy model, and that the diversity of objects and parameters found in nature is of course vastly greater than the one adopted in this study. Other works have for instance adopted different prescriptions~\citep{dwarkadas2005,dwarkadas2007,sarmah2022} and more complex SNR modeling. However the simple model adopted is sufficient to illustrate the statistical method presented in this paper.\\
Physical parameters for each SN type are the total explosion energy, the mass of the ejecta, the mass--loss rate of the progenitor, and the velocity of the progenitor. These parameter values are chosen for each type according to \cite{cristofari2013} and provided in Tab. \ref{snr_table}. For remnants from thermonuclear SNe, the shock is assumed to evolve in the unperturbed inter-stellar-medium (ISM) \citep{chevalier1982}. The SNR shock from core-collapse SNe \citep{bisnovatyi1995} is assumed to expand in a structured medium, which is shaped by the history of the progenitor massive star~\citep{weaver1977,cardillo2015}. During its main sequence, the wind of the massive star inflates a hot cavity with a typical temperature of $10^{6}$ K and a low density of typically $10^{-2}$ cm$^{-3}$. Close to the end of its life, the star enters a late sequence phase.
Thus, when the SN explodes, the SNR shock will successively expand through the dense wind of the progenitor for a few hundred years, and through the cavity for a few kyrs until it finally reaches the unperturbed ISM~\citep{weaver1977}.

\begin{table}
\large
\caption{The parameters used in the simulations for each SN type to compute the SNR dynamical evolution. $\epsilon_{51}$ is the total explosion explosion energy in units of 10$^{51}$ erg.  M$_{\text{ej},\odot}$ and $\dot{M}_{-5}$ are the mass of ejecta in solar masses and the wind mass-loss rate in 10$^{-5}$ M$_{\odot}$ yr$^{-1}$, respectively. The wind speed in units of 10 km s$^{-1}$ and the relative explosion rates are given in $u_{\text{w},6}$ and Rel. rate columns, respectively.
}              %
\centering                                      
\renewcommand{\arraystretch}{1.3}
\begin{tabular}{c c c c c c}          
\\
\hline\hline                        
SNR Type & $\epsilon_{51}$ & M$_{\text{ej},\odot}$ & $\dot{M}_{-5}$ & $u_{\text{w},6}$ & Rel. rate \\    
\hline                                   
    TN     & 1 & 1.4 & $-$ & $-$ & 0.32 \\      
    CC-HEM & 1 & 8   & 1   & 1   & 0.44 \\
    CC-LEM & 1 & 2   & 1   & 1   & 0.22 \\
    CC-HEE & 3 & 1   & 10  & 1   & 0.02 \\
\hline
\label{snr_table}
\end{tabular}
\end{table}

A central parameter for the calculation of the detection probability of SNR PeVatrons is the maximum energy of the accelerated particles $E_{\rm max}$, which is assumed to be equal to the cutoff energy of the respective particle spectrum. The values and temporal evolution of $E_{\rm max}$ in SNR are under current debate, see e.g. \cite{bell2013,schure2013,blasi2019,gabici2019,eMaxSNR2}. Independently of the type of SNR, in the following, it is assumed that $E_\mathrm{max}$ changes during the free expansion phase as detailed in \cite{cristofari2018} and reaches a value of $E_\mathrm{max}=3$ PeV at the transition between the free expansion and Sedov--Taylor phase. This choice of $E_\mathrm{max}=3$~PeV assures that simulated SNRs can accelerate particles up to the knee feature of the CR spectrum. For SNRs from thermonuclear SNe, $E_\mathrm{max}$ is assumed to decrease after the transition to the Sedov--Taylor while for core-collapse SNRs, a temporal increase of $E_\mathrm{max}$ is expected \citep{cristofari2018}. As detailed in \cite{cristofari2018}, highly efficient magnetic field amplification is necessary to reach this value of $E_\mathrm{max}$. However, it is still in agreement with theoretical works and current observations of SNRs~\citep{HESSSNR,Fermi15}.\\
Given the uncertainties in the modeling of $\Gamma_\mathrm{p}$, four benchmark values for the spectral index $\Gamma_\mathrm{p}$ of the proton population between $\Gamma_\mathrm{p}=2.0$ and $\Gamma_\mathrm{p}=2.3$ with steps of 0.1 are considered. For all simulated SNRs, particles are assumed to be accelerated until the end of the Sedov--Taylor phase, i.e. for typically $15-20$ kyr. A total of 50 samples of Galaxies with their SNR are simulated for each benchmark proton spectral index. In total, there are therefore 200 Galaxy samples. On average, there are 450 SNR in a simulated Galaxy.\\
The $\gamma$-ray emission from SNRs is calculated as in \cite{cristofari2013}. The stationary transport equation gives the distribution of CRs inside the SNR, and the gas continuity equation is used to infer the density profile inside the SNR. The $\gamma$-ray luminosity is then calculated as in \cite{kelner_aharonian}. To calculate the $\gamma$-ray emissivity, hadronic interactions are assumed between accelerated protons and nuclei in the ISM. Possible enhancements of the $\gamma$-ray flux due chance associations between SNRs and molecular clouds, as discussed e.g. in \cite{molecular_cloud}, are not considered. Because the focus is on the assessment of the ability of CTA to detect hadronic $\gamma$-ray emission from SNR PeVatrons, leptonic emission mechanisms \citep{snrLepton} and complex SNR spectra which can vary as a function of SN type, age and magnetohydrodynamics profile \citep{complexSNR1,complexSNR2} are also not considered.

Out of all simulated SNRs, a preselection based on the properties of simulated SNRs is performed to reduce subsequent computational efforts. Only true PeVatron SNRs, i.e. only SNRs for which the true energy cutoff of the proton population is larger
than 1 PeV, are further analyzed. Due to the preselection of true SNR PeVatrons, effects resulting from the confusion of false--positive SNR PeVatron detections are not investigated. Additionally, SNRs are only preselected when the true $\gamma$-ray flux at 1 TeV is smaller than one Crab, and the angular extension is smaller than 0.75$^{\circ}$. The motivation for this preselection is that, at an energy of 1 TeV, there are no brighter sources than 1 Crab found in the Galactic Plane and 99$\%$ of the sources detected in the HGPS have an angular extension below 0.75$^{\circ}$ \citep{hgps}. No preselection on the Galactic longitude $l$ is performed. The following discussion is focused on the detectability of SNR PeVatrons with CTA south, which can mainly observe the inner Galaxy, i.e. $|l|<60^\circ$, in which $71\%$ of the simulated SNRs are located. Averaged over all simulated SNR samples, $28$ SNRs per sample are preselected, i.e. the preselection efficiency is typically $6\%$ and depends on the true proton index.\\
Compared to the references, in particular \cite{cristofari2013,cristofari2017,cristofari2018}, the only difference in the simulation is in the choice of $E_\mathrm{max}$, the values for the rates of SNe types and the neglecting of leptonic emission processes. As discussed, the scope of the simulation is only to allow the assessment of the general ability of CTA to find SNR PeVatrons. Although quantitative results are derived in the following, only qualitative conclusions are drawn from the simplified simulation of Galactic SNR PeVatrons.

\subsection{Data analysis}
Given the radial extension and differential $\gamma$-ray flux for each preselected SNR, the $\gamma$-ray signal expected for CTA is computed as detailed in Sec. \ref{cta_sim_ana}. An observation time of $10$ h, which matches the point-like source equivalent exposure expected from the CTA GPS \citep{science_with_cta}, is assumed. As discussed in more detail in \cite{science_with_cta} and \cite{Remy:2021QS}, the aims of the CTA GPS include an unprecedented census of Galactic VHE $\gamma$-ray emitting objects through the detection of hundreds of sources. Therefore, the CTA GPS is a unique opportunity to search for Galactic PeVatrons.\\
The likelihood ratio test statistic $\mathrm{TS}_\mathrm{Det}$, as defined in Eq. \ref{ts_detection}, is calculated for each preselected SNR to test whether the $\gamma$-ray signal is detected. A Gaussian spatial model and power-law spectral model were used for each calculation of $\mathrm{TS}_\mathrm{Det}$. Eventually, the PTS, as detailed in Sec. \ref{s2_models}, is calculated to test whether a source is confirmed as PeVatron.\\
A threshold of $\mathrm{TS}_\mathrm{Det}>30$ is used for the detection of sources. This threshold is adopted from \cite{hgps} and leads to an estimated false--positive fraction of $3\%$ for the source detection in the H.E.S.S. GPS.
It is likely that the false--positive rate with this detection threshold is larger than $3\%$ for the CTA GPS, since the CTA point spread function is narrower than for the H.E.S.S. instrument, while the size of the scanned sky-patch is similar for both analyses resulting in more independent tests for the CTA GPS when compared to the H.E.S.S. GPS. The more conservative detection threshold of $\mathrm{TS}>33.9$ is used as the threshold for the PeVatron confirmation with the PTS when multiple tests are performed on the same sample, which ensures a global significance level of $5\sigma$ when $100$ independent tests (i.e. survey trial factors) are performed. In addition to confirmed PeVatrons, sources are of interest for which a strong constraint on the hadronic spectral energy cutoff can be derived. Such sources, for which the $95\%$ CL lower limit $E_\mathrm{c,\;p}^\mathrm{LL}$ on the spectral energy cutoff in a hadronic emission model is larger than 1 PeV, are in the following called PeVatron candidates. Table \ref{snr_naming} summarizes the different terms and definitions, which are used to describe the following analysis steps.\\
Depending on the true proton spectral index, typically less than $20$ independent tests are performed per simulated sample, as detailed in Tab.~\ref{snr_num_table}. However, for real CTA GPS data, the number of sources is expected to be larger than the number of detected SNRs considered here. Therefore, it is assumed that $100$ PTS tests are to be performed with CTA GPS data. More details on the treatment of multiple hypotheses tests are found in \ref{multiple_testing_appendix}.

\begin{figure*}[ht!]
\centering
\includegraphics[trim={5cm 0 0 0}, width=20cm]{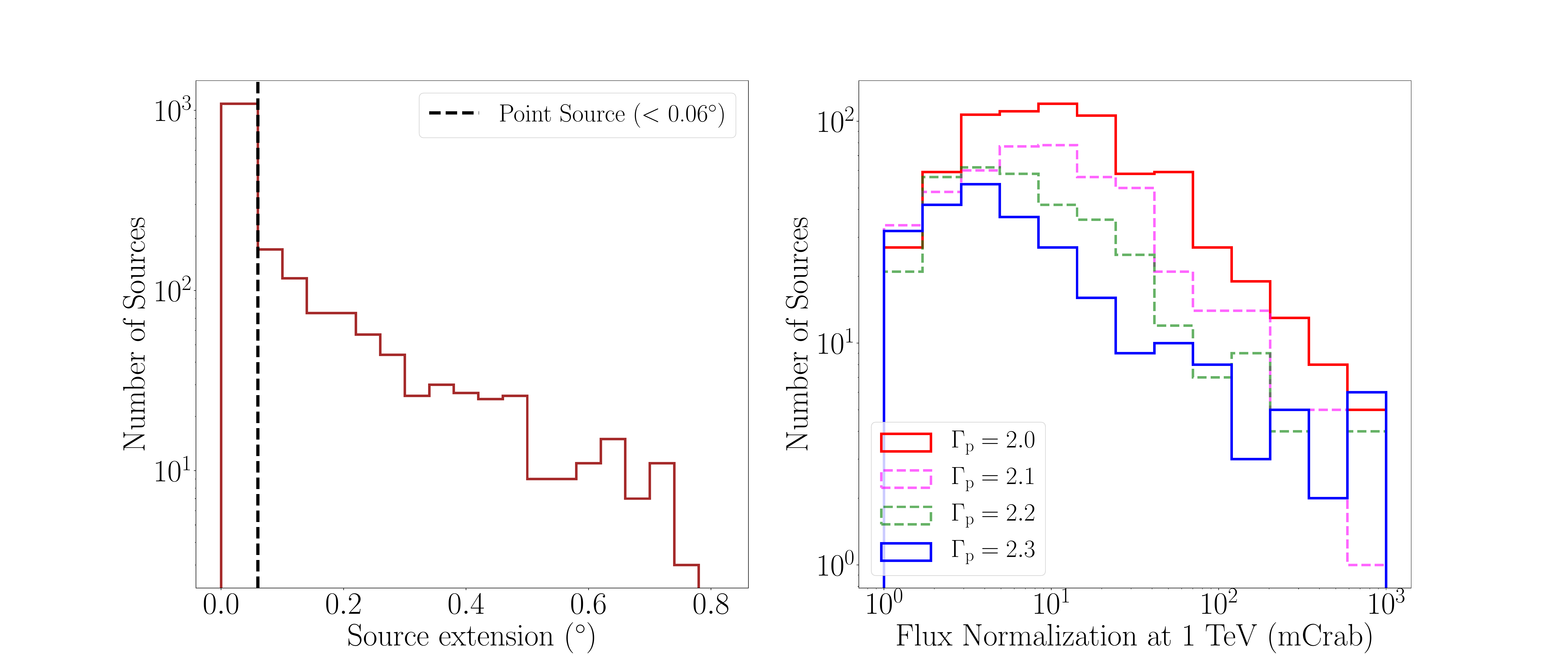}
\caption{Properties of detected SNRs from 200 simulated Galaxy samples. Left panel: Distribution of the angular extension of SNRs shown for all $\Gamma_\mathrm{p}$ values between 2.0 and 2.3. The vertical line marks the size up to which sources are detected as point-like by CTA at energies around $800$ GeV. Right panel: Distributions of the true $\gamma$-ray flux normalization at 1 TeV, originating from true SNR proton spectral indices of $\Gamma_{p}$ = 2.0 (solid red), $\Gamma_{p}$ = 2.1 (dashed magenta), $\Gamma_{p}$ = 2.2 (dashed green) and $\Gamma_{p}$ = 2.3 (solid blue), are shown. }
\label{snr_sim_flux_extension}
\end{figure*}

The left panel of Fig. \ref{snr_sim_flux_extension} shows the distribution of the angular extension for detected SNRs. The angular resolution of CTA south is energy dependent, with values of $\sim$0.06$^\circ$ at $\sim$800\,GeV, $\sim$0.04$^\circ$ at $\sim$5\,TeV, and $\sim$0.02$^\circ$ at $\sim$100\,TeV \citep{cta_mc}. Close to $60\%$ of the detected SNRs appear point-like. Since the majority of detected SNRs are point-like, a first hint on the ability of CTA to confirm SNR PeVatrons can be derived from PeVatron detection probability maps, i.e. Fig. \ref{pts_power_maps} for $10$ h of observation time and Fig. \ref{followup_pts_maps} for $50$ h and $100$ h. The right panel of Fig. \ref{snr_sim_flux_extension} shows the distribution of the flux normalization at $1$ TeV of detected sources for different proton spectral indices. For all considered spectral indices, more than $90\%$, i.e. the vast majority, of detected SNRs have a flux normalization smaller than $100$ mCrab. The derived PeVatron detection probability maps show that a PeVatron confirmation is unlikely with $10$ h CTA exposure but, depending on the proton spectral index, is realistic for larger observation times. A full simulation of the CTA response, which takes the extension of the simulated SNRs into account, is discussed in the next section.
\begin{table*}[ht!]
\large
\caption{Classification criteria applied to preselected SNRs for the analysis of simulated CTA data. Preselected SNRs are true PeVatron SNRs with $\phi_0<1$ Crab and a $\gamma$-ray extension smaller than $0.75^\circ$ at $100$ GeV.}

\centering
\renewcommand{\arraystretch}{1.3}
\begin{tabular}{ll}
\\
\hline\hline
Term & Criteria\\
\hline
Detected SNR & Preselection and $\mathrm{TS}_\mathrm{Det}>30$ \\
PeVatron candidate & $95\%$ CL lower limit on the proton cutoff $E_\mathrm{c,\;p}^\mathrm{LL}>1$ PeV \\
Confirmed PeVatron & $\mathrm{PTS}>25$ (without trials) or $\mathrm{PTS}>33.9$ (with 100 independent trials)\\
\hline
\end{tabular}
\label{snr_naming}
\end{table*}

\subsection{Search for PeVatrons with CTA GPS data}
\label{pevatron_search_gps}
\begin{figure*}[ht!]
\centering
\includegraphics[trim={5cm 0 0 0},width=20cm]{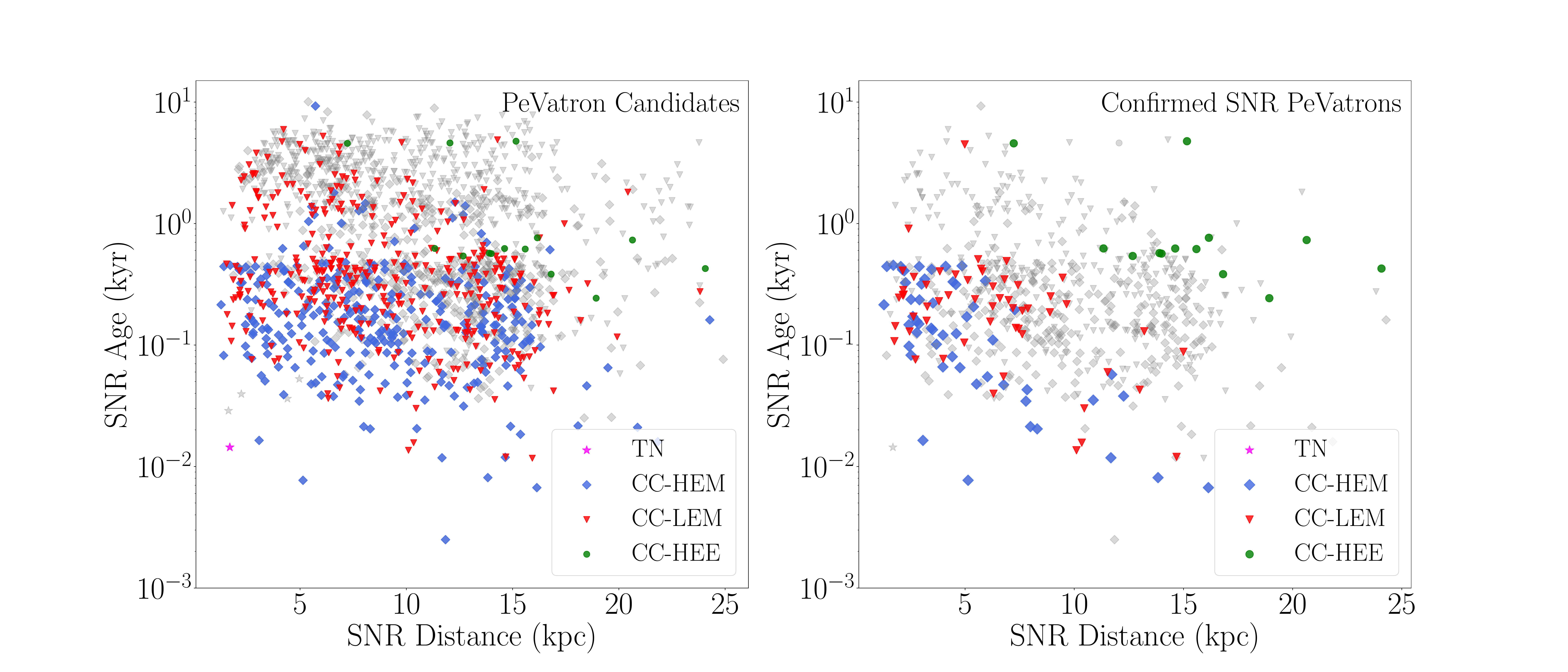}
\caption{Distance--age relation of SNRs from 200 Galaxy samples, combining all $\Gamma_\mathrm{p}$ values between 2.0 and 2.3. Left panel: Grey points mark all detected SNRs giving $E_\mathrm{c,\;p}^\mathrm{LL} < 1$ PeV, while PeVatron candidates ($E_\mathrm{c,\;p}^\mathrm{LL} > 1$ PeV) with their corresponding SNR types are given in different colors. As detailed in the text, the gap around 600--700 years corresponds to the 'transition age' where the evolution of core--collapse SNRs evolves from dense wind to low density cavity. Right panel: Confirmed SNR PeVatrons ($\mathrm{PTS}>33.9$) and their progenitor types are given in different colors. Grey points mark, for comparison, all PeVatron candidates given with different colors in the left plot.}
\label{snr_detection}
\end{figure*}

Table \ref{snr_num_table} shows the dependence of the estimated mean number of detected SNRs on the true proton spectral index. All numbers quoted below refer to preselected SNR, i.e. SNR that are true\,PeVatrons and satisfying additional criteria stated earlier. The total number of detected SNR, including those with sub-PeV cutoffs, therefore is expected to be significantly larger.\\
For a hard spectral index, i.e. $\Gamma_\mathrm{p}=2$, CTA is expected to detect the $\gamma$-ray signal of on average $14.4$ PeVatron SNRs per sample, with Poisson-distributed variations. For $\Gamma_\mathrm{p}=2.3$, the number decreases to $5.2$. However, it is also shown in Tab. \ref{snr_num_table}, that the expected number of PeVatron candidates is lower. On average $6.2$ PeVatron candidates are expected for $\Gamma_\mathrm{p}=2$ and only $1.3$ PeVatron candidates are, on average, expected for $\Gamma_\mathrm{p}=2.3$. The fraction of PeVatron candidates that exhibit point-like characteristics varies between $70\%$ and $80\%$ depending on $\Gamma_\mathrm{p}$. This value is greater than the point-like fraction found for detected SNRs, which is approximately $\%$60, because of the additional $E_\mathrm{c,\;p}^\mathrm{LL}>1$ PeV criterion mentioned in Tab.~\ref{snr_naming}. It is concluded that, the detection of $\gamma$-ray emission of multiple PeVatron SNRs with CTA GPS data is possible. In particular, detection of PeVatron candidates is more likely for hard proton spectra sources than for soft ones. If PeVatron candidates are found, they are likely to be point-like (see Fig. \ref{snr_sim_flux_extension}).\\
The left panel of Fig. \ref{snr_detection} shows the distribution of preselected and detected SNRs as a function of the SNR age and distance. The majority of detected SNRs are of core collapse type, i.e. of type CC-HEM, CC-LEM and CC-HEE. As detailed in Sec. \ref{sec_modeling}, these SNRs evolve in a dense wind for typically a few hundred years, succeeded by an expansion in a low density cavity for typically a few kyr. Around the transition age of $\sim$600--700 years, Fig. \ref{snr_detection} shows a gap, meaning that the detection probability for a PeVatron SNR with CTA GPS data is reduced because, as detailed in \cite{cristofari2018}, the maximal particle energy $E_\mathrm{max}$ is, for core--collapse SNRs, smaller in this phase than during either the free expansion and parts of the Sedov--Taylor phase. The median age and distance, 250 years and 8.5 kpc, of PeVatron candidates are lower than the respective values for detected SNRs, i.e. 440 years and 10.0 kpc. All detected SNRs with SN progenitors of the rare type CC-HEE are found to be PeVatron candidates. The same is true for $25\%$ of the detected SNRs with a type TN progenitor and, respectively, $43\%$ and $33\%$ of SNRs with type CC-HEM and CC-LEM SNe.\\
The confirmation of PeVatrons with the limited exposure provided by CTA GPS data is very challenging, as discussed in Sec. \ref{pev_det_power_maps} and in particular with Fig. \ref{pts_power_maps}. Table \ref{snr_num_table} shows that, with only data acquired in the CTA GPS, the confirmation of on average $1.4$ PeVatrons is expected when the true proton spectral index is hard, i.e. $\Gamma_\mathrm{p}=2$. However, for $\Gamma_\mathrm{p}=2.3$, the confirmation of on average only $0.24$ PeVatrons is expected. This shows that a\,PeVatron confirmation is very unlikely based on CTA GPS data when the true spectral index is soft, i.e. $\Gamma_\mathrm{p}=2.3$. The right panel of Fig. \ref{snr_detection} shows the distance--age relation for confirmed\,PeVatrons from all 200 Galaxy simulations. The median distance to confirmed\,PeVatrons is 5.5 kpc, which is closer than for PeVatron candidates. Notable exceptions are the rare but energetic SNRs resulting from SNe of type CC-HEE. The median age of confirmed PeVatrons is 210 years.
\begin{table*}[ht!]
\large
\caption{Average number of detected SNRs per sample, as well as the median number of PeVatron candidates and confirmed PeVatrons for different proton spectral indices $\Gamma_\mathrm{p}$. A point-like source equivalent CTA exposure of $10$ h, as expected to be acquired in the CTA GPS, is assumed. The confirmation of PeVatrons is based on a PTS threshold of $33.9$, which corresponds to 5$\sigma$ accounting for $100$ independent trials. Errors are the standard error of the mean.}
\centering
\renewcommand{\arraystretch}{1.3}
\begin{tabular}{llllll}
\\
\hline\hline
& $\Gamma_\mathrm{p}$ = 2.0 & $\Gamma_\mathrm{p}$ = 2.1 & $\Gamma_\mathrm{p}$ = 2.2 & $\Gamma_\mathrm{p}$ = 2.3 \\
\hline
Preselected SNRs & $37 \pm 1$ & $35.0 \pm 0.8$ & $25.7 \pm 0.6$ & $15.8 \pm 0.5$ \\
Detected SNRs (among the preselected) & $14.4\pm0.6$ & $9.3\pm0.4$ & $7.0\pm0.4$  & $5.2\pm0.3$  \\
PeVatron candidates & $6.2\pm 0.4$ & $3.9\pm0.3$ & $2.1\pm0.2$ & $1.3\pm0.2$ \\
Confirmed PeVatrons ($\mathrm{PTS}>33.9$) & $1.4\pm0.1$ & $0.5\pm0.1$ & $0.28\pm 0.07$ & $0.24\pm0.08$ \\
\hline
\end{tabular}
\label{snr_num_table}
\end{table*}

\subsection{Discussion of the PTS}
\label{pts_discussion}
\begin{figure*}[ht!]
\centering
\includegraphics[trim={5cm 0 0 0},width=20cm]{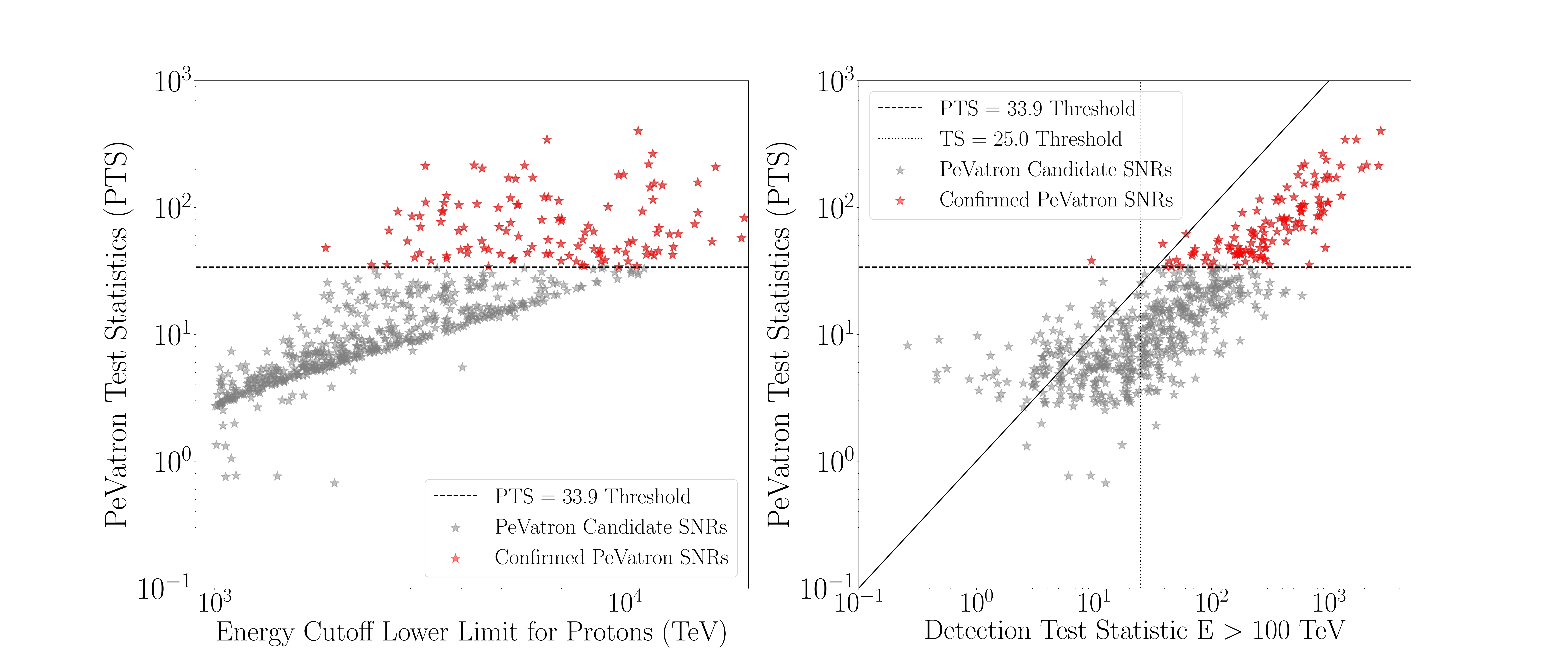}
\caption{Relation between the PTS and the lower 95\% CL limit on the proton spectral cutoff (left) and detection test statistic, TS$_\mathrm{Det}^\mathrm{100\; TeV}$, of the $\gamma$-ray flux above $100$ TeV (right), combining all $\Gamma_\mathrm{p}$ values between 2.0 and 2.3. Confirmed PeVatrons are marked in red and PeVatron candidates are marked in grey. The PTS
threshold of 33.9, corresponding to a global 5$\sigma$ significance for 100 independent tests, is indicated with horizontal dashed black lines. In the right panel, the vertical line indicates the $\gamma$-ray detection threshold TS$_\mathrm{Det}^{\mathrm{100\; TeV}}$=25 and the diagonal shows the equality PTS=TS$_\mathrm{Det}$.}
\label{pev_snr_pts}
\end{figure*}

The simulation of the CTA response to the $\gamma$-ray emission of true PeVatrons provides an opportunity to test the relation between the PTS and more traditional measures for the characterization of properties of source spectra at high energies. The left panel of Fig. \ref{pev_snr_pts} shows the relation of the PTS to the $95\%$ CL lower limit on the proton energy cutoff. Although the PTS correlates with the lower limit on the proton energy cutoff, a PeVatron confirmation cannot be claimed when only the $95\%$ CL lower limit on the proton energy cutoff is larger than 1 PeV. The right panel of Fig. \ref{pev_snr_pts} shows the correlation between the PTS and the source detection test statistic $\mathrm{TS}_\mathrm{Det}$ above a $\gamma$-ray energy threshold of $100$\,TeV. Again, the significant detection of a $\gamma$-ray flux at energies larger than $100$ TeV alone is insufficient for the confirmation of a PeVatron.

\subsection{Deep observations of selected SNR PeVatrons}
\label{deep_pevatron_obs}
Due to the low average exposure of the CTA GPS, a\,PeVatron confirmation is unlikely, except for hard proton spectra, i.e. $\Gamma_\mathrm{p} \leq 2$. In the following it is discussed whether CTA will be able to confirm true SNR PeVatrons with long exposures acquired in deep observations. In practice, a selection of a few promising sources must be performed to schedule deep observations. For each of the 200 simulated SNR samples, the source with the largest 95$\%$ CL lower limit on the proton spectral cutoff is selected. This selection strategy leads to a list of 199 sources because for one simulated SNR sample with $\Gamma_\mathrm{p}=2.3$, there is no SNR detected in the sense of Tab. \ref{snr_naming}. As it was discussed in Sect. \ref{pevatron_search_gps}, the majority of selected PeVatron candidates are point-like. The median age of selected SNRs is 220 years, i.e. typically young SNRs are selected for follow-up observations, and the median distance is 8.2 kpc. Deep observations with exposures of $50$ h and $250$ h are simulated for selected sources and the fraction of PeVatrons which are confirmed is quantified. Since only one source is considered for deep observations per simulated SNR sample, no trial correction is applied and the confirmation threshold of 25 is used for the PTS as indicated in Tab. \ref{snr_naming}.\\
Table \ref{deep_snr_table} summarizes the results. An exposure of $50$ h is likely to be insufficient for a PeVatron confirmation when the proton spectral index is not hard. While $80\%$ of the selected\,PeVatrons are confirmed when the true proton spectral index is hard, i.e. $\Gamma_\mathrm{p}=2$, only $24\%$ can be confirmed for $\Gamma_\mathrm{p}=2.3$. However, for an exposure of $250$ h, the prospects for the confirmation of a SNR PeVatron are excellent. At least $86\%$ of the selected\,PeVatrons are confirmed when $\Gamma_\mathrm{p}\leq 2.3$.\\
This result clearly shows that CTA will be able to confirm SNR PeVatrons when good candidates are selected for deep observations. A selection of candidates can be performed based on data acquired in the CTA GPS or, for example, based on measurements with different experiments such as LHAASO or the planned SWGO \citep{swgo2022}. 

\begin{table*}[ht!]
\large
\caption{Percentage of simulated Galaxy samples where the selected PeVatron is confirmed in deep observations. Different columns correspond to different assumptions on the SNR proton spectral index $\Gamma_\mathrm{p}$ and different rows correspond to different exposures. For each simulated Galaxy sample, the SNR for which the most constraining limit on the proton energy cutoff can be derived with 10 h of CTA data is selected for deep observations. No trials are considered, i.e. the PTS threshold is $25$. The last two rows correspond to a combination of observations with CTA and data with the SST subarray of CTA under moonlight conditions and are discussed in Sec.~\ref{moonlight_sec}. The quoted errors are $68\%$ CL Wilson score intervals \cite{binomial_wilson}.}

\centering
\renewcommand{\arraystretch}{1.4}
\begin{tabular}{llllll}
\\
\hline\hline
Total observation time & $\Gamma_\mathrm{p}$ = 2.0 & $\Gamma_\mathrm{p}$ = 2.1 & $\Gamma_\mathrm{p}$ = 2.2 & $\Gamma_\mathrm{p}$ = 2.3 \\
\hline
50 h nominal & 80$\%^{+5}_{-6}$ & (62$\pm$7)$\%$  & (46$\pm$7)$\%$   & 24$\%^{+7}_{-6}$ \\
100 h nominal & 92$\%^{+3}_{-5}$ & 82$\%^{+5}_{-6}$ & 64$\%^{+6}_{-7}$ & (47$\pm$7)$\%$    \\ 
250 h nominal & $100\%^{+0}_{-2}$ & $96\%^{+2}_{-4}$ & $92\%^{+3}_{-5}$ & $86\%^{+4}_{-6}$ \\
\hline
50 h (10 h nominal NSB + 40 h HNSB)  & 68$\%^{+6}_{-7}$ & (44$\pm$7)$\%$   & 34$\%^{+7}_{-6}$ & 20$\%^{+6}_{-5}$ \\
100 h  (10 h nominal NSB + 90 h HNSB) & 88$\%^{+4}_{-5}$ & 64$\%^{+6}_{-7}$ & (50$\pm$7)$\%$    & 31$\%^{+7}_{-6}$ \\
\hline
\end{tabular}
\label{deep_snr_table}

\end{table*}
 
\section{PeVatron searches with CTA under moonlight conditions}
\label{moonlight_sec}

CTA aims to address many key questions in the field of very high energy astrophysics \citep{science_with_cta} and the optimization of observation time available for each individual key science topic has to be controlled to maximize the overall scientific return. Traditionally, IACT observations are only carried out during periods of astronomical darkness, i.e. with little to no moonlight. This is due to the sensitivity of the photo-multiplier tubes (PMTs) used in the IACT cameras, which degrade when exposed to high--intensity incident light. The southern CTA site will include a large SST array, which will provide excellent sensitivity above energies of $\sim$~10 TeV. These dual-mirror Schwarzschild-Couder telescopes will use silicon photomultipliers (SiPMs). A clear advantage of the SiPM sensor is that it can sustain long periods of exposure to very strong moonlight conditions without substantial changes in its properties. This has already been demonstrated by the excellent long-term stability of the First Geiger-mode avalanche photodiodes (G-APD) Cherenkov Telescope (FACT) camera, which has been in operation at La Palma since 2012 \citep{FACT}. Work by the VERITAS Collaboration has demonstrated that observing with up to 30 times the nominal night sky background (NSB) light, corresponding to observations of a source located 90 degrees from an 80$\%$ illuminated Moon, can provide up to 30$\%$ more observation time per year \citep{VERITASNSB}. Similarly, work by the MAGIC Collaboration has shown that the maximal duty cycle of MAGIC can be increased from 18$\%$ to up to 40$\%$ in total with only moderate performance degradation and without any significant worsening of the angular resolution \citep{MAGICNSB}. The actual NSB level during any given observation depends very strongly upon the Moon phase, the angular distance of the source from the Moon, and the presence of clouds or other reflective material in the atmosphere. These constraints make it difficult to estimate the additional observing yield for any given source, as well as the impact of dramatically varying observing conditions on the sensitivity of the array during these observations. In this work, 30 times the nominal NSB is chosen as a conservative value for typical observations, which is referred to as High NSB (HNSB).\\

\begin{figure}[!ht]
\centering
\includegraphics[width=\hsize]{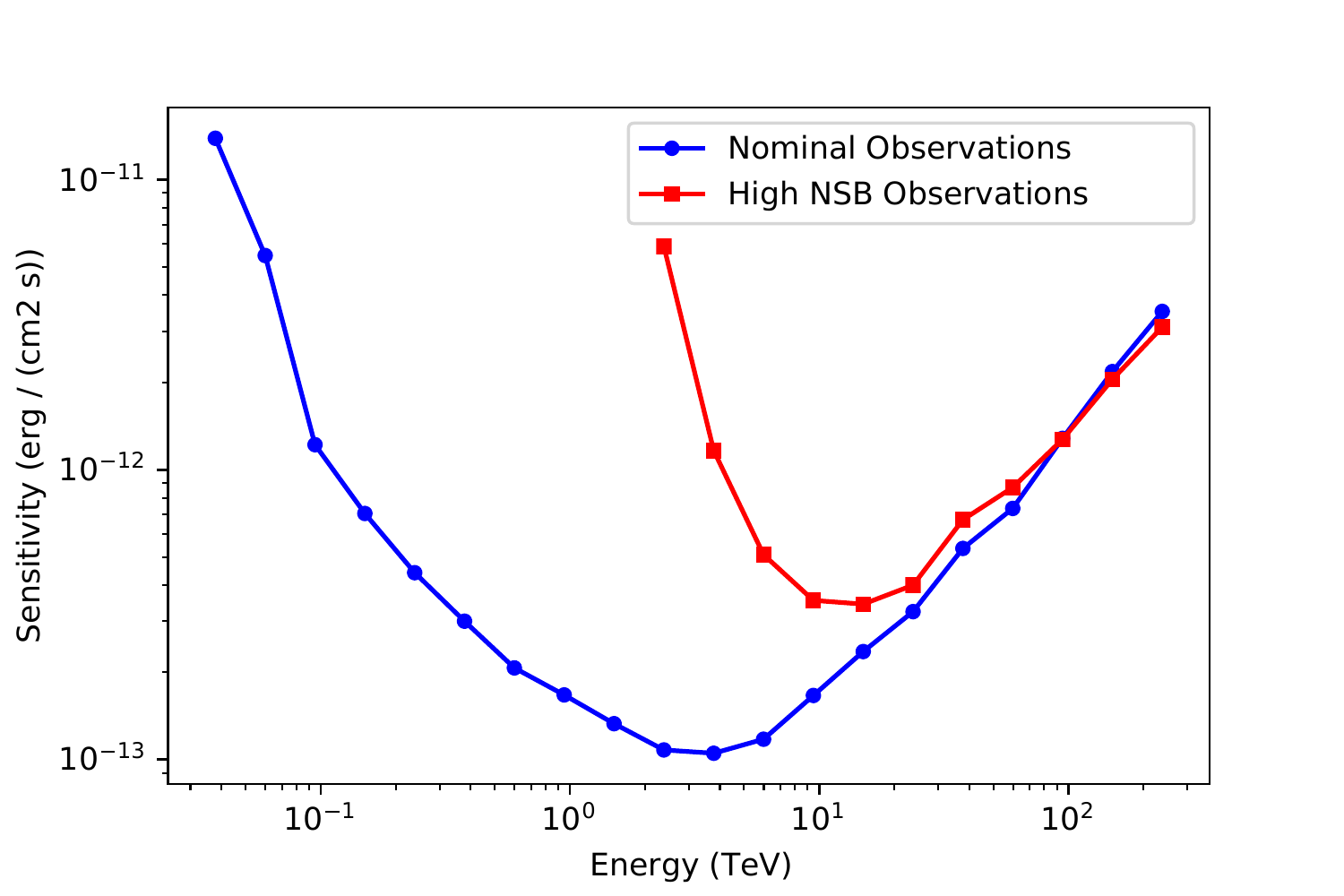}
\caption{Comparison of the CTA point-like source sensitivity for 10 hours observation for different observations conditions. Nominal observations correspond to the CTA Omega array under conditions of astronomical darkness. High NSB observations correspond to the SST sub-array of CTA under conditions of $80\%$ full moon illumination.}
\label{fig:sens}
\end{figure}

Figure \ref{fig:sens} compares the point-like source differential sensitivity of the SST array for HNSB conditions to the sensitivity of the CTA Omega array for nominal conditions. The figure shows that observations under moonlight conditions with only the SST array can provide a similar sensitivity as the CTA Omega array above energies of a few 10's of TeV.\\
The PeVatron detection performance of different strategies for deep PeVatron observations is compared in the following. For all strategies, it is assumed that 10 h point-like source equivalent exposure with the full CTA south array data is available, e.g. as a result of the CTA GPS. The baseline is to perform deep follow-up observations with the full CTA south array under nominal NSB conditions. The alternative is to observe under HNSB conditions with the SST subarray of CTA.

\subsection{Confirmation of SNR PeVatrons} 
\label{moonlight_snr}
\begin{figure*}[ht!]
\centering
\includegraphics[trim={5cm 0 0 0},width=20cm]{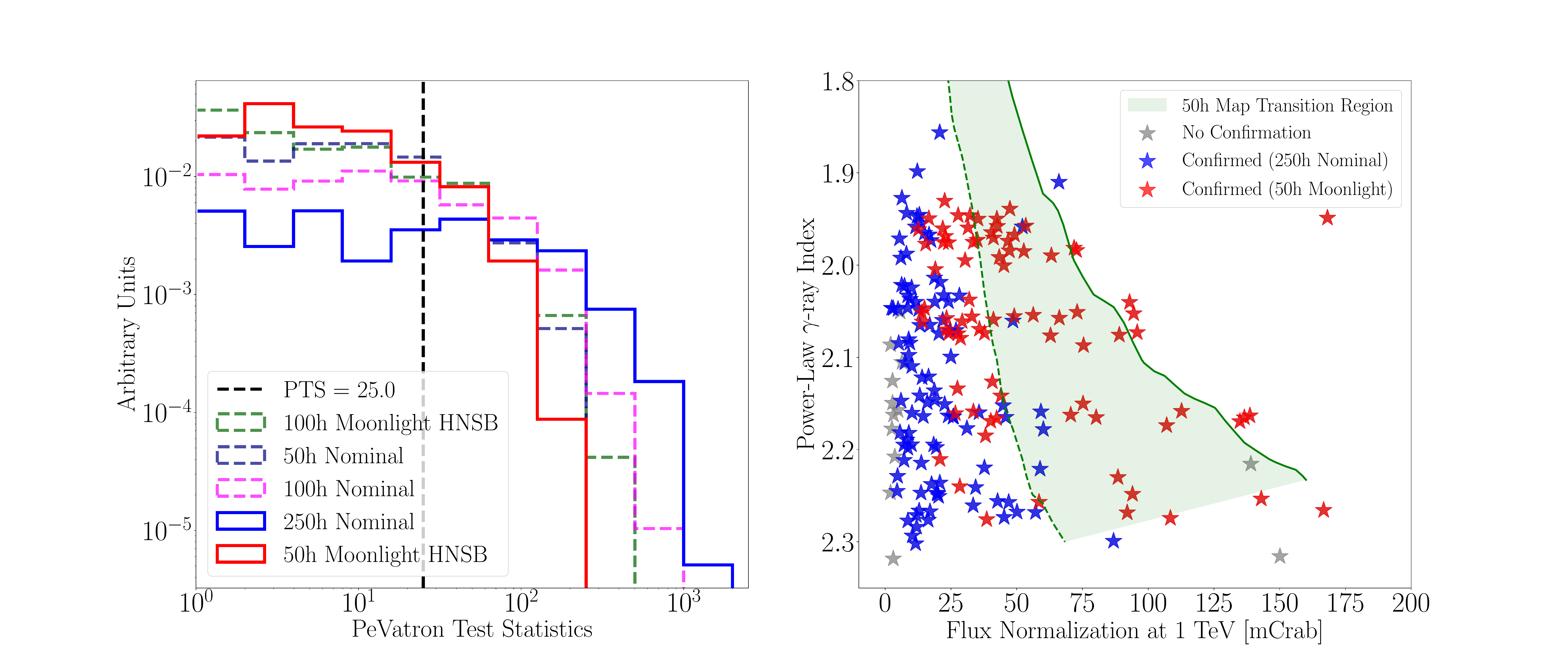}
\caption{The left panel shows the distribution of PTS values obtained from different observation strategies. The PTS identification threshold of 25 is used and shown as black dashed line. The right panel shows the spectral $\gamma$-ray parameters of simulated PeVatron SNRs as reconstructed from 10 h of CTA observations. Blue stars mark PeVatrons that can be confirmed after 250 h of follow-up observations. Red stars mark PeVatrons which can be confirmed with 50 h of moonlight observations. Grey stars indicate the PeVatron SNRs that cannot be confirmed with either follow-up observation strategies. The green area is the "transition region" for 50 h exposure shown in Fig. \ref{followup_pts_maps} (bottom left)}

\label{snr_followup_results}
\end{figure*}

As discussed in Sec. \ref{deep_pevatron_obs}, SNR PeVatrons can be identified with data acquired in deep exposures with CTA. The performance of follow-up observations with the SST subarray of CTA under moonlight conditions is compared to the performance of CTA under nominal conditions in Tab. \ref{deep_snr_table}. For example, one strategy is to combine $10$ h of CTA exposure under nominal conditions and $90$ h of SST subarray exposure under HNSB conditions. In total $100$ h of data is acquired in this strategy. Table \ref{deep_snr_table} shows that the performance with respect to the confirmation of selected SNR PeVatrons is similar to $50$ h of CTA data acquired under nominal conditions. Also shown in Tab. \ref{deep_snr_table} is the expected performance when only $40$ h of SST subarray data under HNSB conditions is combined with $10$ h of CTA data under nominal conditions to a total of $50$ h data. In this case, the performance with respect to the confirmation of selected SNR PeVatrons is significantly worse than for $50$ h data acquired with CTA under nominal conditions. The left panel of Fig. \ref{snr_followup_results} details the distribution of PTS values for the three observation strategies. \\
The right panel of Fig. \ref{snr_followup_results} shows that the PeVatron detection probability maps as introduced in Sec. \ref{section4} can be used to decide whether follow-up observations of sources detected with $10$ h of CTA exposure are promising. Shown are the $\gamma$-ray spectral parameters of SNRs selected as reconstructed with $10$ h of simulated CTA data. The "transition region" for 50 h exposure shown in Fig. \ref{followup_pts_maps}, is overlaid on the figure. Since the "transition region" is defined for hadronic spectral indices $\Gamma_\mathrm{p}$, following \citep{celli2020}, the contour lines are shifted as $\Gamma=\Gamma_\mathrm{p}-0.1$. Figure \ref{snr_followup_results} shows that SNRs with $\phi_0> 50$ mCrab are likely to be confirmed with the combination of 10 h full array CTA and 40 h SST-subarray exposure under moonlight conditions.

\subsection{The source confusion case: HESS J1641$-$463}
\label{moonlight_followup_J1641}

\begin{figure*}[ht!]
\centering
\begin{tabular}{cc}
\includegraphics[width=18cm]{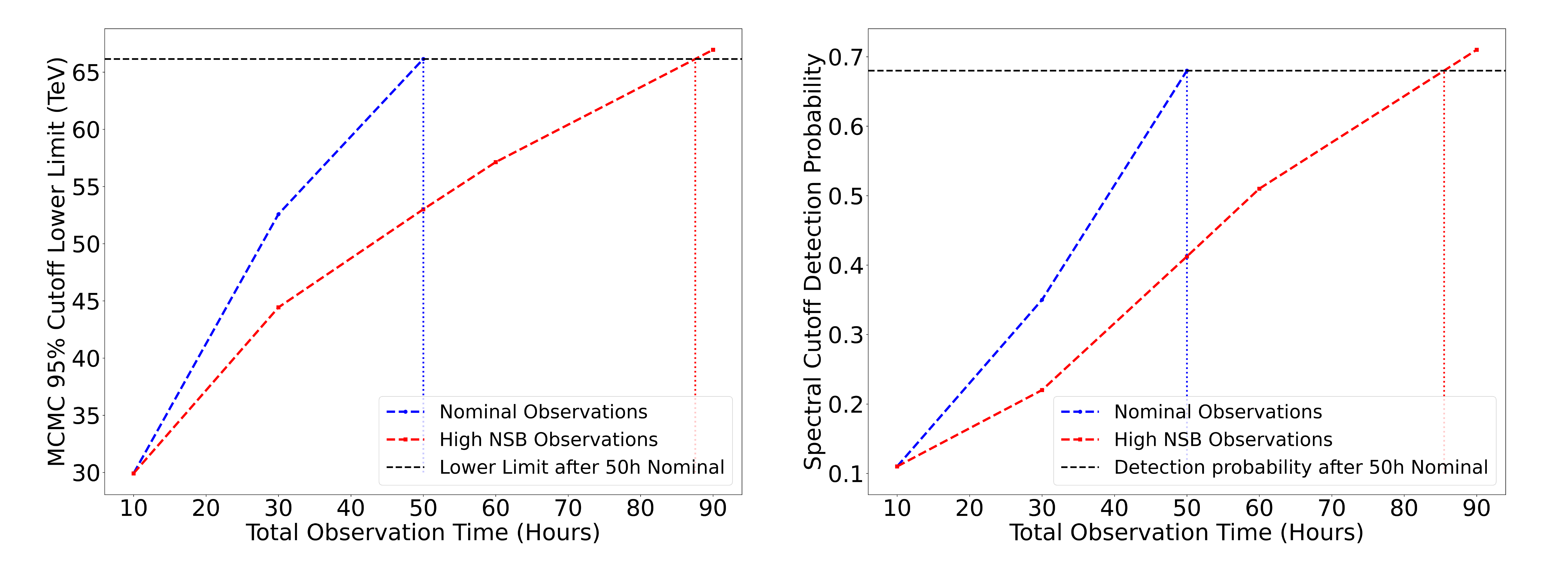}
\end{tabular}
\caption{Performance of SST subarray observations of HESS J1641--463 under moonlight conditions. Left Panel: $95\%$ CL lower limit on the energy cutoff for sources without detection of a cutoff, i.e. $\mathrm{TS}_\lambda< 9$. To obtain the same limit as the 50 h nominal NSB follow up, the total observation time needed under moonlight conditions would be just under 90 h. Right Panel: Spectral cutoff detection probability as a function of total observation time. It is found that a detection probability of 68\% obtained with 50 h nominal NSB can be achieved with just under 90 h high NSB, including 10 h from the GPS.}
\label{fig:nsbres}
\end{figure*}

So far only isolated SNRs are simulated in this study. It is expected that source confusion will be an important issue with CTA, as it is for currently operating instruments \citep{j1826,HESS1702,HWCJ1825_134}. While a full study of the effects of source confusion is beyond the scope of this paper, one exemplary case is discussed in the following for the case of HESS J1641$-$463.\\
The source HESS J1641$-$463 has a hard spectrum, which extends up to a few tens of TeV without showing a significant spectral cutoff and is found to be point-like for the H.E.S.S. instrument \citep{hess1641_paper}. The detection of HESS\,J1641$-$463 was initially hindered due to the proximity to the nearby bright extended $\gamma$-ray source HESS\,J1640$-$465, which has an extension of $\sim$0.11$^{\circ}$ and shows a significant spectral cutoff at $\sim$6 TeV. The angular separation between the best fit positions of HESS\,J1641$-$463 and HESS\,J1640$-$465 is $\sim$0.28$^{\circ}$. This spatially and spectrally complex region is considered as good test case for the feasibility and performance of HNSB observations on measured $\gamma$-ray spectral properties. The spectral results obtained from combined HNSB moonlight observations are compared to observations with the full array under nominal conditions in order to judge the performance. Simulations are performed for the configuration described in Sec. \ref{cta_sim_ana}. The spectral parameters of the two sources are set to the best fit values to the data observed with H.E.S.S., as described in \cite{hess1641_paper}, with the addition of an assumed high energy cutoff of the $\gamma$-ray spectrum at $100$ TeV. Again, the performance expected to result from two different observation strategies is compared. The baseline is a total of $50$ h of CTA observation time under nominal NSB conditions. The alternative is to combine $10$ h of observation time at nominal NSB, resulting e.g. from the GPS, and $90$ h of follow-up observations with the SST subarray under HNSB conditions. For both observation strategy cases, 1000 simulations of the HESS\,J1641$-$463 $/$ HESS\,J1640$-$465 source confusion region are performed.\\
Simulated data sets are analyzed as described in Sec. \ref{cta_sim_ana}. The test statistic $\mathrm{TS_\lambda}$ is calculated according to Eq. \ref{ts_lambda} for each simulated source. The results can be seen in Fig. \ref{fig:nsbres}. In the right--hand panel, the spectral cutoff detection probability is shown for the two cases. A $\mathrm{TS}_\lambda$ cutoff detection threshold of 9, corresponding to 3$\sigma$ level, is used in order to highlight the effects. For comparison, if a $\mathrm{TS}_\lambda$ threshold of 25 instead of 9 is used, then the spectral cutoff detection probability is reduced from around $70\%$ to around 5\% for $50$ h of nominal observations and $90$ h of HNSB observations. The additional data using the HNSB observations lead to the same detection probability of $68\%$ at a total of $\sim85$ h observation compared to 50 h with the full array at nominal NSB\footnote{Extrapolation of detection probability curves to higher observation
times gives that robust 95$\%$ spectral cutoff detection probability can be reached after 68.6 h and 119.9 hours for nominal and HNSB observations, respectively}. One drawback of using high NSB observations is that the lack of data provided at lower energies impacts the accuracy of the best fit spectral parameters. For 90 h of high NSB observations, the error on the spectral index is improved by 20\% compared to that obtained with the GPS dataset of $10$ h, whereas for 50 h nominal NSB observations, the same error is reduced by 40\%. For the error on the differential flux (at 1 TeV), an improvement of 8\% and 56\% for observations with high and respectively nominal NSB levels is found.\\
For the derivation of lower limits, a slightly modified simulation was performed in which a limit on the detection of a cutoff, $\mathrm{TS}_\lambda$\,$<$\,9, was required. Otherwise the simulation would be repeated until the condition was met. Lower limits are derived with the Markov Chain Monte Carlo (MCMC) method described in \ref{MCMC_intro} because the likelihood method fails, due to convergence problems during the optimization for the profile likelihood function. The results are shown in the left panel of Fig.\,\ref{fig:nsbres}. The same lower limit can be achieved with just under 90 h of total observation time for the high NSB case, compared to 50 h for nominal NSB.

\section{Conclusion}
\label{conclusions}

The ability of CTA to search for PeVatron sources is discussed in this work. The focus is on the spectral capabilities of CTA. For sources whose extension is resolved with CTA, the spatial correlation of, for example, radio and $\gamma$-ray data will further help to identify the underlying particle acceleration mechanisms. Additionally, methods for PeVatron searches with $\gamma$-ray detectors are discussed. PeVatron detection probability maps, as introduced in Sec. \ref{section4} for CTA, are used to quantify the sensitivity of a $\gamma$-ray detector to PeVatrons. The PTS is introduced in Sec. \ref{section2} as a test to decide whether a given source is a PeVatron by spectral means.\\
With CTA GPS data, i.e. with $10$ h of CTA exposure, only point-like PeVatrons with bright $\gamma$-ray emission and hard proton spectra are likely to be identified as PeVatrons. Given 10 h CTA GPS data alone, it will be impossible for many sources, in particular when $\Gamma_\mathrm{p}>2$, to decide whether it is a PeVatron. CTA must therefore rely on deep observations of selected PeVatron candidates. PeVatrons with a wide range of spectral parameters can be tested with deep observations with $\mathcal{O}(100)$ hours of exposure.\\
One of the leading hypotheses for the origin of Galactic PeV CRs is the acceleration in SNRs. The ability of CTA to test this hypothesis is investigated in detail based on a Monte--Carlo simulation of Galactic PeVatron SNRs and the hadronic $\gamma$-ray emission resulting from interactions between accelerated protons and the ISM. With CTA GPS data, the detection of a $\gamma$-ray signal from multiple SNR PeVatrons, the majority of which are point-like for CTA, is expected. However, with the limited exposure of the CTA GPS, it can only be confirmed that these sources are PeVatrons when the proton spectrum is hard, e.g. for $\Gamma_\mathrm{p}=2$. It is shown in Sec. \ref{deep_pevatron_obs} that CTA has excellent prospects to confirm SNR PeVatrons with $\Gamma_\mathrm{p}\leq 2.3$ with a typical exposure of $250$ h.\\
An alternative follow-up observation strategy with SSTs under moonlight conditions is discussed in Sec. \ref{moonlight_sec}. It is shown that the performance of the full southern CTA array with respect to spectral PeVatron detection metrics can be reached with follow-up observations of the SST subarray of CTA under moonlight conditions in expense of typically twice the observation time. The strategy can save a significant amount of observation dark time of CTA excluding SSTs for other key science topics different from \,PeVatron searches. This result highlights the importance of the SST type telescopes and the potential of the SiPM technology for PeVatron searches.\\
It is concluded that while CTA has limited spectral sensitivity to search for\,PeVatrons in scanning mode with GPS data, the prospects to find PeVatrons are excellent in deep observations.

\section*{Acknowledgements}

We gratefully acknowledge financial support from the following agencies and organizations: State Committee of Science of Armenia, Armenia; The Australian Research Council, Astronomy Australia Ltd, The University of Adelaide, Australian National University, Monash University, The University of New South Wales, The University of Sydney, Western Sydney University, Australia;
Federal Ministry of Education, Science and Research, and Innsbruck University, Austria; Conselho Nacional de Desenvolvimento Cient\'{\i}fico e Tecnol\'{o}gico (CNPq), Funda\c{c}\~{a}o de Amparo \`{a} Pesquisa do Estado do Rio de Janeiro (FAPERJ), Funda\c{c}\~{a}o de Amparo \`{a} Pesquisa do Estado de S\~{a}o Paulo (FAPESP), Funda\c{c}\~{a}o de Apoio \`{a} Ci\^encia, Tecnologia e Inova\c{c}\~{a}o do Paran\'a - Funda\c{c}\~{a}o Arauc\'aria, Ministry of Science, Technology, Innovations and Communications (MCTIC), Brasil; Ministry of Education and Science, National RI Roadmap Project DO1-153/28.08.2018, Bulgaria; The Natural Sciences and Engineering Research Council of Canada and the Canadian Space Agency, Canada; CONICYT-Chile grants CATA AFB 170002, ANID PIA/APOYO AFB 180002, ACT 1406, FONDECYT-Chile grants, 1161463, 1170171, 1190886, 1171421, 1170345, 1201582, Gemini-ANID 32180007, Chile; Croatian Science Foundation, Rudjer Boskovic Institute, University of Osijek, University of Rijeka, University of Split, Faculty of Electrical Engineering, Mechanical Engineering and Naval Architecture, University of Zagreb, Faculty of Electrical Engineering and Computing, Croatia;
Ministry of Education, Youth and Sports, MEYS  LM2015046, LM2018105, LTT17006, EU/MEYS CZ.02.1.01/0.0/0.0/16\_013/0001403, CZ.02.1.01/0.0/0.0/18\_046/0016007 and CZ.02.1.01/0.0/0.0/16\_019/0000754, Czech Republic; Academy of Finland (grant nr.317636 and 320045), Finland;
Ministry of Higher Education and Research, CNRS-INSU and CNRS-IN2P3, CEA-Irfu, ANR, Regional Council Ile de France, Labex ENIGMASS, OCEVU, OSUG2020 and P2IO, France;
Max Planck Society, BMBF, DESY, Helmholtz Association, Germany;
Department of Atomic Energy, Department of Science and Technology, India;
Istituto Nazionale di Astrofisica (INAF), Istituto Nazionale di Fisica Nucleare (INFN), MIUR, Istituto Nazionale di Astrofisica (INAF-OABRERA) Grant Fondazione Cariplo/Regione Lombardia ID 2014-1980/RST\_ERC, Italy;
ICRR, University of Tokyo, JSPS, MEXT, Japan;
Netherlands Research School for Astronomy (NOVA), Netherlands Organization for Scientific Research (NWO), Netherlands;
University of Oslo, Norway;
Ministry of Science and Higher Education, DIR/WK/2017/12, the National Centre for Research and Development and the National Science Centre, UMO-2016/22/M/ST9/00583, Poland;
Slovenian Research Agency, grants P1-0031, P1-0385, I0-0033, J1-9146, J1-1700, N1-0111, and the Young Researcher program, Slovenia; 
South African Department of Science and Technology and National Research Foundation through the South African Gamma-Ray Astronomy Programme, South Africa;
The Spanish groups acknowledge the Spanish Ministry of Science and Innovation and the Spanish Research State Agency (AEI) through grants AYA2016-79724-C4-1-P, AYA2016-80889-P, AYA2016-76012-C3-1-P, BES-2016-076342, FPA2017-82729-C6-1-R, FPA2017-82729-C6-2-R, FPA2017-82729-C6-3-R, FPA2017-82729-C6-4-R, FPA2017-82729-C6-5-R, FPA2017-82729-C6-6-R, PGC2018-095161-B-I00, PGC2018-095512-B-I00, PID2019-107988GB-C22; the “Centro de Excelencia Severo Ochoa” program through grants no. SEV-2016-0597, SEV-2016-0588, SEV-2017-0709, CEX2019-000920-S; the “Unidad de Excelencia Mar\'ia de Maeztu” program through grant no. MDM-2015-0509; the “Ram\'on y Cajal” programme through grants RYC-2013-14511, RYC-2017-22665; and the MultiDark Consolider Network FPA2017-90566-REDC. They also acknowledge the Atracci\'on de Talento contract no. 2016-T1/TIC-1542 granted by the Comunidad de Madrid; the “Postdoctoral Junior Leader Fellowship” programme from La Caixa Banking Foundation, grants no.~LCF/BQ/LI18/11630014 and LCF/BQ/PI18/11630012; the “Programa Operativo” FEDER 2014-2020, Consejer\'ia de Econom\'ia y Conocimiento de la Junta de Andaluc\'ia (Ref. 1257737), PAIDI 2020 (Ref. P18-FR-1580) and Universidad de Ja\'en; “Programa Operativo de Crecimiento Inteligente” FEDER 2014-2020 (Ref.~ESFRI-2017-IAC-12), Ministerio de Ciencia e Innovaci\'on, 15\% co-financed by Consejer\'ia de Econom\'ia, Industria, Comercio y Conocimiento del Gobierno de Canarias; the Spanish AEI EQC2018-005094-P FEDER 2014-2020; the European Union’s “Horizon 2020” research and innovation programme under Marie Skłodowska-Curie grant agreement no. 665919; and the ESCAPE project with grant no. GA:824064;
Swedish Research Council, Royal Physiographic Society of Lund, Royal Swedish Academy of Sciences, The Swedish National Infrastructure for Computing (SNIC) at Lunarc (Lund), Sweden;
State Secretariat for Education, Research and Innovation (SERI) and Swiss National Science Foundation (SNSF), Switzerland;
Durham University, Leverhulme Trust, Liverpool University, University of Leicester, University of Oxford, Royal Society, Science and Technology Facilities Council, UK;
U.S. National Science Foundation, U.S. Department of Energy, Argonne National Laboratory, Barnard College, University of California, University of Chicago, Columbia University, Georgia Institute of Technology, Institute for Nuclear and Particle Astrophysics (INPAC-MRPI program), Iowa State University, the Smithsonian Institution, Washington University McDonnell Center for the Space Sciences, The University of Wisconsin and the Wisconsin Alumni Research Foundation, USA.

The research leading to these results has received funding from the European Union's Seventh Framework Programme (FP7/2007-2013) under grant agreements No~262053 and No~317446.
This project is receiving funding from the European Union's Horizon 2020 research and innovation programs under agreement No~676134.

This research has made use of the CTA instrument response functions provided by the CTA Consortium and Observatory, see https://www.cta-observatory.org/science/cta-performance/ (version prod3b-v2, prod5 v0.1; \cite{prod3b}, \cite{prod5}) for more details.

\newpage

\appendix
\section{Derivation of spectral cutoff lower limits}
\label{lower_limits}
PeVatron searches with CTA rely on the derivation of statistical statements on the inverse energy cutoff parameter $\lambda$. 
In particular when a significant cutoff detection is impossible, frequentist upper limits $\lambda^\mathrm{UL}$ on the inverse cutoff parameter at a given confidence level CL are of high relevance. 
These limits correspond to one-sided confidence intervals $[0,\lambda^\mathrm{UL}]$ which, by means of the invariance of confidence intervals under monotone transformations, translate into one-sided confidence intervals $[E_\mathrm{c,\;\gamma}^\mathrm{LL},\infty)$ on the energy cutoff 
with lower limit $E_\mathrm{c,\;\gamma}^\mathrm{LL}=1/\lambda^\mathrm{UL}$.\\
The discussion in this section is based on binned event counts $\vec{c}=(c_1,\dots,c_N)$ obtained in realistic simulations of the CTA instrument, including the full instrumental response functions and a residual cosmic ray background model for the southern array \citep{cta_mc}. The event simulation and analysis is performed with the open source framework {\tt gammapy} \citep{gammapy}\footnote{The methods which were developed for the limit calculation are available as \href{https://github.com/residualsilence/ecpli}{'ecpli'} open-source package \citep{ecpli}.}. 
The data-binning into a total of $N$ bins can be either one- or three-dimensional. The latter refers to an independent binning 
in event energy and direction while the one-dimensional binning refers to spatially integrated data. 
In either case, the bin size is much smaller than the respective instrumental resolution in space and energy. 
Limits on the inverse spectral cutoff $\lambda$ are investigated within $\gamma$-ray emission models which, after convolution with the instrumental response functions, 
predict $\vec{n}=(n_1,\dots,n_N)$ counts. The investigated $\gamma$-ray emission models are typically the sum of multiple model components, including flux-parameterizations 
of the expected background and the $\gamma$-ray source of interest. Model parameters 
are the inverse energy cutoff $\lambda$ of the source of interest and a set of other nuisance parameters $\boldsymbol{\theta}$ including, for example, 
flux normalizations and background model parameters.\\
The Poisson likelihood, defined as 

\begin{equation}
\label{LimitLik}
	L(\lambda, \boldsymbol{\theta}|\vec{c}):=\prod_{i=1}^N \exp(-n_i)\frac{n_i}{c_i!}^{c_i}
	\;\mathrm{,}
\end{equation}	
is assumed as connection between simulated event counts $\vec{c}$ and counts $\vec{n}=\vec{n}(\lambda,\boldsymbol\theta)$ 
predicted within the assumed model parameterizations. For reasons of numerical stability and computational performance, the Cash statistic 
$C(\lambda,\boldsymbol{\theta})=2\sum_i(n_i-c_i\ln n_i)$ \citep{cash} 
is used frequently in the following instead of the full likelihood $L(\lambda, \boldsymbol{\theta}|\vec{c})$. The Cash statistic is, 
up to a term which is independent of the model parameters $\lambda$ and $\boldsymbol{\theta}$, proportional to the logarithm of the likelihood given in Eq. \ref{LimitLik}.\\
A detailed comparison of statistical methods to derive $\lambda^\mathrm{UL}$ in the context of typical CTA data analyses is discussed in this section. A selection of 
statistical methods is presented in Sec. \ref{DifferentLimitApp}. 
The comparison discussed in Sec. \ref{CompLimitApp} aims primarily towards the
frequentist coverage of the derived limits and the cutoff 
sensitivity of the method.
Cutoff sensitivity is defined in this context as 
the median limit on the energy cutoff
at fixed true energy cutoff and confidence level. A comparison of the robustness of the methods against mis-specifications of the 
likelihood model in Eq. \ref{LimitLik}, e.g. due to un-modeled systematic errors \citep{gerrit}, is beyond the scope of the discussion.
\subsection{Different approaches}
Different statistical approaches to derive the upper limit $\lambda^\mathrm{UL}$ on the inverse energy cutoff are discussed in this section. The frequentist coverage of the 
upper limit $\lambda^\mathrm{UL}$ is motivated for all methods, under suitable conditions. These conditions are, however, typically difficult to be explicitly verified in a concrete situation. 
Whether or not a reasonable coverage is achieved in practical analyses needs to be tested in realistic Monte Carlo simulations as discussed in Sec. \ref{CompLimitApp}.
\label{DifferentLimitApp}
\subsubsection{Profile likelihood}
\begin{figure*}
\centering
\includegraphics[width=17cm]{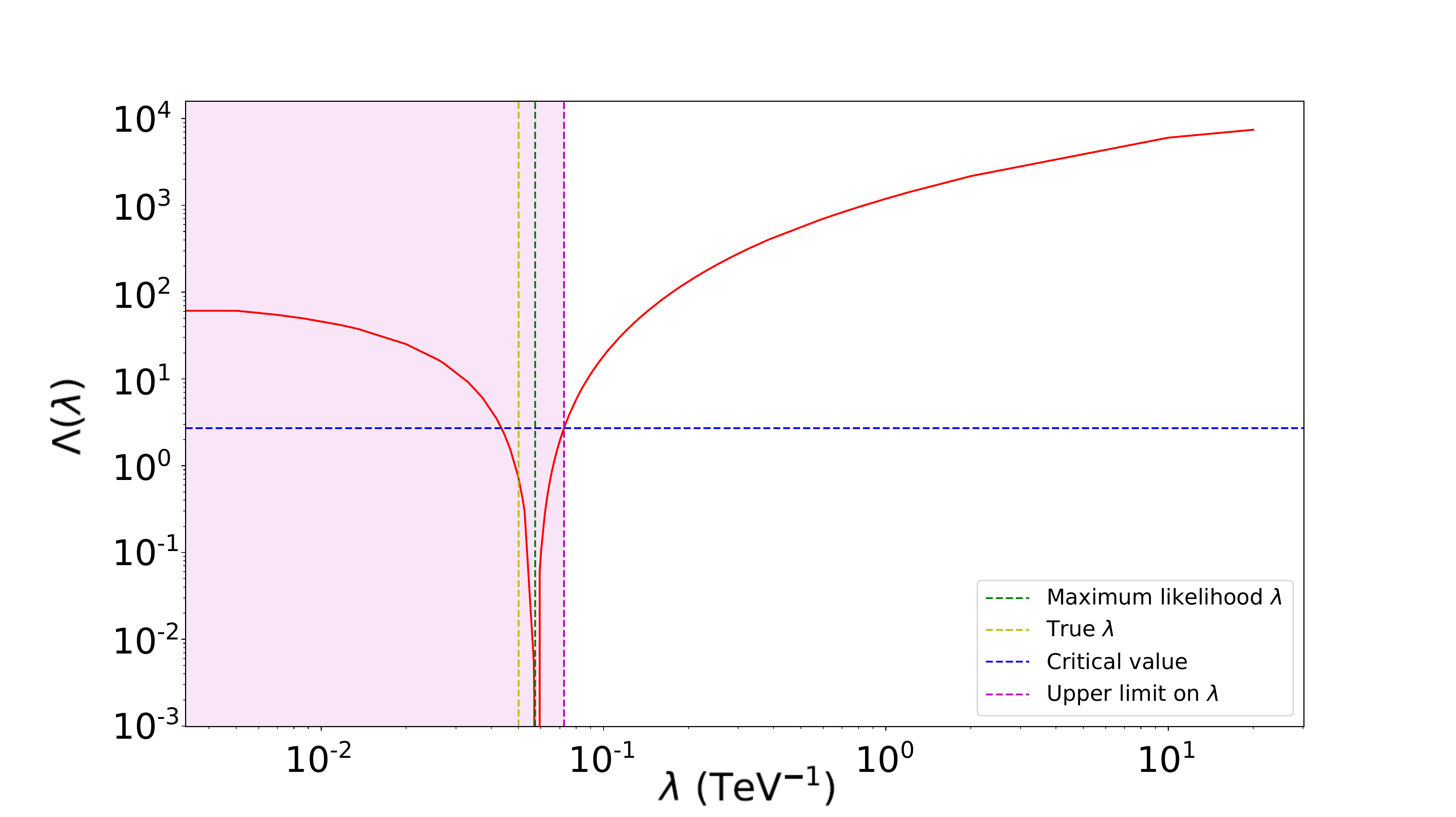}
\caption{Shown in red is the likelihood ratio statistic $\Lambda(\lambda)$ for the analysis of a typical $\gamma$-ray source with true exponential cutoff at $20$ TeV. The maximum likelihood estimator $\hat\lambda=\mathrm{max}(0, \lambda^\mathrm{ML})$ for non-negative $\lambda$ is shown as green vertical line. The yellow line shows the true value of $\lambda$, i.e. $\lambda^\mathrm{true}=1/20\;\mathrm{TeV^{-1}}$. The blue horizontal line indicates the critical value, $\Lambda_\mathrm{crit}\approx 2.71$, for a CL of $\mathrm{CL}=0.95$. The upper limit on $\lambda$ is constructed at $95\%$ CL as intersection of $\Lambda(\lambda)$ with the horizontal critical value in the interval $[\hat\lambda,\infty)$, i.e. as upper end of the acceptance interval of the corresponding likelihood ratio test. The magenta colored area indicates the range of inverse cutoff parameters which is not excluded at $95\%$ CL.}
\label{lr_figure}
\end{figure*}

The profile likelihood method is an example for the inversion of a frequentist hypothesis test. 
Let $L(\lambda)=\mathrm{max}_{\boldsymbol\theta} L(\lambda,\boldsymbol{\theta}|\vec{c})$ be the profile likelihood (see e.g. \cite{pdb})
of Eq. \ref{LimitLik} with respect to the inverse energy cutoff $\lambda$ and $C(\lambda)=\mathrm{max}_{\boldsymbol\theta} C(\lambda,\boldsymbol\theta)$ 
the corresponding Cash statistic.
Let further $\hat\lambda$ be the maximum likelihood estimator for the inverse energy cutoff over the constrained range $\lambda\geq0$. 
To calculate $\hat\lambda$ in practice, the maximum likelihood estimate $\lambda^\mathrm{ML}$ for $\lambda$ is derived over the full range for $\lambda$, including negative values. 
When $\lambda^\mathrm{ML}$ is positive, $\hat\lambda$ is set to $\lambda^\mathrm{ML}$ and otherwise $\hat\lambda=0$. Together, this means $\hat\lambda=\mathrm{max}(\lambda^\mathrm{ML},0)$. 
This method assumes that $L(0)$ is the global maximum likelihood for non-negative $\lambda$ when $\lambda^\mathrm{ML}<0$. 
The latter is true when the profile likelihood $L(\lambda)$ has a unique maximum, which is assumed in the  following. The likelihood ratio test statistic
		
\begin{equation}
\label{LikRat}
\Lambda(\lambda):=-2\ln\frac{L(\lambda)}{L(\hat{\lambda})} = C(\lambda)-C(\hat\lambda)
\end{equation}
enables the comparison of the maximum likelihood $L(\hat{\lambda})$ over $\lambda\geq0$ with the likelihood $L(\lambda)$ for a fixed inverse energy cutoff $\lambda$.
Figure \ref{lr_figure} shows the likelihood ratio statistic $\Lambda(\lambda)$ for a typical analysis together with $\hat\lambda$. 
Given the constraint on non-negative $\lambda$, Eq. \ref{LikRat} is the test statistic for a likelihood ratio test of the null hypothesis $H_0:\;\lambda=\hat\lambda$ against the alternative hypothesis $H_1:\;\lambda=\lambda$. The alternative hypothesis is accepted, i.e. the inverse cutoff parameter $\lambda$ does not give a significantly worse description of the data than the best fitting $\hat\lambda$, when the test statistic $\Lambda(\lambda)$ is smaller than or equal to the critical value $\Lambda_\mathrm{crit}(\mathrm{CL})$ at a given confidence level CL, i.e. $\Lambda(\lambda)\leq\Lambda_\mathrm{crit}(\mathrm{CL})$. The critical value for $95\%$ CL is shown as a horizontal line in Fig. \ref{lr_figure}. 
The acceptance region $A=\{\lambda|\Lambda(\lambda)\leq\Lambda_\mathrm{crit}(\mathrm{CL})\}$ is, due to the assumed unique maximum of the profile likelihood $L(\lambda)$, an interval $A=[\lambda^\mathrm{LL},\lambda^\mathrm{UL}]$. Let $F^{-1}_{\chi_1}$ denote the inverse cumulative probability density function of a $\chi^2$-distributed random variable with one degree of freedom. The choice of $\Lambda_\mathrm{crit}(\mathrm{CL})=F^{-1}_{\chi^2}(2\mathrm{CL}-1)$ is motivated in analogy to the construction in \cite{gerrit}, where a F-test is inverted to derive a limit on 
an exponential cutoff. For this choice of the critical value it is expected that, asymptotically and under suitable regularization conditions, the coverage of the interval $[0,\lambda^\mathrm{UL}]=[0,\lambda^\mathrm{LL})\cup A$, shown as red shaded area in Fig. \ref{lr_figure}, is at least CL \citep{gerrit}. This means that, when the respective conditions apply, $\lambda^\mathrm{UL}$ is a frequentist upper limit on the inverse cutoff parameter at confidence level CL.

\subsubsection{Markov-Chain Monte Carlo}
\label{MCMC_intro}
An upper limit on the inverse cutoff parameter can also be derived when the probability distribution of model parameters is expressed in the framework of Bayesian terminology.
Using an a-priori probability density $p(\lambda,\boldsymbol\theta)$ for the model parameters and 
the ``evidence"

\begin{equation}
	p(\vec{c}):=\int d\lambda d\boldsymbol\theta\; L(\lambda,\boldsymbol\theta|\vec{c})\; p(\lambda,\boldsymbol\theta)\;\mathrm{,}
\end{equation}
the posterior probability density for the model parameters 
$p(\lambda, \boldsymbol\theta|\vec{c})$ given new data $\vec{c}$ can be calculated with Bayes theorem as 

\begin{equation}
	\label{mcmc_eq}
	p(\lambda,\boldsymbol\theta|\vec{c})=\frac{L(\lambda,\boldsymbol\theta|\vec{c})\;p(\lambda,\boldsymbol\theta)}{p(\vec{c})}\;\mathrm{.}
\end{equation}
Markov-Chain Monte Carlo (MCMC) methods refer in the following to techniques which allow the direct sampling from the posterior probability density without the full evaluation of Eq. \ref{mcmc_eq}. 
The techniques avoid, at the price of a correlation between consecutive samples, the computationally complex calculation of the evidence as necessary in traditional Monte Carlo methods for the evaluation of Eq. \ref{mcmc_eq}.\\
The inverse energy cutoff is physically constrained to non-negative values. This translates to the constraint $p(\lambda,\boldsymbol\theta)=0$ when $\lambda<0$ for the prior density. 
Due to this constraint, it holds that the marginal posterior distribution $p(\lambda|\vec{c})=\int d\boldsymbol\theta\; p(\lambda,\boldsymbol\theta|\vec{c})=0$ when $\lambda<0$. An upper limit $\lambda^\mathrm{UL}$ of a credible interval $[0,\lambda^\mathrm{UL}]$ can therefore be calculated as a quantile of the marginal posterior distribution at given credible level $\mathrm{CL}=\int_0^{\lambda^{UL}}d\lambda\;p(\lambda|\vec{c})$. 
Asymptotically and under certain regularization conditions, in particular regarding the smoothness of the prior density close to the true model parameter values, the credibility level $\mathrm{CL}$ is expected to equal the coverage of the interval $[0,\lambda^\mathrm{UL}]$ (see Bernstein-von Mises theorem, 
e.g. \citep{vaart}). The credible interval is then numerically equal to a confidence interval and $\lambda^\mathrm{UL}$ is an upper limit at confidence level CL.\\
MCMC methods differ widely in the concrete technique used to draw samples from the posterior distribution. The {\tt emcee} implementation \citep{emcee_hammer} of the affine invariant ensemble sampler \citep{emcee} is used in this work. The affine invariant ensemble sampler is based on an ensemble of stochastic processes, called walkers, which are set up to efficiently explore the posterior parameter space \citep{emcee, emcee_hammer}. Each ensemble-step of the exploration of the parameter space is affine invariant. This symmetry property of the algorithm allows the handling of highly correlated variables. It was, however, argued that the efficiency of the exploration decreases with increasing number of parameter space dimensions \citep{emcee_high_dimensions} due to the constraint on affine invariant steps. The {\tt emcee} implementation uses the natural logarithm of the posterior density, modulo additive terms which are independent of the model parameters, as input to generate samples from the posterior density. In practice, $-0.5\, C(\lambda,\boldsymbol\theta)+\ln(p(\lambda,\boldsymbol\theta))$ is used in this work for reasons of 
computational efficiency.\\
Following the pragmatic recommendation in \cite{emcee_hammer}, walkers are initialized with random parameters close to the maximum likelihood estimators. When the maximum likelihood estimator 
for the inverse cutoff parameter is negative, a small positive value is chosen. After the initialization of the walkers, a number of Markov-Chain samples are to be discarded until it can be argued that the Markov-Chain samples converged to samples from the posterior. The length of this ``burn-in" period is, following the recommendation in \cite{emcee}, estimated as a multiple of the autocorrelation time of the Markov-Chain. Also following \cite{emcee}, the number of walkers is chosen as a high multiple of the number of free parameters, ensuring a large acceptance fraction for each ensemble step. The statistical error on the derived limit decreases with increasing number of samples. Control over the statistical error is achieved by means of the comparison of multiple
equally set up and independently run Markov-Chains. In practice, the number of samples is selected such that the variation of upper limits on $\lambda$ measured within multiple Markov-Chains is below a predefined precision target. Finally, the prior density $p(\lambda,\boldsymbol\theta)$ is assumed to factor, i.e. $p(\lambda,\boldsymbol\theta)=p(\lambda)\;\prod_i p(\theta_i)$ for $\theta_i\in\boldsymbol\theta$, ignoring a-priori information on parameter correlations. Different concrete choices for the prior density distributions of the respective parameters are discussed in Sec. \ref{CompLimitApp}. 
\subsubsection{Bootstrap}
The bootstrap method relies on the resampling of binned $\gamma$-ray events $\vec{c}$ as ``bootstrap samples" $\vec{c_*}$ \citep{efron1979}. 
A maximum likelihood estimator $\lambda_*$ for the inverse 
energy cutoff parameter is calculated for each bootstrap sample $\vec{c_*}$, resulting in the distribution $f(\lambda_*)$ of the bootstrap estimates. The percentile method \citep{efron1979, bootstrap} is used in the following to infer an upper limit $\lambda^\mathrm{UL}$ as smallest positive $\lambda^\mathrm{UL}$ which satisfies 
$\mathrm{CL}\leq\int_{-\infty}^{\lambda^\mathrm{UL}}d\lambda_*\;f(\lambda_*)$. The interval  $(-\infty,\lambda^\mathrm{UL}]$ and, because the true value of $\lambda$ is always non-negative, also the interval $[0,\lambda^\mathrm{UL}]$ are expected to have frequentist coverage CL under regularization conditions which are, for example, discussed in \cite{efron_computer}.\\ 
The number of bootstrap samples $\vec{c_*}$ to draw from the data $\vec{c}$ is, similar to the discussion of the number of MCMC samples in Sec. \ref{MCMC_intro}, adjusted such that the variation of the limits on $\lambda$ derived from different sets of bootstrap samples agrees within a predefined precision.\\
Different resampling techniques are compared in the following. The non-parametric bootstrap \citep{efron1979, efron_computer} 
resamples binned events as multinomial distributed random vector, $\vec{c_*}\sim\mathrm{Mult}(K,\;\vec{c})$, i.e. $K=\sum_{i=1}^N c_i$ events are drawn with replacement from the $N$ bins in $\vec{c}$. 
Additionally to the unparametric bootstrap, two different parametric bootstrap \citep{efron_computer} resamplings are investigated. 
The "Poisson bootstrap" \citep{poisson_bootstrap} models the number of counts in each bin as independent Poisson-distributed random variable with expectation $c_i$, i.e. $c_{i*}\sim\mathrm{Pois}(c_i)$. A major difference to the unparametric bootstrap is that the total number of 
events $K$ is itself a random variable for the Poisson bootstrap. For large number of bins and large number of total events, the non-parametric resampling agrees with the Poisson bootstrap. The 'best fit bootstrap' makes use of Poisson samples from the maximum likelihood fit to a model, i.e. $c_{i*}\sim\mathrm{Pois}(n_i)$.

\subsection{Performance comparison}
\label{CompLimitApp}
Given that the respective conditions are fulfilled, frequentist limits are expected to result from all of the methods discussed in Sec. \ref{DifferentLimitApp}. However, considering asymptotic conditions as an example, only a partial fulfillment can be expected in practice with finite experimental data. It remains to be tested with realistic simulations whether the expected frequentist coverage of the derived limits holds to a reasonable degree given a finite amount of experimental data. These coverage tests are discussed in the following at $95\%$ confidence level. Different, in particular larger, confidence levels are not studied. Additionally to the limit coverage, the question as to which method provides the best
sensitivity at given true energy cutoff is discussed in this subsection.\\
In Sec. \ref{special_sources}, the comparison of results obtained with different methods is limited to a special set of $\gamma$-ray 
point-like source flux parameters. Generalizations to other point-like source parameters and situations with a complex source morphology, 
consisting of multiple and extended sources, are presented in Sec. \ref{general_sources}. Also discussed in \ref{general_sources} is the lower limit on the hadron spectrum cutoff parameter inferred from $\gamma$-ray measurements.

\subsubsection{Point-like source analysis: Special case}
\label{special_sources}
To facilitate the comparison of methods for a simulated $\gamma$-ray point-like source, given computational restrictions, only a single special case is considered. The model given by Eq. \ref{eq2} was chosen, with a flux normalisation and power--law index of $\phi_0=50$ mCrab and $\Gamma=2.1$. The true energy cutoff was then varied in a range between a few TeV and few hundred TeV.
Reconstructed events from $10$ h of $\gamma$-ray point-like source observations with CTA South at a zenith angle of $20^\circ$ are simulated. Within this setup, lower limits on the energy cutoff are derived with the different methods discussed in Sec. \ref{DifferentLimitApp}. In a first step, only 1-dimensional analyses are discussed.In these analyses, events are binned in reconstructed energy but spatially integrated over a signal region of $0.11^\circ$ radius around the true source position. 
The restriction to 1-dimensional analyses enables a computationally efficient comparison of different methods. 
Results for 3-dimensional analyses, which are computationally much more demanding, are discussed at the end of this subsection.\\
The energy cutoff limits obtained with the bootstrap and MCMC methods are calculated with a precision better than 2\%. Two different sets of prior density distributions 
for the model parameters $\phi_0$, $\Gamma$ and $\lambda$ are investigated for the MCMC method. Uniform prior densities 
have been used in $\gamma$-ray astronomy analyses before, e.g. in \cite{hess_emcee}. However, the uniformity depends on the choice of the parameter. E.g. a uniform prior density in $\lambda$ implies a non-uniform prior density in the energy cutoff $E_\mathrm{c,\;\gamma}=1/\lambda$. Results based on uniform prior densities are therefore compared to results obtained with priors based on gamma distributed random variables with reasonable parameters.

\begin{figure}[ht!]
\includegraphics[width=\hsize]{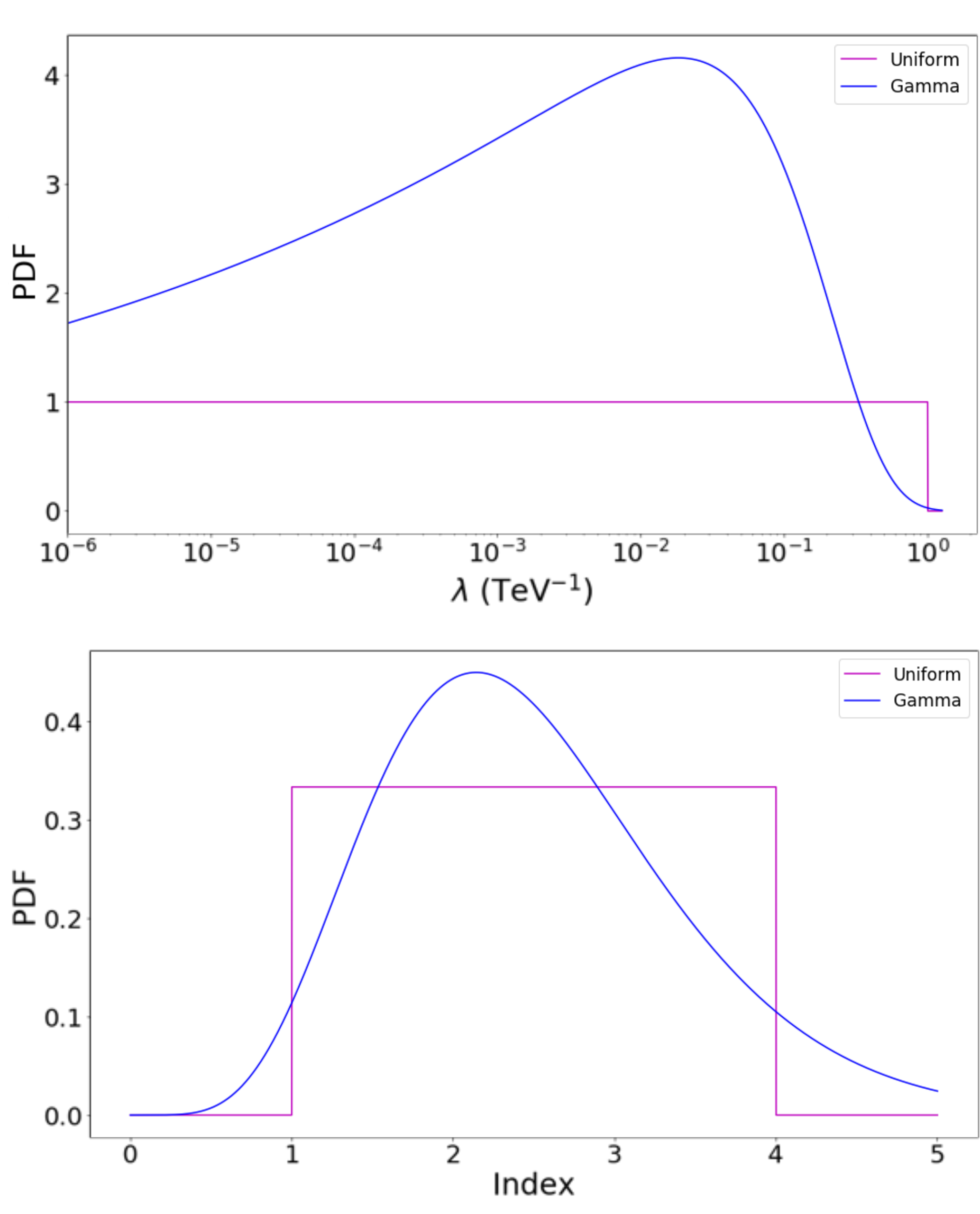}
\caption{Investigated MCMC analysis prior distributions for the inverse energy cutoff $\lambda$ (top) and power-law index $\Gamma$ (bottom). Shown in magenta and blue are the assumed uniform and gamma distributed priors respectively. The uniform prior distributions for $\lambda$ and $\Gamma$ attribute a non-zero density within $[0,1]\;\mathrm{TeV^{-1}}$ and $[1,4]$. The gamma prior distributions have shape (scale) parameters of 1.1 (0.182) and 7 (0.357), respectively for $\lambda$ and $\Gamma$.}
\label{prior_fig}
\end{figure}

\begin{figure*}[ht!]
\centering
\includegraphics[width=17cm]{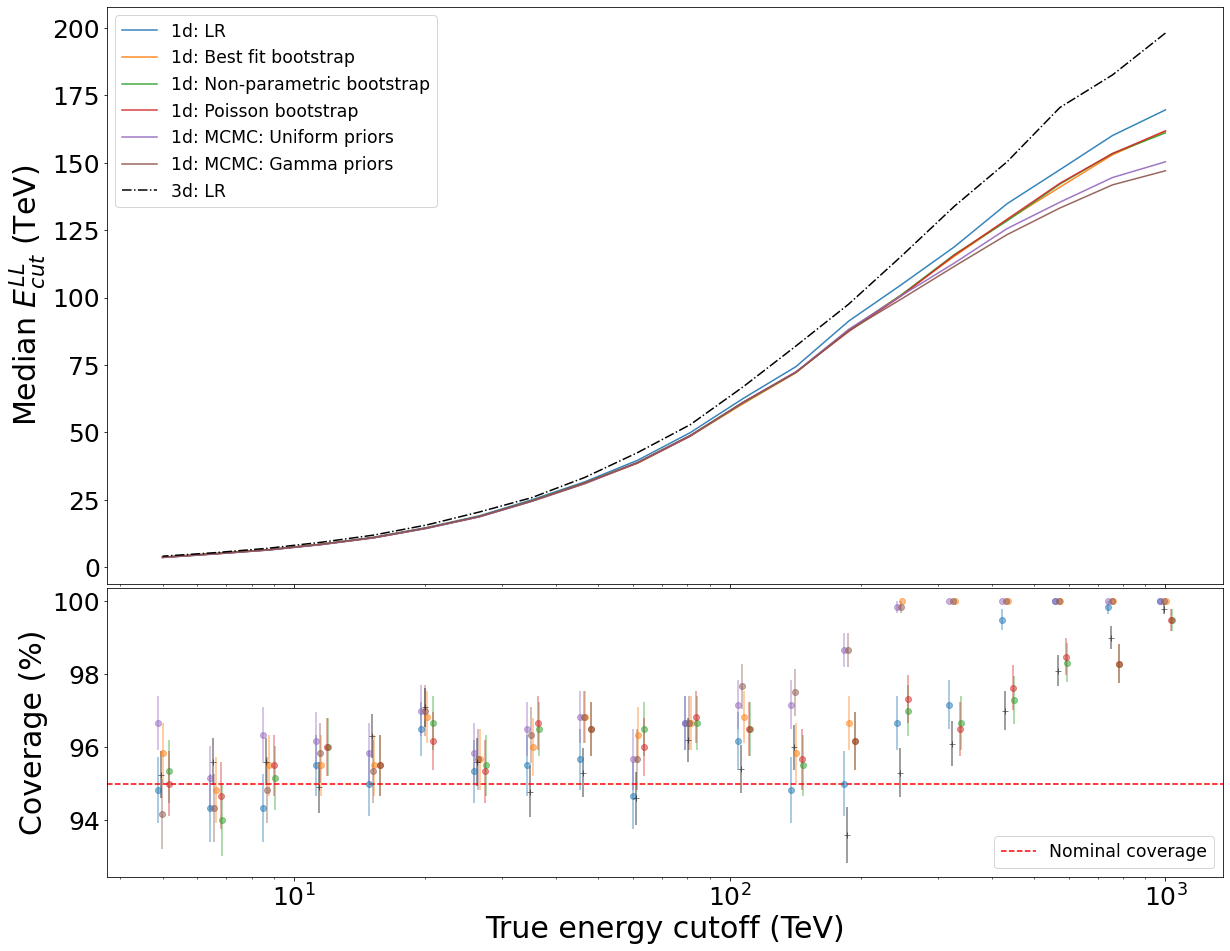}
\caption{Comparison of different methods to derive a lower limit on the energy cutoff for a $\gamma$-ray point-like source with power-law index $\Gamma=2.1$ and flux normalization $\phi_0=50$ mCrab. The instrument response functions of CTA South and $10$ h of observation time at a zenith angle of $20^\circ$ are assumed. Upper panel: Median $95\%$ CL lower limit on the energy cutoff as a function of the true energy cutoff. Color codes for the different methods are indicated in the legend. Solid lines correspond to 1-dimensional analyses. Note that the green, orange and red solid lines overlap. The black dash-dotted line corresponds to profile likelihood limits derived with a 3-dimensional analysis. Lower panel: Coverage of the interval $[E_\mathrm{c,\;\gamma}^\mathrm{LL},\infty)$ for different methods to derive $E_\mathrm{c,\;\gamma}^\mathrm{LL}=1/\lambda^\mathrm{UL}$ as a function of the true energy cutoff. The true energy cutoff is slightly shifted for each method to improve the visibility. Errorbars are approximate $68.3\%$ CL intervals on the coverage, derived under the assumption of the normal approximation to the binomial distribution.}
\label{compar1d}
\end{figure*}

Figure \ref{prior_fig} shows, as an example, the assumed prior distributions for the two parameters $\lambda$ and $\Gamma$.\\
A comparison of the coverage and sensitivity for the specific $\gamma$-ray point-like source considered here are shown in Fig. \ref{compar1d}. All investigated methods lead to upper limits $\lambda^\mathrm{UL}$ on the inverse energy cutoff with reasonable 
frequentist coverage for the respective confidence interval $[0,\lambda^\mathrm{UL}]$. 

The lower panel of Fig. \ref{compar1d} shows that, for the source parameters investigated, excellent coverage is observed for true cutoff energies smaller than $100$ TeV. 
All methods start to over-cover when the true energy cutoff becomes larger than $100$ TeV. The over-coverage for large true cutoff energies can, in analogy to the discussion in 
\cite{gerrit}, be related to the decreasing cutoff detection probability.\\
Given the reasonable coverage of the limits derived with all different methods discussed in Sec. \ref{DifferentLimitApp}, a comparison of their respective sensitivity, defined 
as median lower limit, is motivated. Shown as solid lines in the upper panel of Fig. \ref{compar1d} are the median lower limits on the energy cutoff derived in 1-dimensional analyses as a function of the true energy cutoff. The median limit obtained with different methods relative to the sensitivity of the 1-dimensional profile likelihood analysis is shown in the upper panel of Fig. \ref{relative_comparison}. The profile likelihood method leads, among all the methods investigated with 1-dimensional analyses, 
to the most constraining limits for true cutoff energies between $10$ TeV and $1$ PeV. However, the improvement in sensitivity of the 1-dimensional analysis profile likelihood method over other 1-dimensional analysis methods is below $10\%$ for true $\gamma$-ray energy cutoffs below $100$ TeV.\\
The choice of the prior distribution for the MCMC analysis has an impact on the sensitivity when the true energy cutoff is either within $5$ TeV and $20$ TeV or within $200$ TeV and $1$ PeV, as can be seen in the upper panel of Fig. \ref{relative_comparison}. For example, the MCMC analysis with uniform prior distributions leads to the worst sensitivity among all investigated methods when the true energy cutoff is within $5$ and $10$ TeV. The MCMC analysis with gamma prior distributions leads in the same range of true energy cutoffs to the best sensitivity among all investigated 1-dimensional analyses. It is obvious that the sensitivity of the MCMC method depends strongly on the match between true values and prior distributions.\\
The sensitivity of all investigated bootstrap methods is compatible. The overall best sensitivity is obtained with the 3-dimensional profile likelihood method. However, as shown in the upper panel of Fig. \ref{relative_comparison}, even this method only provides an improvement of less than $20\%$ when compared to the sensitivity of the 1-dimensional profile likelihood method.
\subsubsection{General analyses}
\label{general_sources}
\begin{figure*}[ht!]
\centering
\includegraphics[width=17cm]{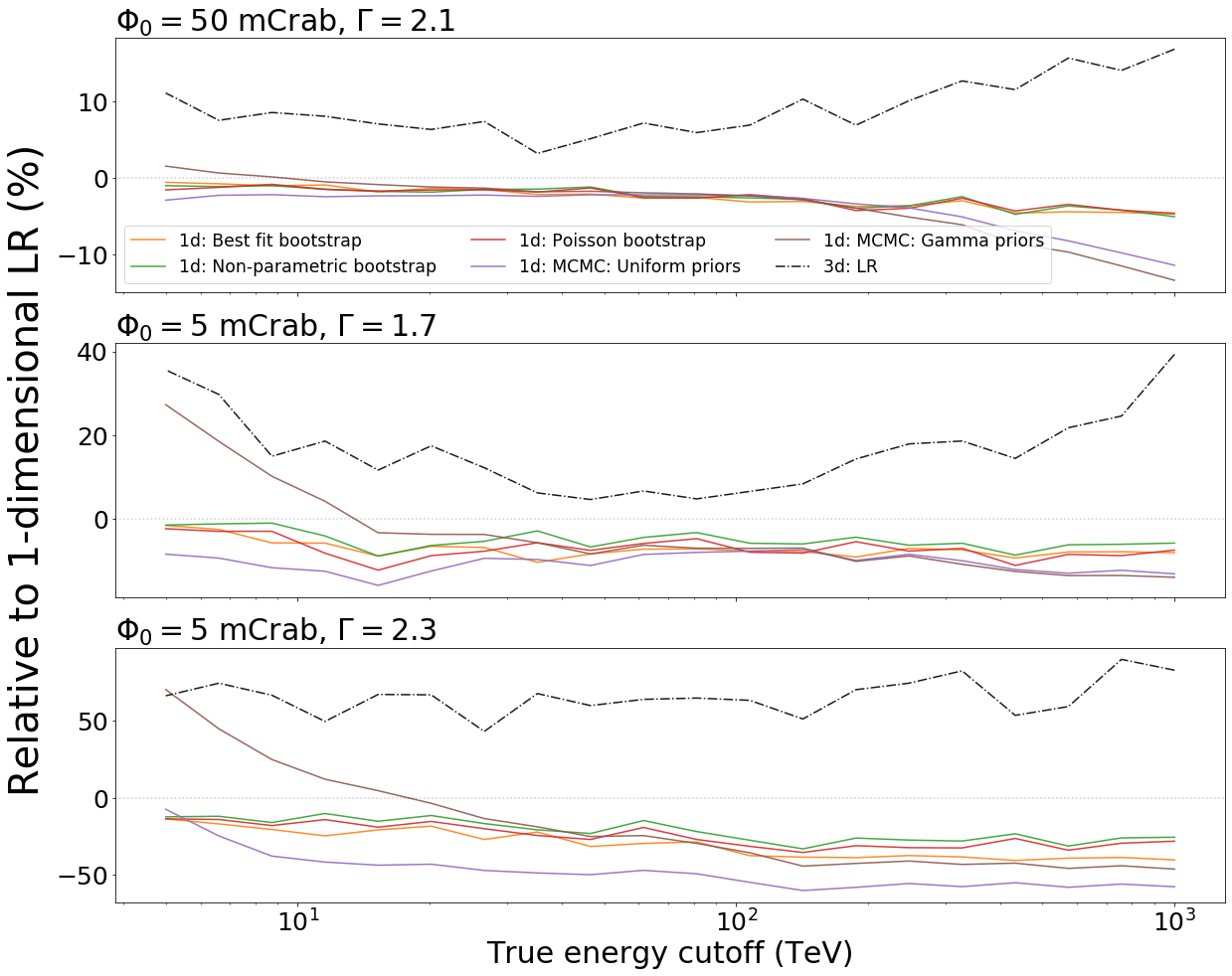}
\caption{Point-like source energy cutoff sensitivity as a function of the true energy cutoff for different methods indicated by the legend in the upper panel relative to the respective sensitivity achievable with a 1-dimensional profile likelihood analysis. The different panels show the relative sensitivity for different point-like source parameters in terms of flux normalization $\phi_0$ and index $\Gamma$, specified in the panel title. The instrument response functions of CTA south and $10$ h of observation time at a zenith angle of $20^\circ$ are assumed.}
\label{relative_comparison}
\end{figure*}

In Sec. \ref{special_sources}, the performance of different methods to derive a lower limit on the energy cutoff for a special point-like source with parameters $\phi_0=50\;\mathrm{mCrab}$ and $\Gamma=2.1$ is discussed. More general point-like sources with parameters $\phi_0\in[5,50]\;\mathrm{mCrab}$ and $\Gamma\in[1.7,2.3]$ are investigated in the following. Limits derived with all methods discussed in Sec. \ref{DifferentLimitApp} lead to a reasonable coverage of the respective intervals $[0,\lambda^\mathrm{UL}]$ and $[E_\mathrm{c,\;\gamma}^\mathrm{LL},\infty)$ 
with over-coverage towards large true energy cutoffs. Significant undercoverage is not observed. Figure \ref{relative_comparison} shows the limit sensitivity relative to that of the 1-dimensional profile likelihood analysis for representative $\gamma$-ray point-like source parameters. Among the investigated 1-dimensional analyses, the profile likelihood method leads typically to the most constraining limit. The exception are small true energy cutoffs where the profile likelihood limits are less constraining than the limits derived with the MCMC method assuming gamma distributed priors. Depending on the source parameters and the true energy cutoff, MCMC results can significantly depend on the exact choice of the prior parameter distributions. For example, for a weak ($\phi_0=5\;\mathrm{mCrab}$) and soft ($\Gamma=2.3$) point-like source, the lower limit on the energy cutoff derived under the assumptions of uniform and gamma distributed priors varies by $20\%$ to $50\%$. The inclusion of spatial fit parameters in 3-dimensional profile likelihood analyses leads to energy cutoff constraint 
improvements over the respective 1-dimensional profile likelihood analyses which are in between a few percent and $60\%$.

Further tests were performed to guarantee the frequentist coverage of lower limits on the energy cutoff derived with the 3-dimensional profile likelihood method for point-like sources with a hadronic particle population whose $\gamma$-ray emission is modeled with the framework described in \cite{naima}.\\
In practice, $\gamma$-ray sources can show a more complicated morphology than the point-like sources discussed up to this point. A realistic example is the $\gamma$-ray source HESS\,J1641$-$463. This source is observed as point-like with H.E.S.S. and the $\gamma$-ray energy spectrum is compatible with a power-law without indications of an exponential cutoff until at least $20$ TeV \citep{hess1641_paper, hess1641_proc}. However, the analysis is complicated by the nearby extended $\gamma$-ray source HESS\,J1640$-$465 \citep{hess1640_paper} for which an exponential cutoff is measured around $6$ TeV. A realistic simulation of this double source setup of HESS\,J1640$-$465 and HESS\,J1641$-$463, assuming model parameters as derived from the measurements discussed in \cite{hess1640_paper, hess1641_paper} and varying true energy cutoff parameters for the source HESS\,J1641$-$463, is performed.
Due to convergence problems during the optimization for the profile likelihood function, it is in general not possible to derive a frequentist limit on the energy cutoff parameter of HESS\,J1641$-$463 with the profile likelihood method. However, a lower limit on the energy cutoff with reasonable frequentist coverage can be derived with the MCMC method. A detailed study of the possible dependence between limits on the energy cutoff parameter and prior information is beyond the scope of this work.

\subsection{Conclusion}
The profile likelihood method provides a computationally very efficient way to derive lower limits on the energy cutoff. Other methods are typically less sensitive or, in case of the MCMC 
method, a possible sensitivity improvement in restricted parameter ranges results from the choice of the prior distributions. The computational effort to derive reasonably precise limits with the described bootstrap and MCMC implementations is also larger than for the profile likelihood method. However, bootstrap and MCMC methods can provide an important alternative in cases where the profile likelihood method cannot be applied, e.g. due to non-converging optimizations for the profile likelihood function.
\section{Notes on the treatment of multiple hypotheses testing}
\label{multiple_testing_appendix}
Different likelihood ratio tests, for example to test for the presence of a $\gamma$-ray source or the presence of a spectral
cutoff, are applied in this work to each source in a source sample. 
The treatment of multiple hypotheses tests is described in the following.
\subsection{Multiple testing for the presence of PeVatrons and spectral cutoffs}
\label{multiple_testing_PTS}
Given a $\gamma$-ray source, the presence of a spectral cutoff is tested with the likelihood ratio test statistic $\mathrm{TS}_\lambda$ defined in Eq. \ref{ts_lambda}. Similarly, the PTS defined in Eq. \ref{eq_PTS} is used to test whether or not the source is a PeVatron. Both tests are used to identify a subset of sources with a certain property in a sample with $\mathcal{O}(100)$ sources. A type I error, i.e. a false positive classification for one source, for example when a source is misclassified as PeVatron, is a severe error which must be avoided when real data are analyzed. Tests for the presence of PeVatrons or spectral cutoffs are therefore based on a fixed family-wise error rate (FWER), 
i.e. on a fixed probability for at least one false positive per sample. Each hypothesis test for the presence of a PeVatron or a spectral cutoff is performed independently. The independence results from the simulation of source samples as isolated sources without
spatial overlap. The FWER is controlled with the Bonferroni correction \citep{wasserman} at a global significance level corresponding to $5\sigma$ or a local test statistic threshold of $\mathrm{TS}=33.9$ for $100$ sources per population. The Bonferroni
correction is a conservative approach which sacrifices the source 
detection probability, or equivalently increases the likelihood for a false negative, to control the FWER. Alternatives with equal control over the FWER but a smaller type II error, i.e. a false negative, such as the methods discussed in \cite{holm} or \cite{hochberg} were tested without significant changes. This conclusion regarding the equivalence of the simple Bonferroni and e.g. the methods discussed in \cite{holm} and \cite{hochberg} must, however, not hold in a setup with more than $\mathcal{O}(100)$ sources per sample.

\subsection{Multiple testing for the presence of a source}
Similar to the presence of a PeVatron and a spectral cutoff, 
the presence of a $\gamma$-ray source is also tested with a likelihood ratio test as described in Sec. \ref{cta_sim_ana}. The source model used in the definition of $\mathrm{TS}_\mathrm{Det}$ in Eq. \ref{ts_detection} has parameters for the flux normalization $\phi_0$ and for the spatial and spectral setup $\vec{\theta}$. The null hypothesis is equivalent to $\phi_0=0$ while the alternative hypothesis is equivalent to $\phi_0>0$. The parameters $\vec{\theta}$ are in this context nuisance parameters which are undefined under the null hypothesis. For example, the nuisance 
parameters which describe the location of a source can take any value when $\phi_0=0$. The optimization with regard to nuisance
parameters which are undefined under the null hypothesis can be considered as multiple correlated hypotheses tests \citep{jan_knut, jan2015}. In this case, Monte Carlo simulations can be used to quantify the type I error probability.A similar analysis setup and corresponding Monte Carlo simulations for the type I error rate are discussed in \cite{hgps} for the H.E.S.S. GPS analysis. The local test statistic threshold of $\mathrm{TS}=30$ is found to correspond to a global type I error probability of $3\%$ per source population. This local test statistic threshold value is used in this work as threshold for $\mathrm{TS}_\mathrm{Det}$. However, the angular resolution of CTA is expected to be better than the angular resolution of H.E.S.S. The spatial correlation of tests with $\mathrm{TS}_\mathrm{Det}$ is therefore different for analyses of CTA data than it is for H.E.S.S. data. As a consequence, the type-I error probability of $3\%$ for the threshold value of $\mathrm{TS}=30$ found in \cite{hgps} is likely to be unrealistic for CTA. Dedicated Monte Carlo simulations, possibly using methods discussed in \cite{gross_vitells} to limit the computational efforts for small type I error probabilities, are to be preformed for the CTA GPS data. However, these simulations are beyond the scope of this work.

\section{Systematic uncertainty of the energy cutoff reconstruction}
\label{systematics}
Simulations with modified IRFs are performed to quantify the effect of systematic errors on the determination of the spectral energy cutoff. Modified IRFs, $X^\prime$, are obtained from standard IRFs, \textit{X}, by re-scaling with $\pm1\sigma_{sys}$, i.e. 

\begin{equation}
X^{\prime} = X \cdot (1\pm \epsilon  B(E))\,\mathrm{,}
\end{equation}
where $\epsilon$ is the relative systematic uncertainty for a given IRF and $B(E)$ is an energy dependent function with $|B(E)|\leq 1$, which defines the shape of the IRF modification. A constant shift, i.e. $B(E)=1$, and a gradient shift over energy are investigated. The latter is defined as 

\begin{equation}
B(E) = \frac{\ln(E/E_\mathrm{min}) + \ln(E/E_\mathrm{max})}{\ln(E_\mathrm{max}/E_\mathrm{min})}.
\end{equation}
The energy range considered for the gradient shift is $E_\mathrm{min}=$100~GeV to $E_\mathrm{max}=$160~TeV. Only IRFs for the effective area and the energy bias are considered. The expected maximal relative systematic uncertainties, $\epsilon$, for CTA are 5\% and 6\%, respectively for the effective area and the energy bias. Three different source types are investigated: A faint source ($\Gamma=2.3$, $\phi_0=10$ mCrab), a medium source ($\Gamma=2.0$, $\phi_0=30$ mCrab) and a bright source ($\Gamma=2.0$, $\phi_0=50$ mCrab). The true spectral $\gamma$-ray cutoff for all simulated sources is assumed to be $100$ TeV. For each source, 1000 simulations are performed with and without IRF modification. Results for the inverse spectral cutoff parameter, $\lambda$, are compared in Tab. \ref{tab:syst}.\\
It is concluded that the effect on the spectral cutoff resulting from a systematic error in the effective area is negligible. However, due to a systematic error on the energy bias, an uncertainty of up to $\mathcal{O}(20\%)$ is expected for the inverse energy cutoff.\\

\begin{table*}
\centering
\caption{Expected systematic errors for the determination of the spectral cutoff parameter. Faint source: $\Gamma=2.3$, $\phi_0=10$ mCrab, medium source: $\Gamma=2.0$, $\phi_0=30$ mCrab and bright source: $\Gamma=2.0$, $\phi_0=50$ mCrab.}
\begin{tabular}{ccccccc}
\\
\hline\hline

 & \multicolumn{2}{c}{Faint}  & \multicolumn{2}{c}{Medium}  & \multicolumn{2}{c}{Bright}  \\
 & Constant & Gradient & Constant & Gradient & Constant & Gradient \\
\hline

Effective Area $\pm5$\%  & $<1$\% & $<1$\% & $<1$\% & $<1$\% & $<1$\% & $<1$\% \\
\hline
Energy Bias +6\% & 16\% & 21\%  & 13\%  & 15\%  & 11\% & 12\% \\
\hline
Energy Bias -6\% & 6\% & 12\%  & 8\%  & 11\%  & 7\% & 10\% \\

\hline\hline

\end{tabular}

\label{tab:syst}
\end{table*}

\section{Comparison between CTA IRFs: "Prod 3b-v2" vs. "Prod 5-v0.1"}
\label{newIRFs}
The CTA IRFs used throughout the paper are "Prod 3b-v2", which represent the southern CTA configuration, including 70 SSTs. Updated "Prod 5-v0.1" CTA IRFs, which take changes in the southern CTA array layout into account, were published while this work was completed. The updated “Alpha” CTA south configuration foresees only 37 SSTs. Therefore, it is expected that the high energy performance of CTA, and in particular its power to detect high energy spectral cutoff features, is reduced when the updated IRFs are used.\\

\begin{figure*}
\centering
\includegraphics[width=18cm]{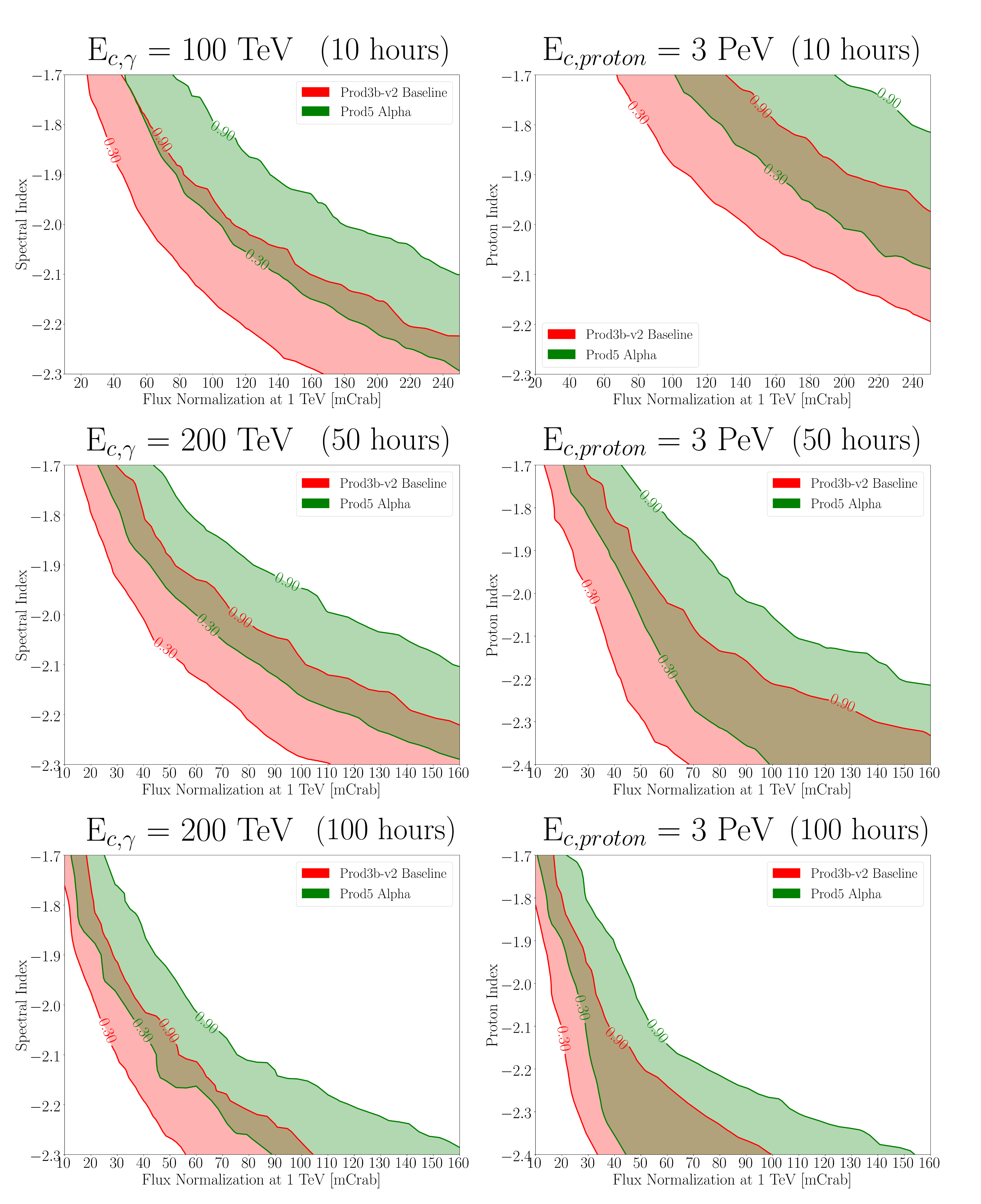}
\caption{Comparison of "transition regions", i.e. the regions between $30\%$ and $90\%$ detection probability, for the detection of a true spectral $\gamma$-ray cutoff (left panels) and the detection of a true PeVatron with a hadronic spectral cutoff at $3$ PeV (right panels) after 10 (top), 50 (middle) and 100 (bottom) hours of observation. The "transition regions" shown as red and green shaded areas are obtained by using "Prod 3b-v2" (Baseline) and "Prod 5-v0.1" (Alpha) IRFs, respectively.}
\label{irf_comp}
\end{figure*}

To quantify the expected high energy performance reduction, probability maps discussed in Sec. \ref{section4} are reproduced using "Prod 5-v0.1" CTA IRFs and the results are compared. Figure \ref{irf_comp} shows comparison between "transition regions" obtained from different CTA IRFs for the detection of true spectral $\gamma$-ray cutoffs and true PeVatrons with a hadronic cutoff at $3$ PeV. Compared to the "Prod 3b-v2" configuration, the detection probabilities for both, PeVatron and high energy $\gamma$-ray cutoffs, are clearly degraded for the "Prod 5-v0.1" IRFs.\\
With the updated configuration of CTA represented by "Prod 5-v0.1" IRFs, the confirmation of an SNR PeVatron with data from the CTA GPS is even less likely than discussed in Sec. \ref{snr_pevatrons} but deep observations of PeVatron remain promising. Follow-up observations of PeVatron candidates under moonlight conditions with the CTA-SST array, as discussed in Sec. \ref{moonlight_sec}, also remain an efficient option to increase the PeVatron follow-up observation time without effect on other key science topics.\\




{
{\bf \large List of Acronyms}\\
\\

\bf{CC-HEE :} Core-collapse High-explosion-energy\\
\bf{CC-HEM :} Core-collapse High-ejecta-mass\\
\bf{CC-LEM :} Core-collapse Low-ejecta-mass\\
\bf{CL :} Confidence Level\\
\bf{CR :} Cosmic Ray\\
\bf{CTA :} Cherenkov Telescope Array\\
\bf{ECPL :} Power-Law with Exponential Cutoff\\
\bf{FACT :} First G-APD Cherenkov Telescope\\
\bf{FWER :} Family-Wise Error Rate\\
\bf{G-APD :} Geiger-mode avalanche photodiods\\
\bf{GPS :} Galactic Plane Survey\\
\bf{H.E.S.S. :} High Energy Stereoscopic System\\
\bf{HAWC :} High Altitude Water Cherenkov Observatory\\
\bf{HNSB :} High-NSB\\
\bf{IACTs :} Imaging Atmospheric Cherenkov Telescopes\\
\bf{IRF :} Instrument Response Function\\
\bf{ISM :} Inter Stellar Medium\\
\bf{LHAASO :} Large High Altitude Air Shower Observatory\\
\bf{MAGIC :} Major Atmospheric Gamma-Ray Imaging Cherenkov\\
\bf{MCMC :} Markov Chain Monte Carlo\\
\bf{NSB :} Night Sky Background\\
\bf{PL :} Power-Law\\
\bf{PMT :} Photo-Multiplier Tube\\
\bf{PTS :} PeVatron Test Statistics\\
\bf{SiPM :} Silicon Photo-Multiplier\\
\bf{SN :} Supernova\\
\bf{SNR :} Supernova Remnant\\
\bf{SST :} Small Sized Telescope\\
\bf{SWGO :} Southern Wide-field Gamma-ray Observatory\\
\bf{TN :} Thermonuclear\\
\bf{TS :} Test Statistics\\
\bf{UHE :} Ultra High Energy\\
\bf{VERITAS :} Very Energetic Radiation Imaging Telescope Array System\\
\bf{VHE :} Very High Energy\\
}

\newpage

\bibliographystyle{elsarticle-harv} 
\bibliography{ap_submission_CTAPeV_revised}

{\onecolumn
{\large \textbf{The Cherenkov Telescope Array Consortium{~}}}

{}~\\

{F.~Acero}{\textsuperscript{1}} 
{, A.~Acharyya}{\textsuperscript{2}}
{, R.~Adam}{\textsuperscript{3}}
{, A.~Aguasca-Cabot}{\textsuperscript{4}}
{, I.~Agudo}{\textsuperscript{5}}
{, A.~Aguirre-Santaella}{\textsuperscript{6}}
{, J.~Alfaro}{\textsuperscript{7}}
{, R.~Aloisio}{\textsuperscript{8}}
{, N.~Álvarez Crespo}{\textsuperscript{9}}
{, R.~Alves Batista}{\textsuperscript{6}}
{, L.~Amati}{\textsuperscript{10}}
{, E.~Amato}{\textsuperscript{11}}
{, G.~Ambrosi}{\textsuperscript{12}}
{, E.O.~Angüner}{\textsuperscript{13,14}}
{, C.~Aramo}{\textsuperscript{15}}
{, C.~Arcaro}{\textsuperscript{16}}
{, T.~Armstrong}{\textsuperscript{14}}
{, K.~Asano}{\textsuperscript{17}}
{, Y.~Ascasibar}{\textsuperscript{6}}
{, J.~Aschersleben}{\textsuperscript{18}}
{, M.~Backes}{\textsuperscript{19,20}}
{, A.~Baktash}{\textsuperscript{21}}
{, C.~Balazs}{\textsuperscript{22}}
{, M.~Balbo}{\textsuperscript{23}}
{, J.~Ballet}{\textsuperscript{1}}
{, A.~Baquero Larriva}{\textsuperscript{24,146}}
{, V.~Barbosa Martins}{\textsuperscript{25}}
{, U.~Barres de Almeida}{\textsuperscript{26}}
{, J.A.~Barrio}{\textsuperscript{24}}
{, D.~Bastieri}{\textsuperscript{16}}
{, J.R.~Baxter}{\textsuperscript{17}}
{, J.~Becker Tjus}{\textsuperscript{27}}
{, W.~Benbow}{\textsuperscript{28}}
{, M.I.~Bernardos-Martín}{\textsuperscript{28}}
{, J.~Bernete}{\textsuperscript{64}}
{, A.~Berti}{\textsuperscript{30}}
{, B.~Bertucci}{\textsuperscript{12}}
{, V.~Beshley}{\textsuperscript{29}}
{, P.~Bhattacharjee}{\textsuperscript{30}}
{, S.~Bhattacharyya}{\textsuperscript{31}}
{, A.~Biland}{\textsuperscript{32}}
{, E.~Bissaldi}{\textsuperscript{33}}
{, J.~Biteau}{\textsuperscript{34}}
{, O.~Blanch}{\textsuperscript{35}}
{, P.~Bordas}{\textsuperscript{4}}
{, E.~Bottacini}{\textsuperscript{16}}
{, J.~Bregeon}{\textsuperscript{36,37}
{, R.~Brose}{\textsuperscript{38}}
{, N.~Bucciantini}{\textsuperscript{11}}
{, A.~Bulgarelli}{\textsuperscript{10}}
{, M.~Capasso}{\textsuperscript{39}}
{, R.A.~Capuzzo Dolcetta}{\textsuperscript{40,41}}
{, P.~Caraveo}{\textsuperscript{42}}
{, M.~Cardillo}{\textsuperscript{43}}
{, R.~Carosi}{\textsuperscript{44}}
{, S.~Casanova}{\textsuperscript{45,46}}
{, E.~Cascone}{\textsuperscript{47}}
{, F.~Cassol}{\textsuperscript{14}}
{, F.~Catalani}{\textsuperscript{48}}
{, M.~Cerruti}{\textsuperscript{49}}
{, P.~Chadwick}{\textsuperscript{50}}
{, S.~Chaty}{\textsuperscript{49}}
{, A.~Chen}{\textsuperscript{51}}
{, M.~Chernyakova}{\textsuperscript{52}}
{, A.~Chiavassa}{\textsuperscript{53,54}}
{, J.~Chudoba}{\textsuperscript{55}}
{, C.~Coimbra-Araujo}{\textsuperscript{56}}
{, V.~Conforti}{\textsuperscript{10}}
{, J.L.~Contreras}{\textsuperscript{24}}
{, A.~Costa}{\textsuperscript{57}}
{, H.~Costantini}{\textsuperscript{14}}
{, P.~Cristofari}{\textsuperscript{58}}
{, R.~Crocker}{\textsuperscript{59}}
{, G.~D'Amico}{\textsuperscript{60}}
{, F.~D'Ammando}{\textsuperscript{61}}
{, A.~De Angelis}{\textsuperscript{16}}
{, V.~De Caprio}{\textsuperscript{47}}
{, E.M.~de Gouveia Dal Pino}{\textsuperscript{62}}
{, E.~de Ona Wilhelmi}{\textsuperscript{25}}
{, V.~de Souza}{\textsuperscript{63}}
{, C.~Delgado}{\textsuperscript{64}}
{, D.~della Volpe}{\textsuperscript{65}}
{, D.~Depaoli}{\textsuperscript{45}}
{, T.~Di Girolamo}{\textsuperscript{15,66}}
{, F.~Di Pierro}{\textsuperscript{53}}
{, R.~Di Tria}{\textsuperscript{67}}
{, L.~Di Venere}{\textsuperscript{68}}
{, S.~Diebold}{\textsuperscript{69}}
{, J.I.~Djuvsland}{\textsuperscript{60}}
{, A.~Donini}{\textsuperscript{40}}
{, M.~Doro}{\textsuperscript{16}}
{, R.d.C.~Dos Anjos}{\textsuperscript{56}}
{, V.V.~Dwarkadas}{\textsuperscript{70}}
{, S.~Einecke}{\textsuperscript{71}}
{, D.~Elsässer}{\textsuperscript{72}}
{, G.~Emery}{\textsuperscript{65}}
{, C.~Evoli}{\textsuperscript{8}}
{, D.~Falceta-Goncalves}{\textsuperscript{73}}
{, E.~Fedorova}{\textsuperscript{74}}
{, S.~Fegan}{\textsuperscript{3}}
{, G.~Ferrand}{\textsuperscript{75}}
{, E.~Fiandrini}{\textsuperscript{12}}
{, M.~Filipovic}{\textsuperscript{76}}
{, V.~Fioretti}{\textsuperscript{10}}
{, M.~Fiori}{\textsuperscript{77}}
{, L.~Foffano}{\textsuperscript{43}}
{, G.~Fontaine}{\textsuperscript{3}}
{, S.~Fukami}{\textsuperscript{17}}
{, G.~Galanti}{\textsuperscript{42}}
{, G.~Galaz}{\textsuperscript{7}}
{, V.~Gammaldi}{\textsuperscript{6}}
{, C.~Gasbarra}{\textsuperscript{78}}
{, A.~Ghalumyan}{\textsuperscript{79}}
{, G.~Ghirlanda}{\textsuperscript{80}}
{, M.~Giarrusso}{\textsuperscript{81}}
{, G.~Giavitto}{\textsuperscript{25}}
{, N.~Giglietto}{\textsuperscript{33}}
{, F.~Giordano}{\textsuperscript{67}}
{, M.~Giroletti}{\textsuperscript{61}}
{, A.~Giuliani}{\textsuperscript{42}}
{, L.~Giunti}{\textsuperscript{49}}
{, N.~Godinovic}{\textsuperscript{82}}
{, J.~Goulart Coelho}{\textsuperscript{56}}
{, L.~Gréaux}{\textsuperscript{34}}
{, D.~Green}{\textsuperscript{83}}
{, M-H.~Grondin}{\textsuperscript{84}}
{, O.~Gueta}{\textsuperscript{25}}
{, S.~Gunji}{\textsuperscript{85}}
{, T.~Hassan}{\textsuperscript{64}}
{, M.~Heller}{\textsuperscript{65}}
{, S.~Hernández-Cadena}{\textsuperscript{86}}
{, J.~Hinton}{\textsuperscript{45}}
{, B.~Hnatyk}{\textsuperscript{74}}
{, R.~Hnatyk}{\textsuperscript{74}}
{, D.~Hoffmann}{\textsuperscript{14}}
{, W.~Hofmann}{\textsuperscript{45}}
{, J.~Holder}{\textsuperscript{87}}
{, D.~Horan}{\textsuperscript{3}}
{, P.~Horvath}{\textsuperscript{88}}
{, M.~Hrabovsky}{\textsuperscript{88}}
{, D.~Hrupec}{\textsuperscript{89}}
{, T.~Inada}{\textsuperscript{17}}
{, F.~Incardona}{\textsuperscript{57}}
{, S.~Inoue}{\textsuperscript{75}}
{, K.~Ishio}{\textsuperscript{90}}
{, M.~Jamrozy}{\textsuperscript{91}}
{, P.~Janecek}{\textsuperscript{55}}
{, I.~Jiménez Martínez}{\textsuperscript{64}}
{, W.~Jin}{\textsuperscript{2}}
{, I.~Jung-Richardt}{\textsuperscript{92}}
{, J.~Jurysek}{\textsuperscript{23}}
{, P.~Kaaret}{\textsuperscript{93}}
{, V.~Karas}{\textsuperscript{94}}
{, U.~Katz}{\textsuperscript{92}}
{, D.~Kerszberg}{\textsuperscript{35}}
{, B.~Khélifi}{\textsuperscript{49}}
{, D.B.~Kieda}{\textsuperscript{95}}
{, R.~Kissmann}{\textsuperscript{96}}
{, T.~Kleiner}{\textsuperscript{25}}
{, G.~Kluge}{\textsuperscript{97}}
{, W.~Kluzniak}{\textsuperscript{98}}
{, J.~Knödlseder}{\textsuperscript{99}}
{, Y.~Kobayashi}{\textsuperscript{17}}
{, K.~Kohri}{\textsuperscript{100}}
{, N.~Komin}{\textsuperscript{51}}
{, P.~Kornecki}{\textsuperscript{58}}
{, H.~Kubo}{\textsuperscript{17}}
{, N.~La Palombara}{\textsuperscript{42}}
{, M.~Láinez}{\textsuperscript{24}}
{, A.~Lamastra}{\textsuperscript{40}}
{, J.~Lapington}{\textsuperscript{101}}
{, M.~Lemoine-Goumard}{\textsuperscript{84}}
{, J.-P.~Lenain}{\textsuperscript{102}}
{, F.~Leone}{\textsuperscript{81,103}}
{, G.~Leto}{\textsuperscript{57}}
{, F.~Leuschner}{\textsuperscript{69}}
{, E.~Lindfors}{\textsuperscript{104}}
{, I.~Liodakis}{\textsuperscript{104}}
{, T.~Lohse}{\textsuperscript{105}}
{, S.~Lombardi}{\textsuperscript{40}}
{, F.~Longo}{\textsuperscript{106}}
{, R.~López-Coto}{\textsuperscript{5}}
{, M.~López-Moya}{\textsuperscript{24}}
{, A.~López-Oramas}{\textsuperscript{107}}
{, S.~Loporchio}{\textsuperscript{67}}
{, P.L.~Luque-Escamilla}{\textsuperscript{108}}
{, O.~Macias}{\textsuperscript{109}}
{, J.~Mackey}{\textsuperscript{38}}
{, P.~Majumdar}{\textsuperscript{17,110}}
{, D.~Mandat}{\textsuperscript{55}}
{, M.~Manganaro}{\textsuperscript{111}}
{, G.~Manicò}{\textsuperscript{81}}
{, M.~Marconi}{\textsuperscript{47}}
{, J.~Martí}{\textsuperscript{108}}
{, G.~Martínez}{\textsuperscript{64}}
{, M.~Martinez}{\textsuperscript{35}}
{, O.~Martinez}{\textsuperscript{9}}
{, A.J.T.S.~Mello}{\textsuperscript{56}}
{, S.~Menchiari}{\textsuperscript{112}} 
{, D. M.-A.~Meyer}{\textsuperscript{113}}
{, S.~Micanovic}{\textsuperscript{111}}
{, D.~Miceli}{\textsuperscript{16}}
{, M.~Miceli}{\textsuperscript{114,115}}
{, J.~Michalowski}{\textsuperscript{46}}
{, T.~Miener}{\textsuperscript{24}}
{, J.M.~Miranda}{\textsuperscript{9}}
{, A.~Mitchell}{\textsuperscript{92}}
{, B.~Mode}{\textsuperscript{116}}
{, R.~Moderski}{\textsuperscript{98}}
{, L.~Mohrmann}{\textsuperscript{45}}
{, E.~Molina}{\textsuperscript{4}}
{, T.~Montaruli}{\textsuperscript{65}}
{, D.~Morcuende}{\textsuperscript{24}}
{, G.~Morlino}{\textsuperscript{11}}
{, A.~Morselli}{\textsuperscript{78}}
{, M.~Mosè}{\textsuperscript{16}}
{, E.~Moulin}{\textsuperscript{117}}
{, R.~Mukherjee}{\textsuperscript{39}}
{, K.~Munari}{\textsuperscript{57}}
{, T.~Murach}{\textsuperscript{25}}
{, A.~Nagai}{\textsuperscript{65}}
{, S.~Nagataki}{\textsuperscript{75}}
{, R.~Nemmen}{\textsuperscript{62}}
{, J.~Niemiec}{\textsuperscript{46}}
{, D.~Nieto}{\textsuperscript{24}}
{, M.~Nievas Rosillo}{\textsuperscript{107}}
{, M.~Nikolajuk}{\textsuperscript{118}}
{, K.~Nishijima}{\textsuperscript{119}}
{, K.~Noda}{\textsuperscript{17}}
{, B.~Novosyadlyj}{\textsuperscript{120}}
{, S.~Nozaki}{\textsuperscript{83}}
{, M.~Ohishi}{\textsuperscript{17}}
{, S.~Ohm}{\textsuperscript{25}}
{, Y.~Ohtani}{\textsuperscript{17}}
{, A.~Okumura}{\textsuperscript{121,122}}
{, B.~Olmi}{\textsuperscript{11}}
{, R.A.~Ong}{\textsuperscript{123}}
{, M.~Orienti}{\textsuperscript{61}}
{, R.~Orito}{\textsuperscript{124}}
{, M.~Orlandini}{\textsuperscript{10}}
{, E.~Orlando}{\textsuperscript{106}}
{, S.~Orlando}{\textsuperscript{114}}
{, M.~Ostrowski}{\textsuperscript{91}}
{, I.~Oya}{\textsuperscript{125}}
{, F.R.~Pantaleo}{\textsuperscript{33}}
{, J.M.~Paredes}{\textsuperscript{4}}
{, B.~Patricelli}{\textsuperscript{40,126}}
{, M.~Pecimotika}{\textsuperscript{111}}
{, M.~Peresano}{\textsuperscript{54}}
{, J.~Pérez-Romero}{\textsuperscript{6}}
{, M.~Persic}{\textsuperscript{77}}
{, O.~Petruk}{\textsuperscript{29}}
{, G.~Piano}{\textsuperscript{43}}
{, E.~Pietropaolo}{\textsuperscript{127}}
{, G.~Pirola}{\textsuperscript{83}}
{, C.~Pittori}{\textsuperscript{40}}
{, M.~Pohl}{\textsuperscript{25,113}}
{, G.~Ponti}{\textsuperscript{80}}
{, E.~Prandini}{\textsuperscript{16}}
{, G.~Principe}{\textsuperscript{106}}
{, C.~Priyadarshi}{\textsuperscript{35}}
{, E.~Pueschel}{\textsuperscript{25}}
{, G.~Pühlhofer}{\textsuperscript{69}}
{, M.L.~Pumo}{\textsuperscript{81,103}}
{, A.~Quirrenbach}{\textsuperscript{128}}
{, R.~Rando}{\textsuperscript{16}}
{, S.~Razzaque}{\textsuperscript{129}}
{, P.~Reichherzer}{\textsuperscript{27}}
{, A.~Reimer}{\textsuperscript{96}}
{, O.~Reimer}{\textsuperscript{96}}
{, M.~Renaud}{\textsuperscript{36}}
{, T.~Reposeur}{\textsuperscript{84}}
{, M.~Ribó}{\textsuperscript{4}}
{, T.~Richtler}{\textsuperscript{130}}
{, J.~Rico}{\textsuperscript{35}}
{, F.~Rieger}{\textsuperscript{45}}
{, M.~Rigoselli}{\textsuperscript{42}}
{, L.~Riitano}{\textsuperscript{116}}
{, V.~Rizi}{\textsuperscript{127}}
{, E.~Roache}{\textsuperscript{28}}
{, P.~Romano}{\textsuperscript{80}}
{, G.~Romeo}{\textsuperscript{57}}
{, J.~Rosado}{\textsuperscript{24}}
{, G.~Rowell}{\textsuperscript{71}}
{, B.~Rudak}{\textsuperscript{98}}
{, I.~Sadeh}{\textsuperscript{25}}
{, S.~Safi-Harb}{\textsuperscript{132}}
{, L.~Saha}{\textsuperscript{28}}
{, S.~Sailer}{\textsuperscript{45}}
{, M.~Sánchez-Conde}{\textsuperscript{6}}
{, S.~Sarkar}{\textsuperscript{133}}
{, K.~Satalecka}{\textsuperscript{104}}
{, F.G.~Saturni}{\textsuperscript{40}}
{, A.~Scherer}{\textsuperscript{7}}
{, P.~Schovánek}{\textsuperscript{55}}
{, F.~Schussler}{\textsuperscript{117}}
{, U.~Schwanke}{\textsuperscript{105}}
{, S.~Scuderi}{\textsuperscript{42}}
{, M.~Seglar-Arroyo}{\textsuperscript{30}}
{, O.~Sergijenko}{\textsuperscript{74}}
{, M.~Servillat}{\textsuperscript{58}}
{, R-Y.~Shang}{\textsuperscript{39}}
{, P.~Sharma}{\textsuperscript{34}}
{, H.~Siejkowski}{\textsuperscript{134}}
{, V.~Sliusar}{\textsuperscript{23}}
{, A.~Słowikowska}{\textsuperscript{135}}
{, H.~Sol}{\textsuperscript{58}}
{, A.~Specovius}{\textsuperscript{92}}
{, S.T.~Spencer}{\textsuperscript{92}}
{, G.~Spengler}{\textsuperscript{105}}
{, A.~Stamerra}{\textsuperscript{40}}
{, S.~Stanič}{\textsuperscript{31}}
{, T.~Starecki}{\textsuperscript{136}}
{, R.~Starling}{\textsuperscript{101}}
{, T.~Stolarczyk}{\textsuperscript{1}}
{, L.A.~Stuani Pereira}{\textsuperscript{63}}
{, Y.~Suda}{\textsuperscript{137}}
{, T.~Suomijarvi}{\textsuperscript{34}}
{, I.~Sushch}{\textsuperscript{20,120}}
{, H.~Tajima}{\textsuperscript{121,122}}
{, P-H.T.~Tam}{\textsuperscript{17}}
{, S.J.~Tanaka}{\textsuperscript{138}}
{, F.~Tavecchio}{\textsuperscript{80}}
{, V.~Testa}{\textsuperscript{40}}
{, W.~Tian}{\textsuperscript{17,131}}
{, L.~Tibaldo}{\textsuperscript{99}}
{, D.F.~Torres}{\textsuperscript{139}}
{, N.~Tothill}{\textsuperscript{76}}
{, B.~Vallage}{\textsuperscript{140}}
{, P.~Vallania}{\textsuperscript{53,141}}
{, C.~van Eldik}{\textsuperscript{92}}
{, J.~van Scherpenberg}{\textsuperscript{83}}
{, J.~Vandenbroucke}{\textsuperscript{116}}
{, M.~Vazquez Acosta}{\textsuperscript{107}}
{, M.~Vecchi}{\textsuperscript{18}}
{, S.~Vercellone}{\textsuperscript{80}}
{, G.~Verna}{\textsuperscript{14}}
{, A.~Viana}{\textsuperscript{63}}
{, J.~Vignatti}{\textsuperscript{142}}
{, V.~Vitale}{\textsuperscript{78}}
{, V.~Vodeb}{\textsuperscript{31}}
{, S.~Vorobiov}{\textsuperscript{31}}
{, T.~Vuillaume}{\textsuperscript{30}}
{, S.J.~Wagner}{\textsuperscript{128}}
{, R.~Walter}{\textsuperscript{23}}
{, M.~White}{\textsuperscript{131}}
{, A.~Wierzcholska}{\textsuperscript{46}}
{, M.~Will}{\textsuperscript{83}}
{, D.~Williams}{\textsuperscript{143}}
{, L.~Yang}{\textsuperscript{129,144}}
{, T.~Yoshida}{\textsuperscript{145}}
{, T.~Yoshikoshi}{\textsuperscript{17}}
{, G.~Zaharijas}{\textsuperscript{31}}
{, L.~Zampieri}{\textsuperscript{77}}
{, D.~Zavrtanik}{\textsuperscript{31}}
{, M.~Zavrtanik}{\textsuperscript{31}}
{, V.I.~Zhdanov}{\textsuperscript{74}}
{, M.~Živec}{\textsuperscript{31}}
}
}

{\twocolumn

{1 : Université Paris Saclay, Université Paris Cité, CEA, CNRS, AIM, F-91191 Gif-sur-Yvette Cedex, France{~} \label{ins1}} 

{2 : University of Alabama, Tuscaloosa, Department of Physics and Astronomy, Gallalee Hall, Box 870324 Tuscaloosa, AL 35487-0324, USA{~} \label{ins2}}

{3 : Laboratoire Leprince-Ringuet, École Polytechnique (UMR 7638, CNRS/IN2P3, Institut Polytechnique de Paris), 91128 Palaiseau, France{~} \label{ins3}} 

{4 : Departament de Física Quàntica i Astrofísica, Institut de Ciències del Cosmos, Universitat de Barcelona, IEEC-UB, Martí i Franquès, 1, 08028, Barcelona, Spain{~} \label{ins4}} 

{5 : Instituto de Astrofísica de Andalucía-CSIC, Glorieta de la Astronomía s/n, 18008, Granada, Spain{~} \label{ins5}} 

{6 : Instituto de Física Teórica UAM/CSIC and Departamento de Física Teórica, Universidad Autónoma de Madrid, c/ Nicolás Cabrera 13-15, Campus de Cantoblanco UAM, 28049 Madrid, Spain{~} \label{ins6}} 

{7 : Pontificia Universidad Católica de Chile, Av. Libertador Bernardo O'Higgins 340, Santiago, Chile{~} \label{ins7}} 

{8 : Gran Sasso Science Institute (GSSI), Viale Francesco Crispi 7, 67100 L'Aquila, Italy and INFN-Laboratori Nazionali del Gran Sasso (LNGS), via G. Acitelli 22, 67100 Assergi (AQ), Italy{~} \label{ins8}} 

{9 : Grupo de Electronica, Universidad Complutense de Madrid, Av. Complutense s/n, 28040 Madrid, Spain{~} \label{ins9}} 

{10 : INAF - Osservatorio di Astrofisica e Scienza dello spazio di Bologna, Via Piero Gobetti 93/3, 40129 Bologna, Italy{~} \label{ins10}} 

{11 : INAF - Osservatorio Astrofisico di Arcetri, Largo E. Fermi, 5 - 50125 Firenze, Italy{~} \label{ins11}} 

{12 : INFN Sezione di Perugia and Università degli Studi di Perugia, Via A. Pascoli, 06123 Perugia, Italy{~} \label{ins12}} 

{13 : TÜBİTAK Research Institute for Fundamental Sciences, 41470 Gebze, Kocaeli, Turkey{~} \label{ins13}} 

{14 : Aix-Marseille Université, CNRS/IN2P3, CPPM, 163 Avenue de Luminy, 13288 Marseille cedex 09, France{~} \label{ins14}} 

{15 : INFN Sezione di Napoli, Via Cintia, ed. G, 80126 Napoli, Italy{~} \label{ins15}} 

{16 : INFN Sezione di Padova and Università degli Studi di Padova, Via Marzolo 8, 35131 Padova, Italy{~} \label{ins16}} 

{17 : Institute for Cosmic Ray Research, University of Tokyo, 5-1-5, Kashiwa-no-ha, Kashiwa, Chiba 277-8582, Japan{~} \label{ins17}} 

{18 : Kapteyn Astronomical Institute, University of Groningen, Landleven 12, 9747 AD Groningen, The Netherlands{~} \label{ins18}} 

{19 : University of Namibia, Department of Physics, Chemistry \& Material Science, 340 Mandume Ndemufayo Ave., Pioneerspark, Windhoek, Namibia{~} \label{ins19}} 

{20 : Centre for Space Research, North-West University, Potchefstroom, 2520, South Africa{~} \label{ins20}} 

{21 : Universität Hamburg, Institut für Experimentalphysik, Luruper Chaussee 149, 22761 Hamburg, Germany{~} \label{ins21}} 

{22 : School of Physics and Astronomy, Monash University, Melbourne, Victoria 3800, Australia{~} \label{ins22}} 

{23 : Department of Astronomy, University of Geneva, Chemin d'Ecogia 16, CH-1290 Versoix, Switzerland{~} \label{ins23}} 

{24 : IPARCOS-UCM, Instituto de Física de Partículas y del Cosmos, and EMFTEL Department, Universidad Complutense de Madrid, E-28040 Madrid, Spain{~} \label{ins24}} 

{25 : Deutsches Elektronen-Synchrotron, Platanenallee 6, 15738 Zeuthen, Germany{~} \label{ins25}} 

{26 : Centro Brasileiro de Pesquisas Físicas, Rua Dr. Xavier Sigaud 150, RJ 22290-180, Rio de Janeiro, Brazil{~} \label{ins26}} 

{27 : Institut für Theoretische Physik, Lehrstuhl IV: Plasma-Astroteilchenphysik, Ruhr-Universität Bochum, Universitätsstraße 150, 44801 Bochum, Germany{~} \label{ins27}} 

{28 : Center for Astrophysics \textbar{} Harvard \& Smithsonian, 60 Garden St, Cambridge, MA 02138, USA{~} \label{ins28}} 

{29 : Pidstryhach Institute for Applied Problems in Mechanics and Mathematics NASU, 3B Naukova Street, Lviv, 79060, Ukraine{~} \label{ins29}} 
{30 : Univ. Savoie Mont Blanc, CNRS, Laboratoire d'Annecy de Physique des Particules - IN2P3, 74000 Annecy, France{~} \label{ins30}} 

{31 : Center for Astrophysics and Cosmology, University of Nova Gorica, Vipavska 13, 5000 Nova Gorica, Slovenia{~} \label{ins31}} 

{32 : ETH Zurich, Institute for Particle Physics and Astrophysics, Otto-Stern-Weg 5, 8093 Zurich, Switzerland{~} \label{ins32}}  

{33 : INFN Sezione di Bari and Politecnico di Bari, via Orabona 4, 70124 Bari, Italy{~} \label{ins33}} 

{34 : Laboratoire de Physique des 2 infinis, Irene Joliot-Curie,IN2P3/CNRS, Université Paris-Saclay, Université de Paris, 15 rue Georges Clemenceau, 91406 Orsay, Cedex, France{~} \label{ins34}}  

{35 : Institut de Fisica d'Altes Energies (IFAE), The Barcelona Institute of Science and Technology, Campus UAB, 08193 Bellaterra (Barcelona), Spain{~} \label{ins35}} 

{36 : Laboratoire Univers et Particules de Montpellier, Université de Montpellier, CNRS/IN2P3, CC 72, Place Eugène Bataillon, F-34095 Montpellier Cedex 5, France{~} \label{ins36}} 

{37 : Univ. Grenoble Alpes, CNRS, Laboratoire de Physique Subatomique et de Cosmologie, 53 avenue des Martyrs, 38026 Grenoble cedex, France{~} \label{ins37}} 

{38 : Dublin Institute for Advanced Studies, 31 Fitzwilliam Place, Dublin 2, Ireland{~} \label{ins38}} 

{39 : Department of Physics, Columbia University, 538 West 120th Street, New York, NY 10027, USA{~} \label{ins39}} 

{40 : INAF - Osservatorio Astronomico di Roma, Via di Frascati 33, 00078, Monteporzio Catone, Italy{~} \label{ins40}} 

{41 : Dipartimento di Fisica - Sapienza, Università di Roma, piazzale A. Moro 5 -00151 Roma, Italy{~} \label{ins41}} 

{42 : INAF - Istituto di Astrofisica Spaziale e Fisica Cosmica di Milano, Via A. Corti 12, 20133 Milano, Italy{~} \label{ins42}} 

{43 : INAF - Istituto di Astrofisica e Planetologia Spaziali (IAPS), Via del Fosso del Cavaliere 100, 00133 Roma, Italy{~} \label{ins43}} 
{44 : INFN Sezione di Pisa, Largo Pontecorvo 3, 56217 Pisa, Italy{~} \label{ins44}} 

{45 : Max-Planck-Institut für Kernphysik, Saupfercheckweg 1, 69117 Heidelberg, Germany{~} \label{ins45}} 

{46 : The Henryk Niewodniczański Institute of Nuclear Physics, Polish Academy of Sciences, ul. Radzikowskiego 152, 31-342 Cracow, Poland{~} \label{ins46}} 

{47 : INAF - Osservatorio Astronomico di Capodimonte, Via Salita Moiariello 16, 80131 Napoli, Italy{~} \label{ins47}} 

{48 : Escola de Engenharia de Lorena, Universidade de São Paulo, Área I - Estrada Municipal do Campinho, s/n°, CEP 12602-810, Pte. Nova, Lorena, Brazil{~} \label{ins48}} 

{49 : Universit{\'e} Paris Cité, CNRS, Astroparticule et Cosmologie, F-75013 Paris, France{~} \label{ins49}} 

{50 : Centre for Advanced Instrumentation, Dept. of Physics, Durham University, South Road, Durham DH1 3LE, United Kingdom{~} \label{ins50}} 

{51 : University of the Witwatersrand, 1 Jan Smuts Avenue, Braamfontein, 2000 Johannesburg, South Africa{~} \label{ins51}} 

{52 : Dublin City University, Glasnevin, Dublin 9, Ireland{~} \label{ins52}} 

{53 : INFN Sezione di Torino, Via P. Giuria 1, 10125 Torino, Italy{~} \label{ins53}} 

{54 : Dipartimento di Fisica - Universitá degli Studi di Torino, Via Pietro Giuria 1 - 10125 Torino, Italy{~} \label{ins54}} 

{55 : FZU - Institute of Physics of the Czech Academy of Sciences, Na Slovance 1999/2, 182 21 Praha 8, Czech Republic{~} \label{ins55}} 
{56 : Universidade Federal Do Paraná - Setor Palotina, Departamento de Engenharias e Exatas, Rua Pioneiro, 2153, Jardim Dallas, CEP: 85950-000 Palotina, Paraná, Brazil{~} \label{ins56}} 

{57 : INAF - Osservatorio Astrofisico di Catania, Via S. Sofia, 78, 95123 Catania, Italy{~} \label{ins57}} 

{58 : LUTH, GEPI and LERMA, Observatoire de Paris, CNRS, PSL University, 5 place Jules Janssen, 92190, Meudon, France{~} \label{ins58}} 

{59 : Research School of Astronomy and Astrophysics, Australian National University, Canberra ACT 0200, Australia{~} \label{ins59}} 

{60 : Department of Physics and Technology, University of Bergen, Museplass 1, 5007 Bergen, Norway{~} \label{ins60}} 

{61 : INAF - Istituto di Radioastronomia, Via Gobetti 101, 40129 Bologna, Italy{~} \label{ins61}} 

{62 : Instituto de Astronomia, Geofísico, e Ciências Atmosféricas - Universidade de São Paulo, Cidade Universitária, R. do Matão, 1226, CEP 05508-090, São Paulo, SP, Brazil{~} \label{ins62}} 

{63 : Instituto de Física de São Carlos, Universidade de São Paulo, Av. Trabalhador São-carlense, 400 - CEP 13566-590, São Carlos, SP, Brazil{~} \label{ins63}}  

{64 : CIEMAT, Avda. Complutense 40, 28040 Madrid, Spain{~} \label{ins64}}

{65 : University of Geneva - Département de physique nucléaire et corpusculaire, 24 rue du Général-Dufour, 1211 Genève 4, Switzerland{~} \label{ins65}} 

{66 : Universitá degli Studi di Napoli "Federico II" - Dipartimento di Fisica "E. Pancini", Complesso universitario di Monte Sant'Angelo, Via Cintia - 80126 Napoli, Italy{~} \label{ins66}} 

{67 : INFN Sezione di Bari and Università degli Studi di Bari, via Orabona 4, 70124 Bari, Italy{~} \label{ins67}} 

{68 : INFN Sezione di Bari, via Orabona 4, 70126 Bari, Italy{~} \label{ins68}} 

{69 : Institut für Astronomie und Astrophysik, Universität Tübingen, Sand 1, 72076 Tübingen, Germany{~} \label{ins69}} 

{70 : University of Chicago, 5640 South Ellis Avenue, Chicago, IL 60637, USA{~} \label{ins70}} 

{71 : School of Physical Sciences, University of Adelaide, Adelaide SA 5005, Australia{~} \label{ins71}} 

{72 : Department of Physics, TU Dortmund University, Otto-Hahn-Str. 4, 44221 Dortmund, Germany{~} \label{ins72}} 

{73 : Escola de Artes, Ciências e Humanidades, Universidade de São Paulo, Rua Arlindo Bettio, CEP 03828-000, 1000 São Paulo, Brazil{~} \label{ins73}} 

{74 : Astronomical Observatory of Taras Shevchenko National University of Kyiv, 3 Observatorna Street, Kyiv, 04053, Ukraine{~} \label{ins74}} 

{75 : RIKEN, Institute of Physical and Chemical Research, 2-1 Hirosawa, Wako, Saitama, 351-0198, Japan{~} \label{ins75}} 

{76 : Western Sydney University, Locked Bag 1797, Penrith, NSW 2751, Australia{~} \label{ins76}} 

{77 : INAF - Osservatorio Astronomico di Padova, Vicolo dell'Osservatorio 5, 35122 Padova, Italy{~} \label{ins77}} 

{78 : INFN Sezione di Roma Tor Vergata, Via della Ricerca Scientifica 1, 00133 Rome, Italy{~} \label{ins78}} 

{79 : Alikhanyan National Science Laboratory, Yerevan Physics Institute, 2 Alikhanyan Brothers St., 0036, Yerevan, Armenia{~} \label{ins79}} 

{80 : INAF - Osservatorio Astronomico di Brera, Via Brera 28, 20121 Milano, Italy{~} \label{ins80}} 

{81 : INFN Sezione di Catania, Via S. Sofia 64, 95123 Catania, Italy{~} \label{ins81}} 

{82 : University of Split - FESB, R. Boskovica 32, 21 000 Split, Croatia{~} \label{ins82}} 

{83 : Max-Planck-Institut für Physik, Föhringer Ring 6, 80805 München, Germany{~} \label{ins83}} 

{84 : LP2I Bordeaux, Univ. Bordeaux, CNRS-IN2P3, UMR 5797, 19 Chemin du Solarium, CS 10120, F-33175 Gradignan Cedex, France{~} \label{ins84}} 

{85 : Department of Physics, Yamagata University, Yamagata, Yamagata 990-8560, Japan{~} \label{ins85}} 

{86 : Universidad Nacional Autónoma de México, Delegación Coyoacán, 04510 Ciudad de México, Mexico{~} \label{ins86}} 

{87 : Department of Physics and Astronomy and the Bartol Research Institute, University of Delaware, Newark, DE 19716, USA{~} \label{ins87}} 

{88 : Palacky University Olomouc, Faculty of Science, RCPTM, 17. listopadu 1192/12, 771 46 Olomouc, Czech Republic{~} \label{ins88}} 

{89 : Josip Juraj Strossmayer University of Osijek, Trg Ljudevita Gaja 6, 31000 Osijek, Croatia{~} \label{ins89}} 

{90 : Faculty of Physics and Applied Computer Science, University of Lódź, ul. Pomorska 149-153, 90-236 Lódź, Poland{~} \label{ins90}} 

{91 : Astronomical Observatory, Jagiellonian University, ul. Orla 171, 30-244 Cracow, Poland{~} \label{ins91}} 

{92 : Friedrich-Alexander-Universit{\"a}t Erlangen-N{\"u}rnberg, Erlangen Centre for Astroparticle Physics (ECAP), Nikolaus-Fiebiger-Str. 2, 91058 Erlangen, Germany{~} \label{ins92}} 

{93 : University of Iowa, Department of Physics and Astronomy, Van Allen Hall, Iowa City, IA 52242, USA{~} \label{ins93}} 

{94 : Astronomical Institute of the Czech Academy of Sciences, Bocni II 1401 - 14100 Prague, Czech Republic{~} \label{ins94}} 

{95 : Department of Physics and Astronomy, University of Utah, Salt Lake City, UT 84112-0830, USA{~} \label{ins95}} 

{96 : Institut für Astro- und Teilchenphysik, Leopold-Franzens-Universität, Technikerstr. 25/8, 6020 Innsbruck, Austria{~} \label{ins96}} 

{97 : University of Oslo, Department of Physics, Sem Saelandsvei 24 - PO Box 1048 Blindern, N-0316 Oslo, Norway{~} \label{ins97}} 

{98 : Nicolaus Copernicus Astronomical Center, Polish Academy of Sciences, ul. Bartycka 18, 00-716 Warsaw, Poland{~} \label{ins98}} 

{99 : Institut de Recherche en Astrophysique et Planétologie, CNRS-INSU, Université Paul Sabatier, 9 avenue Colonel Roche, BP 44346, 31028 Toulouse Cedex 4, France{~} \label{ins99}} 

{100 : Institute of Particle and Nuclear Studies, KEK (High Energy Accelerator Research Organization), 1-1 Oho, Tsukuba, 305-0801, Japan{~} \label{ins100}} 

{101 : School of Physics and Astronomy, University of Leicester, University Road, Leicester LE1 7RH, United Kingdom{~} \label{ins101}} 

{102 : Sorbonne Université, CNRS/IN2P3, Laboratoire de Physique Nucléaire et de Hautes Energies, LPNHE, 4 place Jussieu, 75005 Paris, France{~} \label{ins102}} 

{103 : Dipartimento di Fisica e Astronomia, Sezione Astrofisica, Universitá di Catania, Via S. Sofia 78, I-95123 Catania, Italy{~} \label{ins103}} 

{104 : Finnish Centre for Astronomy with ESO, University of Turku, Finland, FI-20014 University of Turku, Finland{~} \label{ins104}} 

{105 : Department of Physics, Humboldt University Berlin, Newtonstr. 15, 12489 Berlin, Germany{~} \label{ins105}} 

{106 : INFN Sezione di Trieste and Università degli Studi di Trieste, Via Valerio 2 I, 34127 Trieste, Italy{~} \label{ins106}} 

{107 : Instituto de Astrofísica de Canarias and Departamento de Astrofísica, Universidad de La Laguna, La Laguna, Tenerife, Spain{~} \label{ins107}} 

{108 : Escuela Politécnica Superior de Jaén, Universidad de Jaén, Campus Las Lagunillas s/n, Edif. A3, 23071 Jaén, Spain{~} \label{ins108}} 

{109 : Anton Pannekoek Institute/GRAPPA, University of Amsterdam, Science Park 904 1098 XH Amsterdam, The Netherlands{~} \label{ins109}} 

{110 : Saha Institute of Nuclear Physics, Bidhannagar, Kolkata-700 064, India{~} \label{ins110}} 

{111 : University of Rijeka, Faculty of Physics, Radmile Matejčić 2, 51000 Rijeka, Croatia{~} \label{ins111}} 

{112 : INFN and Università degli Studi di Siena, Dipartimento di Scienze Fisiche, della Terra e dell'Ambiente (DSFTA), Sezione di Fisica, Via Roma 56, 53100 Siena, Italy{~} \label{ins112}} 

{113 : Institut für Physik \& Astronomie, Universität Potsdam, Karl-Liebknecht-Strasse 24/25, 14476 Potsdam, Germany{~} \label{ins113}} 

{114 : INAF - Osservatorio Astronomico di Palermo "G.S. Vaiana", Piazza del Parlamento 1, 90134 Palermo, Italy{~} \label{ins114}} 

{115 : Dipartimento di Fisica e Chimica E. Segrè, Università degli Studi di Palermo, Piazza del Parlamento 1, 90134, Palermo, Italy{~} \label{ins115}} 

{116 : University of Wisconsin, Madison, 500 Lincoln Drive, Madison, WI, 53706, USA{~} \label{ins116}} 

{117 : IRFU, CEA, Université Paris-Saclay, Bât 141, 91191 Gif-sur-Yvette, France{~} \label{ins117}} 

{118 : University of Białystok, Faculty of Physics, ul. K. Ciołkowskiego 1L, 15-245 Białystok, Poland{~} \label{ins118}} 

{119 : Department of Physics, Tokai University, 4-1-1, Kita-Kaname, Hiratsuka, Kanagawa 259-1292, Japan{~} \label{ins119}} 

{120 : Astronomical Observatory of Ivan Franko National University of Lviv, 8 Kyryla i Mephodia Street, Lviv, 79005, Ukraine{~} \label{ins120}} 

{121 : Institute for Space-Earth Environmental Research, Nagoya University, Chikusa-ku, Nagoya 464-8601, Japan{~} \label{ins121}} 

{122 : Kobayashi-Maskawa Institute (KMI) for the Origin of Particles and the Universe, Nagoya University, Chikusa-ku, Nagoya 464-8602, Japan{~} \label{ins122}} 

{123 : Department of Physics and Astronomy, University of California, Los Angeles, CA 90095, USA{~} \label{ins123}} 

{124 : Graduate School of Technology, Industrial and Social Sciences, Tokushima University, Tokushima 770-8506, Japan{~} \label{ins124}} 

{125 : Cherenkov Telescope Array Observatory, Saupfercheckweg 1, 69117 Heidelberg, Germany{~} \label{ins125}} 

{126 : University of Pisa, Largo B. Pontecorvo 3, 56127 Pisa, Italy{~} \label{ins126}} 

{127 : Dipartimento di Scienze Fisiche e Chimiche - Universit{\`a} degli Studi dell'Aquila, Via Vetoio 1, 67100 L'Aquila, Italy and INFN-Laboratori Nazionali del Gran Sasso (LNGS), via G. Acitelli 22, 67100 Assergi (AQ), Italy{~} \label{ins127}} 

{128 : Landessternwarte, Zentrum für Astronomie der Universität Heidelberg, Königstuhl 12, 69117 Heidelberg, Germany{~} \label{ins128}} 

{129 : Centre for Astro-Particle Physics (CAPP) and Department of Physics, University of Johannesburg, PO Box 524, Auckland Park 2006, South Africa{~} \label{ins129}} 

{130 : Departamento de Astronomía, Universidad de Concepción, Barrio Universitario S/N, Concepción, Chile{~} \label{ins130}}

{131 : National Astronomical Observatories, CAS, Beijing, China{~} \label{ins131}}

{132 : The University of Manitoba, Dept of Physics and Astronomy, Winnipeg, Manitoba R3T 2N2, Canada{~} \label{ins132}} 

{133 : University of Oxford, Department of Physics, Clarendon Laboratory, Parks Road, Oxford OX1 3PU, United Kingdom{~} \label{ins133}} 

{134 : Academic Computer Centre CYFRONET AGH, ul. Nawojki 11, 30-950 Cracow, Poland{~} \label{ins134}} 

{135 : Institute of Astronomy, Faculty of Physics, Astronomy and Informatics, Nicolaus Copernicus University in Toruń, ul. Grudziądzka 5, 87-100 Toruń, Poland{~} \label{ins135}} 

{136 : Warsaw University of Technology, Institute of Electronic Systems, Nowowiejska 15/19, 00-665 Warsaw, Poland{~} \label{ins136}}

{137 : Department of Physical Science, Hiroshima University, Higashi-Hiroshima, Hiroshima 739-8526, Japan{~} \label{ins137}} 

{138 : Department of Physical Sciences, Aoyama Gakuin University, Fuchinobe, Sagamihara, Kanagawa, 252-5258, Japan{~} \label{ins138}} 

{139 : Institute of Space Sciences (ICE-CSIC), and Institut d'Estudis Espacials de Catalunya (IEEC), and Institució Catalana de Recerca I Estudis Avançats (ICREA), Campus UAB, Carrer de Can Magrans, s/n 08193 Cerdanyola del Vallés, Spain{~} \label{ins139}} 

{140 : IRFU / DIS, CEA, Université de Paris-Saclay, Bat 123, 91191 Gif-sur-Yvette, France{~} \label{ins140}} 

{141 : INAF - Osservatorio Astrofisico di Torino, Strada Osservatorio 20, 10025 Pino Torinese (TO), Italy{~} \label{ins141}} 

{142 : CCTVal, Universidad Técnica Federico Santa María, Avenida España 1680, Valparaíso, Chile{~} \label{ins142}} 

{143 : Santa Cruz Institute for Particle Physics and Department of Physics, University of California, Santa Cruz, 1156 High Street, Santa Cruz, CA 95064, USA{~} \label{ins143}} 

{144 : School of Physics and Astronomy, Sun Yat-sen University, Zhuhai, China{~} \label{ins144}} 

{145 : Faculty of Science, Ibaraki University, Mito, Ibaraki, 310-8512, Japan{~} \label{ins145}} 

{146 : Faculty of Science and Technology, Universidad del Azuay, Cuenca, Ecuador{~} \label{ins146}}
}





\end{document}